\documentclass[a4paper,11pt]{article}
\usepackage{amsmath,amssymb}
\usepackage{graphics,graphicx}
\usepackage{booktabs}
\usepackage{hyperref}
\usepackage{a4wide}
\usepackage{cite}

\def\pt{\mathsf{t}}
\def\pG{\mathsf{G}}

\numberwithin{equation}{section}

\begin{document}
\thispagestyle{empty}
\begin{flushright}
UWThPh 2020-20
\end{flushright}
\vspace{8mm}
\begin{center}
{\LARGE\bf Sphere partition function of Calabi-Yau GLSMs}
\end{center}
\vspace{8mm}
\begin{center}
  {\large David Erkinger\footnote{{\tt
        david.josef.erkinger@univie.ac.at}}${}^{*}$, Johanna Knapp\footnote{{\tt
        johanna.knapp@unimelb.edu.au}}${}^{\dagger}$}
\end{center}
\vspace{3mm}
\begin{center}
  ${}^*$ {\em Mathematical Physics Group, University of Vienna\\ \vskip 0.1cm
  Boltzmanngasse 5, 1090 Vienna, Austria}\\\vskip 0.1cm
${}^{\dagger}$ {\em School of Mathematics and Statistics, University of Melbourne \\\vskip 0.1 cm
  Parkville 3010 VIC, Australia}
\end{center}
\vspace{15mm}
\begin{abstract}
  \noindent The sphere partition function of Calabi-Yau gauged linear sigma models (GLSMs) has been shown to compute the exact K\"ahler potential of the K\"ahler moduli space of a Calabi-Yau. We propose a universal expression for the sphere partition function evaluated in hybrid phases of Calabi-Yau GLSMs that are fibrations of Landau-Ginzburg orbifolds over some base manifold. Special cases include Calabi-Yau complete intersections in toric ambient spaces and Landau-Ginzburg orbifolds. The key ingredients that enter the expression are Givental's $I/J$-functions, the Gamma class and further data associated to the hybrid model. We test the proposal for one- and two-parameter abelian GLSMs, making connections, where possible, to known results from mirror symmetry and FJRW theory. 
\end{abstract}
\newpage
\setcounter{tocdepth}{1}
\tableofcontents
\setcounter{footnote}{0}
%%%%%%%%%%%%%%%%%%%%%%%%%%%%%%%%%%%%%%%%%%%%%%%%%%%%%%%%%%%%%%%%%%%%%%%%%%%%%
\section{Introduction and summary}
The richness of moduli spaces of string compactifications manifests itself in highly non-trivial dualities and correspondences and intricate underlying mathematical structures. The swampland program has shown that there is a deep connection between the mathematical properties of stringy moduli spaces and consistency requirements of theories of quantum gravity. This has provided new motivation to explore parameter spaces associated to string compactifications.

Naturally, the focus is on loci in the moduli space $\mathcal{M}$ where string compactifications are geometric. This is due to the fact that in geometric regions of the moduli space the tools to study string theory are best developed. This includes toric geometry, mirror symmetry, topological string theory, etc. However, geometric regions are very special and one may ask if the structures we know very well in geometry also exist elsewhere in $\mathcal{M}$. There are many reasons for the answer to be ``yes''. One of them is the worldsheet CFT of string theory that does not care whether it has a geometric space-time realisation or not, and structures such as those encoded in the $tt^*$-equations \cite{Cecotti:1991me} hold anywhere in $\mathcal{M}$. 
Also the fundamental structures responsible for the swampland constraints should be visible in all regions of the moduli space.  

In order to test such statements, in particular at the quantum level, one requires a concrete realisation of the worldsheet CFT that is valid at a specific locus in $\mathcal{M}$ and some neighbourhood parameterised by marginal deformations. Furthermore one needs the tools to carry out concrete calculations. In most regions of the moduli space quantum corrections are large, and suitable realisations of the CFT are unknown. Exceptions are certain limiting regions such as geometric ones where the worldsheet CFT is realised in terms of non-linear sigma models. Other loci of the moduli space that are fairly well-studied are Landau-Ginzburg (orbifold) points. The majority of limiting points will be neither geometric nor Landau-Ginzburg but some kind of hybrids thereof, or something even more general. If we are after structures that are the same everywhere in the moduli space the diversity of these models poses a problem. For instance, the mathematics and physics of a Landau-Ginzburg theory is very different from the mathematics and physics of a non-linear sigma model. To make connections between different loci of the moduli space, one requires suitable methods to connect well-studied geometric regions to non-geometric ones. 

The main focus of this work will be the K\"ahler moduli space $\mathcal{M}_K$ of a type II string compactification on a Calabi-Yau threefold. The K\"ahler moduli space is ``difficult'' in the sense that the physical observables receive quantum corrections through worldsheet instantons. Furthermore $\mathcal{M}_K$ decomposes into chambers. %Going from one chamber to another amounts to non-trivial analytic continuation.
Going from one chamber to another allows one to establish a connection between these observables at different limiting points.  

The stringy K\"ahler moduli space can be probed making use of the gauged linear sigma model (GLSM) \cite{Witten:1993yc} that provides a common UV description of the CFTs parameterised by $\mathcal{M}_K$. The different chambers in $\mathcal{M}_K$ correspond to different phases, i.e.~low-energy configurations, of the GLSM. The tools to compute quantum corrected observables in different regions of $\mathcal{M}_K$ come from supersymmetric localisation. It has been shown that the path integral evaluated in different (curved) backgrounds computes exact (instanton-corrected) quantities in Calabi-Yau compactifications. This includes the K\"ahler potential (sphere partition function) \cite{Benini:2012ui,Doroud:2012xw,Jockers:2012dk}, the elliptic genus (torus partition function) \cite{Gadde:2013dda,Benini:2013nda,Benini:2013xpa},  D-brane central charge and open Witten index (hemisphere and annulus partition function) \cite{Sugishita:2013jca,Honda:2013uca,Hori:2013ika}, and correlation functions (including Yukawa couplings) \cite{Closset:2015rna}. In geometric regions these results can be checked against results from mirror symmetry. It is expected that the partition functions compute analogous quantities in non-geometric phases of the GLSM. This was for instance shown in the context of the sphere partition function \cite{Gomis:2012wy} which was connected to the K\"ahler potential on $\mathcal{M}_K$ via $tt^*$-geometry. New derivations via anomalies of the $(2,2)$ theory were given in \cite{Gerchkovitz:2014gta,Gomis:2015yaa}. The results from supersymmetric localisation are a strong hint that the structure of these objects must be similar in different phases, because the expressions have the same UV origin. 

In \cite{hrprimitive,Knapp:2020oba} it was proposed that the hemisphere partition function of a Calabi-Yau GLSM, which conjecturally computes the exact central charge of a D-brane, has the same structure in every phase. This was shown to hold for geometric and Landau-Ginzburg phases. The ingredients that enter into the expression for the hemisphere partition function are a state space associated to the phase and a non-degenerate pairing, the Gamma class, Givental's $I$/$J$-functions \cite{Givental:2004ab}, and the Chern character of the brane. The mathematical formalism required to understand the result is FJRW theory \cite{Fan:2007vi,Chiodo:2012qt}. It defines enumerative invariants in Landau-Ginzburg orbifolds and combines Givental's mirror construction with the Landau-Ginzburg/Calabi-Yau correspondence to establish a connection between Gromov-Witten theory and FJRW theory at genus $0$. These mathematical results thus give natural expressions and structures that are valid beyond geometric regions in the moduli space, and the supersymmetric partition functions can be expressed in terms of them. Further note that the FJRW formalism also has been developed for certain classes of hybrid models \cite{MR3590512,MR3842378,clader2018wallcrossing,zhao2019landauginzburgcalabiyau} and general statements about state spaces have been given
in \cite{Chiodo_2018}.

In this work we consider the sphere partition function. Based on the examples we have analysed, we found that in a hybrid-type phase, that is realised as a Landau-Ginzburg orbifold model with superpotential $W$ and orbifold group $G$ fibered over a base manifold $B$, the sphere partition function takes the following universal form:
\begin{equation}
  \label{zs2formula}
  Z_{S^2}^{phase}(\pt,\overline{\pt})=C \sum_{\delta\in G}\int_{B}
  (-1)^{\mathrm{Gr}}\frac{\widehat{\Gamma}_{\delta}(H)}{\widehat{\Gamma}_{\delta}^*(H)}I_{\delta}(u(\pt),H){I}_{\delta}(\overline{u}(\overline{\pt}),H)=\langle \overline{I},I\rangle, 
\end{equation}
where $\pt$ is the FI-theta parameter of the GLSM and $\overline{\pt}$ is its complex conjugate. In the first equality, the sum over $\delta\in G$ is over a subset of twisted sectors of the orbifold group referred to as narrow sectors in the mathematics literature,
$\mathrm{Gr}$ is (the eigenvalue of) a grading operator acting on the narrow state space and its eigenvalues are visible in (\ref{zs2formula}) in orbifold-type phases. It is somewhat hidden in geometric settings, see Section \ref{subsec-geometry}.  
Furthermore, we collectively denote the generators of the K\"ahler cone in $H^2(B)$ by $H$. $\widehat{\Gamma}_{\delta}(H)$ and $\widehat{\Gamma}^*_{\delta}(H)$ denote the component of the Gamma class associated to the twisted sector $\delta$ and its conjugate, and $I_{\delta}(u(\pt),H)$ is the component of Givental's $I$-function associated to the sector $\delta$. There is also a $J$-function $J_{\delta}(u(\pt),H)$ that is related to the $I$-function by a change of frame and coordinates. Both, the $I$-function and the Gamma class can be decomposed further with respect to a basis of $H^2(B)$. The $I$-function depends on the local coordinate $u(\pt)$ of the phase. By ${I}_{\delta}(\overline{u}(\overline{\pt}),H)$ we mean taking the $I$-function and replacing $u\rightarrow\overline u$. Geometric phases and Landau-Ginzburg phases correspond to special cases: in the Landau-Ginzburg case $B$ is a point, whereas in the Calabi-Yau case $B$ is the Calabi-Yau itself and $G$ is trivial. The constant $C$ is a normalisation constant. In geometric phases these structures, and in particular the appearance of the $I$-function, have been observed before \cite{Bonelli:2013mma,Ueda:2016wfa,Kim:2016jye,Gerhardus:2018zwb,Goto:2018bci,Honma:2018fgw}. The quotient of Gamma classes has been analysed in \cite{Halverson:2013qca} at the perturbative level. The final equality in (\ref{zs2formula}) is to be understood as follows: $\langle\cdot,\cdot\rangle$ is the topological pairing on the state space of the theory in the phase, $|I\rangle$ is an expansion of the $I$-function in terms of this basis, $\langle\overline{I}|$ is the complex (CPT) conjugate of $|I\rangle$ in the sense of the $tt^*$-formalism. Further clarification on the the parings, in particular the topological vs.~the hermitean pairing, will be given in Section~\ref{sec-formula}. In the following sections we will give further details on how to understand this expression and collect evidence by considering several classes of examples.

The article is organised as follows. In Section \ref{sec-glsmtt} we recall the basic definitions of the GLSM and the sphere partition function. We furthermore review the definition of the K\"ahler potential of $\mathcal{M}_K$ in the context of $tt^*$-geometry. In Section \ref{sec-formula} we give more details on the proposal (\ref{zs2formula}), in particular in Landau-Ginzburg and geometric settings. In the remaining sections we study examples. Section \ref{sec-onepar} focuses on a well-studied class of fourteen one-parameter GLSMs whose large volume phases are Calabi-Yau hypersurfaces and complete intersections in toric ambient spaces. These models have already played a role in one of our previous work \cite{Erkinger:2019umg} to which we refer for technical details on the sphere partition function. These models are particularly interesting as they have different types of non-geometric phases, including Landau-Ginzburg orbifold and hybrid phases, that we can test (\ref{zs2formula}) for and where we have additional means of cross-checking the result, for instance via mirror symmetry or FJRW theory. There are also more exotic phases, called pseudo-hybrids, where we encounter structures similar to (\ref{zs2formula}). In Section \ref{sec-twopar} we consider a two-parameter model where we in particular conjecture new expressions for the $I$-function and the Gamma class in hybrid phases. Further technical details on the computations can be found in the Appendix.\\\\ 
{\bf Acknowledgements:} We would like to thank Mauricio Romo, Emanuel Scheidegger, and Thorsten Schimannek for discussions and comments on the manuscript. JK would like to thank Ilarion Melnikov for correspondence. DE thanks Urmi Ninad for discussions and the School of Mathematics and Statistics of the University of Melbourne for hospitality during a short-term stay. DE acknowledges financial support by the Vienna Doctoral School in Physics (VDSP). The authors were partially supported by the Austrian Science Fund (FWF): [P30904-N27]. All data generated or analysed during this study are included in this article. 
%%%%%%%%%%%%%%%%%%%%%%%%%%%%%%%%%%%%%%%%%%%%%%%%%%%%%%%%%%%%%%%%%%%%%%%%%%%%%%
\section{Sphere partition function and $tt^*$}
\label{sec-glsmtt}
In this section we review the definition of the sphere partition function and its connection to the exact K\"ahler potential $K(t,\overline{t})$ on $\mathcal{M}_K$ \cite{Jockers:2012dk,Gomis:2012wy} in phases of Calabi-Yau GLSMs. We also recall the worldsheet definition of $K(t,\overline{t})$ in terms of $tt^*$-geometry \cite{Cecotti:1991me}. 
The power of Givental's formalism combined with FJRW theory is that it also applies beyond geometric settings, notably Landau-Ginzburg orbifold phases \cite{Fan:2007vi,Chiodo:2012qt} and certain types of hybrid phases \cite{MR3590512,MR3842378,clader2018wallcrossing,zhao2019landauginzburgcalabiyau}. This provides a framework to define and compute the objects entering (\ref{zs2formula}). First we give more details on Landau-Ginzburg models where explicit expressions for the ingredients of (\ref{zs2formula}) have been given recently \cite{Knapp:2020oba}. Then we comment on geometric and hybrid phases. 
%%%%%%%%%%%%%%%%%%%%%%%%%%%%%%%%%%%%%%%%%%%%%%%%%%%%%%%%%%%%%%%%%%%%%%%%%%%%%%
\subsection{GLSM and sphere partition function}
We consider a GLSM with gauge group $\pG$. The scalar components $\phi_i$ of the chiral multiplets are coordinates on a complex vector space $V$ (i.e.~they take values in $V^{*}$), with $i=1,\ldots,\mathrm{dim}V$. In the case of a Calabi-Yau GLSM they transform in the representation $\rho_V:\pG\rightarrow SL(V)$. We further need the vector $U(1)$ R-symmetry $R:U(1)_V\rightarrow GL(V)$. The gauge and R-charges of the $\phi_i$, denoted by $Q_i$ and $R_i$ respectively, are the weights of these representations. The gauge charges can be organised into a $\mathrm{rk}\pG\times \mathrm{dim}V$-matrix $\mathsf{C}$.  We will consider models with non-vanishing superpotential $W\in\mathrm{Sym}V^*$. The FI-parameters $\zeta$ and the theta angles $\theta$ combine into the complexified K\"ahler parameters $\pt=2 \pi \zeta-i\theta\in \mathfrak{g}^*_{\mathbb{C}}$ where \(\mathfrak{g}\) is the Lie algebra of $\pG$. Furthermore we denote by $\mathfrak{t}$ the Lie algebra of a maximal torus of $\pG$. The scalar components of the vector multiplet are denoted by $\sigma\in\mathfrak{g}_{\mathbb{C}}$. There is a natural pairing $\langle,\rangle:\mathfrak{g}_{\mathbb{C}}\times \mathfrak{g}^*_{\mathbb{C}}\rightarrow\mathbb{C} $. The sphere partition function is defined as
\begin{align}
\begin{split}
 Z_{S^2} ({\zeta}, {\theta}) &=  \frac{1}{(2 \pi)^{\text{dim}~\mathfrak{t}}|\mathcal{W}|} \sum\limits_{m} \int\limits^{i \infty}_{-i \infty} \mathrm{d}^{\text{dim} ~\mathfrak{t}}{\sigma} 
 \prod\limits_{\alpha>0} (-1)^{\langle \alpha, m\rangle} \left( \frac{1}{4} \langle \alpha, m \rangle^2 + \langle \alpha, {\sigma} \rangle^2 \right) \\
 &
 \frac{\Gamma\left(\frac{1}{2} {R}_j - i\langle Q_j, {\sigma} \rangle  - \frac{1}{2} \langle Q_j, m\rangle\right)}
  {\Gamma\left(1-\frac{1}{2} {R}_j +i \langle Q_j, {\sigma} \rangle  - \frac{1}{2} \langle Q_j, m\rangle\right)}
 e^{-4 \pi i \langle {\zeta},{ \sigma} \rangle- i \langle {\theta},m \rangle}
\end{split}
  \end{align}
where $\alpha>0$ denotes the positive roots of $\pG$ and the $m\in\mathbb{Z}^{\mathrm{dim}\mathfrak{t}}$, taking values on the coroot lattice of $\pG$, account for the discrete values of the gauge field strength on the sphere. \(|\mathcal{W}|\) is
the cardinality of the Weyl group.

The convergence of this integral is governed by the factor \(e^{-4 \pi i \langle \zeta,\sigma \rangle}\) and thus by the choice of phase. To evaluate the integral in a given phase, we have to choose an integration contour that does not hit any of the poles and that leads to a convergent result $Z_{S^2}^{phase}$ for the integral. Evaluating integrals of this type can be quite challenging in the multi-dimensional case. A prescription can be found in \cite{MR1631772}, see also \cite{Gerhardus:2015sla} for a review in the context of the sphere partition function. 
%%%%%%%%%%%%%%%%%%%%%%%%%%%%%%%%%%%%%%%%%%%%%%%%%%%%%%%%%%%%%%%%%%%%%%%%%
\subsection{\(t t ^\ast \)-geometry}
Originally  \(t t^\ast\)-geometry  was  studied in
\cite{Cecotti:1991me}. In our discussion we mostly follow
\cite{Bershadsky:1993cx,Cecotti:2013mba,Hori:2003ic,Alim:2012gq}. For a review in the spirit of this paper see \cite{Knapp:2020oba}.   We consider an \(\mathcal{N}=(2,2)\) theory in two 
dimensions with a mass gap. The nilpotency of the supercharges makes it possible to study cohomologies of 
operators and states with respect to certain combinations of the supercharge operators. In total there are four different cohomologies in the NS-sector of the theory denoted by
\begin{align}
(c,c), \quad (a,c), \quad (a,a), \quad (c,a),
\end{align} 
where \(c\) stands for chiral and \(a\) for 
anti-chiral. The charge conjugates of \((c,c)\), 
\((a,c)\) operators are of type  
 \((a,a)\) and \((a,c)\), respectively.  The structures of 
the four different cohomologies are related by 
spectral flow \cite{Schwimmer:1986mf,Lerche:1989uy} and therefore
we focus on \((c,c)\) and the conjugate \((a,a)\).
From these 
operators it is possible to construct deformations of the theory. 
Let \(t^i, \bar{t}^{i}\) be the parameters describing the exactly marginal deformations. % -- the only type of deformations we will consider in the following.
These take values in a coordinate patch of the moduli
space \(\mathcal{M}\) of the theory.
The space of (anti-)chiral operators has a ring structure 
\begin{align}
\phi_i \phi_k &= C_{ik}^l \phi_l, &
\bar{\phi}_{i} \bar{\phi}_k &= \bar{C}_{ik}^l \bar{\phi}_l.
\end{align} 
The \(C_{ik}^l\) (\(\bar{C}^l_{ik}\)) are functions of \(t^i\)
(\(\bar{t}^i\)). The chiral algebra is represented on the ground
states \(|k \rangle\) of the theory: 
\begin{align}
\phi_i | k \rangle = C_{ik}^l | l \rangle.
\end{align} 
If we now change the parameters \(t_i, \bar{t}_i\) the
ground-states will vary in the full Hilbert space
of the theory. This is denoted by  \(|i(t,\bar{t})\rangle\). The ground states are sections of the
ground state bundle \(\mathcal{V}\). We can introduce a connection as follows
\begin{align}
  \frac{\partial}{\partial t^i} |k (t, \bar{t})\rangle
&=\left( A_i\right)^l_k |l(t,\bar{t}) \rangle, &
\frac{\partial}{\partial \bar{t}^i} |k (t, \bar{t})\rangle
  &= \left(\bar{A}_i\right)^l_k  |l(t,\bar{t}) \rangle.
\end{align}
 We will denote the associated covariant derivative by
\begin{align}
D_i &= \frac{\partial}{\partial t^i} - A_i, &
\bar{D}_i &= \frac{\partial}{\partial \bar{t}^i}
- \bar{A}_i.
\end{align} 
To get a basis of ground-states in the Ramond-sector a topological or anti-topological twist of the theory is performed and the path-integral with the respective operator insertion is evaluated on a hemisphere, which is deformed into a cigar-shaped geometry. By
application of a topological twist one gets a
holomorphic basis, which we denote by \(|i \rangle\). In this 
basis the anti-holomorphic part of the connection vanishes
 \begin{align}
(\bar{A}_i)^l_k &= 0.
\end{align}
 An anti-topological twist gives an anti-holomorphic basis
\(|\bar{i} \rangle\), with 
\((A_i)^{\bar{l}}_{\bar{k}} = 0\).
The various ground states are obtained by insertion of
(anti-)chiral operators into the path integral. There is a distinguished ground
state that is denoted by
\(|0 \rangle\) in a topological  theory and \(|\bar{0}\rangle\)
in the anti-topological  theory. There are two possible
pairings on this bundle, depending on the chosen basis, a purely
topological one 
\begin{align}
\eta_{ij} &= \langle j |i \rangle,
\label{eqn:ttbarTopologicalMetric}
\end{align} and a hermitian one 
\begin{align}
g_{i \bar{j}} &= \langle \bar{j}  | i \rangle.
\label{eqn:ttbarHermitianMetric}
\end{align}
In the following we will often write $\langle\cdot,\cdot\rangle$ for the topological pairing and $\langle\cdot|\cdot\rangle$ for the hermitean pairing. Both pairings can be obtained by computing the path integral on the
sphere, with the appropriate operator insertions. The 
topological metric (\ref{eqn:ttbarTopologicalMetric}) is obtained by
sewing two topologically twisted path integrals on the hemisphere and \(g\)
by gluing two path integrals on the hemisphere where in one an
anti-topological twist has been applied. Both, \(|i \rangle\) 
and \(|\bar{j}\rangle\) are a basis of the same
space and therefore they must be related by a change of basis 
\begin{align}
  \label{mmatrix}
|\bar{j} \rangle &= M_{\bar{j}}^i | i \rangle.
\end{align}
 \(M\) encodes the action of CPT conjugation and therefore it must
fulfil
 \begin{align}
M M^\ast = 1.
\end{align}
 The whole structure of the ground state bundle is encoded in the  \(t t^{\ast}\)-equations \cite{Cecotti:1991me}:
 \begin{align}
[D_i, D_j] &= 0, & [D_i,\bar{D}_j] &= - [C_i, \bar{C}_j], &
[\bar{D}_i, \bar{D}_j] &= 0,\\
[D_i, C_j] &= [D_j,C_i], &
%[\bar{D}_i, \bar{C}_j] &=
%[\bar{D}_j, \bar{C}_i], &
[D_i,\bar{C}_j] &= [\bar{D}_i,C_j]=0, &
[\bar{D}_i, \bar{C}_j] &= [\bar{D}_j, \bar{C}_i], \\
[C_i, C_j] &= 0, & & &
[\bar{C}_i, \bar{C}_j] &= 0.
\end{align}
 As one can prove by using the \(t t^\ast\)-equations, it is
possible to introduce a covariant derivative \(\nabla_i, \nabla_{\bar{i}}\) with
vanishing curvature on \(\mathcal{V}\):
\begin{align}
\nabla_i = D_i - C_i.
\end{align} 
%The new connection allows us to identify all fibres of \(\mathcal{V}\) with a fibre over a chosen point by parallel transport. Locally, this allows the interpretation of  \(\mathcal{V}\)  as a product bundle with a fixed fibre \(V\), the vector space of ground states. In this setting \(\nabla_i, \nabla_{\bar{i}}\) are simply
%given by the ordinary derivatives \(\partial_i, \partial_{\bar{i}}\). On
%\(\mathcal{V}\) a real structure is given by declaring states which are
%invariant under CPT as real.

The flatness of the connection allows to identify the fibres of $\mathcal{V}$ with a fixed fibre $V$ at a chosen point by parallel transport. Choose $V$ to be the vector space of ground states. $\nabla_i$, $\nabla_{\bar{i}}$ reduce to the ordinary derivatives $\frac{\partial}{\partial t^i}$, $\frac{\partial}{\partial\bar{r}^i}$ in this setup. CPT provides a real structure on $\mathcal{V}$, by declaring CPT invariant states as real. 

Let us now focus on theories with a \(\mathcal{N}=(2,2)\)
superconformal symmetry with\footnote{This is 
	related to the central charge \(c\) of the superconformal algebra by \(c=3 \hat{c}\).}
 \(\hat{c}=3\). Of particular interest 
are chiral fields with conformal dimension \((\frac{1}{2},\frac{1}{2})\) which are the exactly marginal fields. Deformations constructed 
from these operators thus preserve the conformal symmetry. 
%Additionally, these operators are among the generators of a ring $\mathcal{H}^{def}$ that we will focus on from now on. 
We introduce a fixed basis of real vectors %\(|\alpha \rangle\)
%(\(\alpha = 1, \dots, 2m+2)\)
\begin{equation}
  \label{cftbasis}
  \{|0\rangle,|a_1\rangle,\ldots,|a_m\rangle,|a^1\rangle,\ldots,|a^m\rangle,|\Omega\rangle \}
  \end{equation}
on \(V\), given \(m\) marginal directions. In this basis CPT conjugation is complex conjugation. The basis consists of the unique ground state $|0\rangle$ with no insertion, the states corresponding to the marginal fields, their 
duals with respect to (\ref{eqn:ttbarTopologicalMetric}), and 
the unique ground state $\Omega$ corresponding to the chiral field 
with conformal dimension \((\frac{3}{2},\frac{3}{2})\).  %We thus have $\mathrm{dim}\mathcal{H}^{def}=2m+2$.
In the case of a SCFT with $\hat{c}=3$ the bundle $V$ decomposes into
\begin{equation}
  V=\mathcal{L}\oplus(\mathcal{TM}\otimes \mathcal{L})\oplus \overline{(\mathcal{TM}\otimes \mathcal{L})}\oplus\overline{\mathcal{L}},
  \end{equation}
where $\mathcal{L}$ is the line bundle corresponding to the state $|0\rangle$. The fibres of $(\mathcal{TM}\otimes \mathcal{L})$ are spanned by the $|a_i\rangle$, where $\mathcal{TM}$ is the holomorphic tangent space of $\mathcal{M}$. The conjugate bundles are spanned by the states
\begin{equation}
  |a_{\bar{i}}\rangle=g_{\bar{i}k}|a^k\rangle, \qquad |\bar{0}\rangle=g_{\bar{0}0}|0\rangle,
  \end{equation}
using $(\ref{eqn:ttbarHermitianMetric})$. By restricting the indices $i,j$ to the marginal deformations, we obtain the Zamolodchikov metric \cite{zamolodchikov1986,Cecotti:1991me}:
\begin{align}
G_{i \bar{\jmath}} &= \frac{g_{i \bar{\jmath}}}{\langle \bar{0}| 0 \rangle}.
\end{align}
It follows from the $tt^*$- equations that 
\begin{align}
G_{i \bar{\jmath}} &= - \partial_i \partial_{j} \log \langle\bar{0} |0 \rangle.
\end{align}
This result allows the following interpretation 
\begin{align}
  \label{kahlercft}
e^{-K(t, \bar{t})} = \langle \bar{0} | 0\rangle,
\end{align}
where \(K(t, \bar{t})\) is the K\"ahler potential of \(G_{i \bar{\jmath}}\). The Zamolodchikov metric gives 
the natural metric on the moduli space of \(\mathcal{N}=(2,2)\)
superconformal theories.  

Returning to phases of the GLSM, it was conjectured in \cite{Jockers:2012dk} that the sphere partition function of 
the GLSM calculates the exact K\"ahler potential of 
the moduli space of the Calabi-Yau target space. In
\cite{Jockers:2012dk} the conjecture was tested in examples 
 with the help of mirror symmetry. In \cite{Gomis:2012wy}
the conjecture was verified using $tt^*$-geometry. 
%%%%%%%%%%%%%%%%%%%%%%%%%%%%%%%%%%%%%%%%%%%%%%%%%%%%%%%%%%%%%%%%%%%%%%%%%%%%
We thus have two ways to define the K\"ahler potential on $\mathcal{M}_K$. The first via the GLSM:
\begin{equation}
  \label{zs2kahler}
  Z^{phase}_{S^2}(\pt,\overline{\pt})=e^{-K(\pt,\overline{\pt})}.
\end{equation}
On the other hand we have (\ref{kahlercft}) via $tt^*$-geometry.
Before we conclude
\begin{equation}
   Z^{phase}_{S^2}(\pt,\overline{\pt})=\langle \overline{0}|0\rangle,
  \end{equation}
let us clarify  the meaning of the coordinates $\pt$ and $t$ appearing in (\ref{zs2kahler}) and (\ref{kahlercft}). In the worldsheet CFT the ``flat coordinates'' $t$ correspond to the deformation parameters associated to the marginal deformations. They are required, for instance, to extract the information about enumerative invariants from the K\"ahler potential. These are not the FI-theta parameters $\pt$ of the GLSM. The two choices of coordinates are related by a coordinate change. In geometric phases and Landau-Ginzburg phases it is known how to extract this information from the results of supersymmetric localisation \cite{Jockers:2012dk,Knapp:2020oba}. 
It coincides with the mirror map and exchanges $I$- and $J$-functions. FJRW theory gives prescriptions to compute this map in more general settings. The GLSM is thus a means to compute $\langle \overline{0}|0\rangle$ exactly for different realisations of worldsheet CFTs.

%%%%%%%%%%%%%%%%%%%%%%%%%%%%%%%%%%%%%%%%%%%%%%%%%%%%%%%%%%%%%%%%%%%%%%%%
\section{Universal expression for $Z_{S^2}$ in phases of GLSMs}
\label{sec-formula}
We observe that, given a Calabi-Yau GLSM, the sphere partition function in a phase that is a Landau-Ginzburg orbifold with orbifold group $G$ fibered over a base manifold $B$ can always be written in the form (\ref{zs2formula}) that we repeat here for convenience:
\begin{equation}
  Z_{S^2}^{phase}(\pt,\overline{\pt})=C \sum_{\delta\in G}\int_{B}
  (-1)^{\mathrm{Gr}}\frac{\widehat{\Gamma}_{\delta}(H)}{\widehat{\Gamma}_{\delta}^*(H)}I_{\delta}(u(\pt),H){I}_{\delta}(\overline{u}(\overline{\pt}),H)=\langle \overline{I},I\rangle. 
\end{equation}

To give more details on the last equality, we expand the $I$-function in terms of a basis of the state space. Here we have to make an important restriction. From now on we will focus on ``narrow'' states which belong to a specific subset of the states corresponding to the marginal deformations. Conditions to identify ``narrow'' states in different types of phases will be given in the subsections below. Given $h\leq m$ narrow marginal deformations\footnote{Note that in all our examples $h=m$.}, the state space reduces to a $2h+2$-dimensional space which we will denote by $\mathcal{H}$ and whose basis elements we denote by ${e}_r$. Comparing with (\ref{cftbasis}), there are two distinguished basis elements that are identified with $\{|0\rangle,|\Omega\rangle\}$, respectively, and $2h$ elements associated to those $\{|a_i\rangle,|a^i\rangle\}$ that are narrow. We will further denote by $\mathcal{H}_{nar}$ the $h$-dimensional subspace corresponding to the narrow deformations. Then we can expand the $I$-function as follows:
\begin{equation}
  |I\rangle=\sum_{r}I_re_{r}. 
\end{equation}
In the context of the sphere partition function the question is what is the complex (CPT) conjugate of this expression. Results from geometry \cite{Halverson:2013qca} and the examples discussed below suggest the definition
\begin{equation}
  \label{idual}
  \langle\overline{I}|=\sum_{r}\overline{I}_re^*_{r}, \qquad \overline{I}(\overline{u})=(-1)^{\mathrm{Gr}}\frac{\widehat{\Gamma}}{\widehat{\Gamma}^*}I(\overline{u}),
\end{equation}
where $e^*_r$ is the dual of $e_r$ such that $\langle e^*_{r'},e_r\rangle=c\cdot\delta_{r,r'}$ with some normalisation constant $c$. In the case of hybrid models this may have to be modified depending on the pairing that is used. See Section \ref{sec-hybgeneral} for some comments.
Note that there are two pairings at play: one is the hermitian pairing $\langle\cdot|\cdot\rangle$ induced by (\ref{eqn:ttbarHermitianMetric}) that naturally appears in the definition of $e^{-K(t,\overline{t})}$, the other one is a topological pairing $\langle\cdot,\cdot\rangle$ induced by (\ref{eqn:ttbarTopologicalMetric}). Working with the $I$-function, it is natural to use the topological pairing. This suggests that the relation (\ref{idual}) is a realisation of the matrix $M$ (\ref{mmatrix}) that implements CPT conjugation so that one formally has
\begin{equation}
  \langle\overline{I}|I\rangle:=\langle I(\overline{u})M,I(u)\rangle\equiv \langle\overline{I},I\rangle.
\end{equation}
By the last equivalence we mean that we absorb the action of $M$ in the definition of $\overline{I}$ as indicated in (\ref{idual}) when we write $\langle \overline{I},I\rangle$.  Similar observations have been made in \cite{Knapp:2020oba} in the context of the D-brane central charge, where spectral flow was required to relate the pairing between the $(a,c)$- and $(c,c)$-rings to the topological pairing. 

%%%%%%%%%%%%%%%%%%%%
%Since $I$ can be expanded in terms of a basis of $\mathcal{H}^{def}$ and the Gamma class can be interpreted as an endomorphism on $\mathcal{H}^{def}$ we can write the sphere partition function, where the pairing is evaluated, as follows
%Another way to write $Z_{S^2}^{phase}$ reflects the mirror picture where $e^{-K}=\overline{\varpi} \Sigma\varpi$, where $\varpi$ is a basis of periods of the mirror Calabi-Yau and $\Sigma$ is a symplectic matrix. Since the components of $I$ are nothing but the mirror periods,

Another way to write the information in $Z_{S^2}^{phase}$ is as follows. We interpret $I$ as a  $2h+2$-dimensional column vector. %Furthermore we can define a row vector $\overline{I}$ that is the complex conjugate transpose of $I$ rather than (\ref{idual}).
Then $(-1)^{\mathrm{Gr}}\frac{\widehat{\Gamma}}{\widehat{\Gamma}^*}$ and the pairing can be represented as a $(2h+2)\times (2h+2)$-matrix $M$ and we can write
\begin{equation}
  \label{zs2matrix}
  Z_{S^2}^{phase}=I(\overline{u})^TMI(u),
  \end{equation}
%where $I$ and $\overline{I}$ are understood as $\mathrm{dim}\mathcal{H}^{def}$-dimensional column and row vectors, respectively, whose entries are related by complex conjugation.  and $M$ is a $\mathrm{dim}\mathcal{H}^{def}\times \mathrm{dim}\mathcal{H}^{def}$-matrix that encodes $(-1)^{\mathrm{Gr}}\frac{\widehat{\Gamma}}{\widehat{\Gamma}^*}$. 
We will see in the examples that the structure of the matrix $M$ depends on the type of phase and that its entries are, at least in the examples we have considered, consistent with the components of $(-1)^{\mathrm{Gr}}\frac{\widehat{\Gamma}}{\widehat{\Gamma}^*}$ and the pairing. We note that (\ref{zs2matrix}) has been observed before in the context of mirror symmetry, where the components of $I$ have an interpretation as periods of the mirror Calabi-Yau. Indeed, for the case of the quintic, the matrix $M$ is related, up to a choice of normalisation, to a matrix ``$\sigma_{rs}$'' defined in Section 4 of \cite{Candelas:1990rm}. 

To get to the flat coordinates, we denote by $I_0$ the component of the $I$-function that corresponds to the unique ground state $|0\rangle$ and by $I_j$ $(j\in1,\ldots,h)$ the components that capture the narrow deformations.  %one singles out two subspaces of the state space: a one-dimensional subspace  corresponding to the unique ground state $|0\rangle$, and a subspace that captures the marginal deformations. Call the corresponding components of the $I$-function $I_0$ and $I_{j}$.
Then the flat coordinates are defined by
\begin{equation}
  t_j(u)=\frac{I_j}{I_0}.
\end{equation}
The $J$-function is then defined as
\begin{equation}
  J(t(u))=\frac{I}{I_0}.
\end{equation}
The transition from the $I$-function to the $J$-function thus corresponds to a change of normalisation of the sphere partition function:
\begin{equation}
  \widetilde{Z}_{S^2}^{phases}(t,\overline{t})=C\sum_{\delta\in G}\int_B
(-1)^{\mathrm{Gr}}  \frac{\widehat{\Gamma}_{\delta}(H)}{\widehat{\Gamma}^*_{\delta}(H)}\frac{I_{\delta}(u(t),H)I_{\delta}(\overline{u}(\overline{t}),H)}{I_0(u(t))\overline{I}_0(\overline{u}(\overline{t}))}=\langle\overline{J},J\rangle.
\end{equation}
This amounts to a K\"ahler transformation. These structures can be used to extract enumerative invariants from the GLSM partition functions \cite{Jockers:2012dk,Knapp:2020oba} that encode the $I$-function.

In the following we make the discussion more precise for specific types of phases.
%%%%%%%%%%%%%%%%%%%%%%%%%%%%%%%%%%%%%%%%%%%%%%%%%%%%%%%%%%%%%%%%%%%%%%%%%%%%%%%
\subsection{Landau-Ginzburg orbifolds and FJRW theory}
\label{sec:lgFJRWBackground}
A convenient class of models to test this conjecture are Landau-Ginzburg orbifolds since we can check the results of the sphere partition function against the definitions of the Gamma class and $I$-function that has been defined in FJRW theory \cite{Fan:2007vi,Chiodo:2012qt}. In \cite{Knapp:2020oba} it was shown how this information is encoded in the Landau-Ginzburg data to which we refer to for details.

We consider a Landau-Ginzburg orbifold with orbifold group $G$ with $N$ fields $x_i$ and holomorphic, quasi-homogeneous, $G$-invariant superpotential\footnote{To avoid cluttered notation we denote the superpotantiels in Landau-Ginzburg models, hybrids and GLSMs with the same letter $W$. We hope the distinction is clear from the context.} $W$ satisfying $dW^{-1}(0)=\{0\}$. Let the $x_i$ have left R-charge $q_i$ so that the superpotential has left R-charge $1$: $W(\lambda^{q_i}x_i)=\lambda W(x_i)$. The vector R-charge of $W$ is $2$. If $W$ is of degree $d$ this implies that there is a $\mathbb{Z}_d$-orbifold action   $\langle J\rangle$ with $J=\left(e^{2\pi i q_1},\ldots, e^{2\pi i q_N}\right)$. In this work we restrict ourselves to Landau-Ginzburg orbifolds with $G=\langle J\rangle$, even though the subsequent statements are more general \cite{Knapp:2020oba}.

The state space $\mathcal{H}^{LG}$ consists of $\gamma$-twisted sectors \cite{Vafa:1989xc,Intriligator:1990ua}
\begin{equation}
  \mathcal{H}^{LG}=\bigoplus_{\gamma\in G}\mathcal{H}_{\gamma},
\end{equation}
where each $\mathcal{H}_{\gamma}$ is made up of fields that satisfy untwisted boundary conditions in the $\gamma$-twisted sector. For our choice of $G$ we can write $\gamma=J^{\ell}$ ($\ell=0,\ldots,d-1$). Then the untwisted boundary conditions are given by $x_i(e^{2i\pi}z)=e^{2\pi i q_i\ell}x_i(z)$ with $q_i\ell\in\mathbb{Z}$. One then considers the $G$-invariant states built out of these fields. Among the states of $\mathcal{H}^{LG}$ one can identify the (ground-)states $|0\rangle^{(c,c)}_{\gamma},|0\rangle^{(a,c)}_{\gamma}$ in the $(c,c)$- and $(a,c)$-rings, and the RR ground states $|0\rangle_{\gamma}^R$. They are isomorphic via spectral flow \cite{Lerche:1989uy}: 
\begin{equation}
  \label{ringflow}
  \mathcal{U}_{\left(-\frac{1}{2},-\frac{1}{2}\right)}|0\rangle^{(c,c)}_{\gamma}=|0\rangle_{\gamma}^R, \qquad \mathcal{U}_{(-1,0)}|0\rangle^{(c,c)}_{\gamma}=|0\rangle^{(a,c)}_{\gamma J},
\end{equation}
where $\mathcal{U}_{(r,\overline{r})}$ is the spectral flow operator with R-charges $(\hat{c}r,\hat{c}\overline{r})$ with $\hat{c}=\sum_{i=1}^N(1-2q_i)$. The elements of the $(c,c)$-ring have an explicit expression in terms of $G$-invariant monomials of the Jacobi ring of $W_{\gamma}=W|_{\mathrm{Fix}\gamma}$ where $\mathrm{Fix}\gamma$ is defined as the set of $x_i$ fixed by the action of $\gamma$. Via spectral flow one gets an indirect description of the other states. The left and right R-charges $(q,\overline{q})$ of the vacuum states are the eigenvalues of the generators $F_{L/R}$ of the left and right moving R-symmetries:
\begin{align}
  F_{L}|0\rangle_{\gamma}=&\left(\mathrm{age}(\gamma)-\frac{N}{2}+\sum_{j:\ell q_j\in\mathbb{Z}}q_{j}+\frac{\hat{c}}{2}\right)|0\rangle_{\gamma}\nonumber\\
F_{R}|0\rangle_{\gamma}=&\left(-\mathrm{age}(\gamma)+\frac{N}{2}-n_{\gamma}+\sum_{j:\ell q_j\in\mathbb{Z}}q_{j}+\frac{\hat{c}}{2}\right)|0\rangle_{\gamma},
  \end{align}
with
\begin{equation}
  \mathrm{age}(\gamma)=\sum_{j}q_j, \qquad n_{\gamma}=\mathrm{dim}(\mathrm{Fix}(\gamma)).
  \end{equation}

In the following we will restrict to narrow sectors. We will refer to those sectors of the $(a,c)$-ring as narrow that have $(q,\overline{q})=(-1,1)$ and satisfy $n_{\gamma J^{-1}}=0$ \cite{Knapp:2020oba}. The other sectors are referred to as broad. Being one-dimensional, the narrow sectors are specified by the label $\delta\in G$ and we denote them by $\phi_{\delta}$. One can define the following pairing on the $(c,c)$-ring
\begin{equation}
  \label{lgpairing}
  \langle\phi_{\delta},\phi_{\delta'}\rangle=\frac{1}{|G|}\delta_{\delta,{\delta'}^{-1}}.
\end{equation}
The pairing on the $(a,c)$-ring can be inferred from (\ref{ringflow}).

In order to define the $I$-function and the Gamma class we need to take into account further information about marginal deformations in the narrow sectors. %In our case these are elements of the $(a,c)$-ring that have left/right R-charges $(-1,1)$.
If the space of narrow marginal deformations has dimension $h$ the information about the corresponding marginal deformations can be encoded in a $h\times (h+N)$-matrix $q$ that can be determined from the defining data of the Landau-Ginzburg orbifold \cite{Knapp:2020oba}. In connection to GLSMs with gauge group $U(1)^h$ that have Landau-Ginzburg orbifold phases the matrix $q$ can be obtained as follows. Take the matrix $\mathsf{C}$ of GLSM gauge charges and divide it up into blocks $\mathsf{C}=(L\ S)$, where the $h\times h$ matrix $L$ contains the charges of those fields that obtain a VEV in the Landau-Ginzburg phase. Then $q=L^{-1}\mathsf{C}$. Note, however, that it is possible to define $q$  and $L$ without a GLSM.

The $I$-function and the Gamma class can be defined explicitly in terms of $q$. Before we do that, a word of caution concerning labelling conventions. The Gamma class and the $I$-function are associated to the $(a,c)$-ring and are expressible in terms of basis elements $e_{\delta}^{(a,c)}$. However it turns out that the labelling of FJRW theory which is closer to the labelling of the $(c,c)$-ring is most convenient. The relation between these basis elements is
\begin{equation}
  \label{lgbasis}
  e_{J\delta}^{(a,c)}=e_{\delta}^{(c,c)}=e_{\delta^{-1}},
\end{equation}
where the latter is the FJRW basis. Since in our examples $\delta=J^{\ell}$, $\ell=0,\ldots,d-1$ we will choose the labels $e_{\ell}$. Now we can give the definition of the $I$-function for Landau-Ginzburg orbifolds \cite{Knapp:2020oba}:
\begin{equation}
    \label{lgifun}
  I_{\ell}(u) = -\sum_{\substack{k_1,\dots,k_h \geq 0\\k' \equiv
      \ell \mod d}} 
    \frac{ u^k}{\prod_{a=1}^h\Gamma(k_a+1)}
    \prod_{j=1}^N\frac{(-1)^{\langle-\sum_{a=1}^h
      k_a q_{a,h+j}+q_{j}\rangle}\Gamma(\langle\sum_{a=1}^h
      k_a q_{a,h+j}-q_{j}\rangle)}{\Gamma(1+\sum_{a=1}^h
      k_a q_{a,h+j}-q_{j})},
\end{equation}
where $\langle x\rangle=x- \lfloor x\rfloor$ and $u^k=\prod_i u_i^{k_i}$.  The integers $k_i$ have periodicities encoded in the matrix $L$ associated to the action of the orbifold group $G$:
\begin{equation}
k\sim k+L^Tm\quad\forall m\in\mathbb{Z}^h.
\end{equation}
 From a GLSM standpoint the matrix $L$ encodes how the Landau-Ginzburg orbifold group is embedded in the GLSM gauge group. This allows one to associate different values of $k$ to different sectors labeled by $\ell$. This can be systematised by making use of the Smith normal form of $L$. We refer to \cite{Knapp:2020oba} for details. The Landau-Ginzburg $I$-function is then given by
\begin{equation}
  I_{LG}(u)=\sum_{\delta\in G}I_{\delta}(u)e_{\delta}^{(a,c)}.
  \label{eqn:fjrwIfuncGeneral}
\end{equation}
    The matrix $q$ also encodes the information to define the Gamma class. The Gamma class acts diagonally on $\mathcal{H}^{(a,c)}$ and one defines
    \begin{equation}
      \label{lggamma}
      \widehat \Gamma_{LG} {e}_{\gamma}^{(a,c)}=\widehat{\Gamma}_{\gamma} {e}_{\gamma}^{(a,c)}\qquad \widehat{\Gamma}_{\delta}=\prod_{j=1}^N\Gamma\left(1-\left\langle\sum_{a=1}^h
      k_a q_{a,h+j} - q_j\right\rangle \right).
    \end{equation}
    Note that $\widehat{\Gamma}_{\ell}=\widehat{\Gamma}_{\delta^{-1}J}$.
    The conjugate expression is given by 
     \begin{align}
     \label{lggammaCon}
    \widehat\Gamma_{LG}^* {e}_\gamma^{(a,c)} = \widehat\Gamma_\gamma^\ast  {e}_\gamma^{(a,c)} 
    \qquad
    \widehat\Gamma^\ast_\delta= \prod_{j=1}^N \Gamma \left(\left\langle
    \sum_{a=1}^h k_a q_{a,h+j} - q_j \right\rangle\right).
    \end{align}
Finally we introduce 
\begin{equation}
	\label{lgGrading}
	  \mathrm{Gr}=\sum_{j=1}^N\left\langle -\sum_{a=1}^h k_a q_{a,h+j} + q_j \right\rangle.
\end{equation}
It coincides with the eigenvalues of the grading operator defined on the FJRW state space.

     We find that the sphere partition function in Landau-Ginzburg models has the following form
    \begin{equation}
    \label{lgproposal}
      Z_{S^2}^{LG}(\pt,\overline{\pt})=\frac{1}{|G|}\sum_{\delta}(-1)^{\mathrm{Gr}}\frac{\widehat{\Gamma}_{\delta}}{\widehat{\Gamma}^*_{\delta}}I_{\delta}(u(\pt))I_{\delta}(\overline{u}(\overline{\pt}))=\langle \overline{I}_{LG}(\overline{u}(\overline{\pt})),I_{LG}(u(\pt))\rangle,
    \end{equation}
    The pairing is (\ref{lgpairing}). Here we have defined
    \begin{equation}
      \langle \overline{I}_{LG}(\overline{u}(\overline{\pt}))|=\sum_{\delta}(-1)^{\mathrm{Gr}}\frac{\widehat{\Gamma}_{\delta}}{\widehat{\Gamma}^*_{\delta}}I_{\delta}(\overline{u}(\overline{\pt}))e_{\delta^{-1}}.
    \end{equation}
    To make the connection to the $J$-function and the flat coordinate $t$, we select the element $I_0$ (associated to the basis element $e_{0}^{(a,c)}$) that is the unique element that has left/right R-charges $(q,\overline{q})=(0,0)$. Furthermore we take the elements $I_{\delta_a}$ ($a=1,\ldots,h$) of charges $(q,\overline{q})=(-1,1)$ corresponding to the marginal deformations. Then the flat coordinates are
\begin{equation}
  t_a=\frac{I_{\delta_a}}{I_0}.
\end{equation}
The $J$-function is defined by
\begin{equation}
  J_{LG}(t)=\frac{I_{LG}(u(t))}{I_0(u(t))}.
\end{equation}
%%%%%%%%%%%%%%%%%%%%%%%%%%%%%%%%%%%%%%%%%%%%%%%%%%%%%%%%%%%%%%%%%%%%%%%%%%%%%%%
\subsection{Geometry}
\label{subsec-geometry}
Geometric phases are well-studied and the ingredients to (\ref{zs2formula}) can be found in the literature for many classes of examples. 
The appearance of the $I$-function in the context of the sphere partition function in geometric phases of abelian and non-abelian GLSMs has been noted in \cite{Bonelli:2013mma,Ueda:2016wfa,Kim:2016jye,Gerhardus:2018zwb,Goto:2018bci,Honma:2018fgw}.

A general expression for the $I$-function for Calabi-Yaus that are nef complete intersections in smooth toric varieties can be found in \cite{MR1653024,Cox:2000vi}. We follow \cite{Cox:2000vi} where also the result for the two-parameter example in Section \ref{sec-twopar} has been discussed. Let $X_{\Sigma}$ be a smooth toric variety associated to a toric fan $\Sigma$ and let $\mathcal{L}_1,\ldots,\mathcal{L}_{\ell}$ be line bundles on $X_{\Sigma}$ generated by global sections. We also associate an ($N$-)lattice polytope $\Delta^{*}$ to $X_{\Sigma}$. Let $X\subset X_{\Sigma}$ be a smooth complete intersection defined by a global section of $\mathcal{V}=\oplus_{i=1}^{\ell}\mathcal{L}_i$. Denote by $D_{\rho}\in H^2(X_{\Sigma})$ the cohomology class of the divisor (usually also denoted by $D_{\rho}$) associated to the one-dimensional cones $\rho\in\Sigma(1)$ of $\Sigma$. Furthermore choose an integral basis $H_1,\ldots,H_h$ of $H^2(X_{\Sigma},\mathbb{Z})$, which lies in the closure of the K\"ahler cone. %, and set $\delta=\sum_{i=1}^ht_iH_i$.
Furthermore, $\beta\in H_{2}(X_{\Sigma},\mathbb{Z})$ and we define $\mathcal{L}_i(\beta)=\int_{\beta}c_1(\mathcal{L}_i)$ and $D_{\rho}(\beta)=\int_{\beta}D_{\rho}$. Then the $I$-function $I_{X}$ is given by
\begin{align}
  \label{geomi}
  I_X(u,H)&=
  \prod_iu_i^{H_i}
  \sum_{\beta\in M(X_{\Sigma})}\prod_{i=1}^hu_i^{\int_{\beta}H_i}\frac{\prod_{i=1}^{\ell}\prod_{m=-\infty}^{\mathcal{L}_i(\beta)}\left(c_1(\mathcal{L}_i)-m\right)\prod_{\rho}\prod_{m=-\infty}^0(D_{\rho}-m)}{\prod_{i=1}^{\ell}\prod_{m=-\infty}^{0}\left(c_1(\mathcal{L}_i)-m\right)\prod_{\rho}\prod_{m=-\infty}^{D_{\rho}(\beta)}(D_{\rho}-m) },
\end{align}
where $M(X_{\Sigma})$ is the Mori cone.
In the GLSM context, the generators of the Mori cone coincide with the row vectors of the matrix $\mathsf{C}$ of GLSM charges whose column vectors span the secondary fan of $X_{\Sigma}$. The components of $I_X$ are obtained by expanding $I_X$ as a power series in $H_1,\ldots,H_h$. 

Similarly, the Gamma class of $X$ and its conjugate\footnote{Compared some other works in the literature $\widehat{\Gamma}_{X}(H)$ and $\widehat{\Gamma}^*_{X}(H)$ may be exchanged. We are using the convention used in \cite{Hori:2013ika}.} can be written as
\begin{equation}
  \label{geomgamma}
  \widehat{\Gamma}_{X}(H)=\frac{\prod_{\rho}\Gamma\left(1-D_{\rho}\right)}{\prod_{i=1}^{\ell}\Gamma\left(1-c_1(\mathcal{L}_i) \right)},\qquad \widehat{\Gamma}^*_{X}(H)=\frac{\prod_{\rho}\Gamma\left(1+D_{\rho}\right)}{\prod_{i=1}^{\ell}\Gamma\left(1+c_1(\mathcal{L}_i) \right)}
  \end{equation}
where $H$ collectively denotes $H_1,\ldots,H_h$. The Gamma class is invertible since an expansion in terms of a power series of $H$ begins with a constant term and we can invert the series. This is why expressions like $\frac{\widehat{\Gamma}}{\widehat{\Gamma}^*}$ make sense. 

To define the pairing $\langle\cdot,\cdot\rangle$, consider $\alpha,\beta\in H^{even}(X,\mathbb{C})$. Then the relevant pairing is given by the Mukai pairing \cite{MR2141853,Halverson:2013qca} 
\begin{equation}
  \label{mukai}
  \langle\alpha,\beta\rangle=\int_X\alpha^{\vee}\wedge\beta,
\end{equation}
where in the Calabi-Yau case $\alpha^{\vee}=(-1)^{\mathrm{Gr}}\alpha$. The grading operator $\mathrm{Gr}$ acts as follows on $H^{even}(X,\mathbb{C})$:
\begin{equation}
    \mathrm{Gr}\alpha=k\:\alpha,\qquad\textrm{for}\qquad \alpha\in H^{2k}(X,\mathbb{C}).  
    \end{equation}
This coincides with the definition in \cite{Chiodo:2012qt}. 

Here we have restricted to the cohomology of the Calabi-Yau that descends from the cohomology of the ambient space $X_{\Sigma}$. % and we assume that this space has dimension $h$.
We exclude the primitive cohomology of $X$, i.e~the cohomology associated to divisors on $X$ that do not have no counterpart in the ambient geometry. This is the geometric analogue to the restriction to narrow sectors in the Landau-Ginzburg setting. The pairing is evaluated by making use of the intersection ring of $X$.
In the geometric setting (\ref{zs2formula}) simplifies to
\begin{equation}
  Z_{S^2}^{geom}(\pt,\overline{\pt})=\int_X %(-1)^{\mathrm{Gr}}
  \frac{\widehat{\Gamma}_X(H)}{\widehat{\Gamma}_X^*(H)}I_X(u(\pt),H) {I}_X(\overline{u}(\overline{\pt}),H)=\langle \overline{I}_X,I_X\rangle
  \end{equation}
  
The Gamma class and its relation to perturbative corrections has been discussed in \cite{Halverson:2013qca}, where also the quotient $\frac{\widehat{\Gamma}}{\widehat{\Gamma}^*}$ has first been observed and has been linked to complex conjugation via an indirect argument using $K$-theory. Let us briefly summarise this. There is an isomorphism between $H^{even}(X,\mathbb{C})$ and $K_{hol}(X)\otimes\mathbb{C}$, where $K_{hol}(X)$ is holomorphic $K$-theory \cite{Hosono:2000eb}, which involves the Gamma class \cite{iritani2007real,Katzarkov:2008hs,Iritani:2009ab,MR3112512} 
\begin{equation}
  \mu:[\mathcal{E}]\mapsto \mathrm{ch}(\mathcal{E})\wedge\widehat{\Gamma}_X.
\end{equation}
It has been suggested that complex conjugation for $w \in  H^{even}(X,\mathbb{C})$ works as follows:
\begin{equation}
  w\mapsto \mathrm{ch}^{-1}\left(\frac{w}{\widehat{\Gamma}_X} \right)\mapsto  \mathrm{ch}^{-1}\left(\frac{\overline{w}}{\widehat{\Gamma}_X^*}\right)\mapsto \overline{w}\;\frac{\widehat{\Gamma}_X}{\widehat{\Gamma}_X^*},
\end{equation}
where the map in the middle is complex conjugation on  $K_{hol}(X)$. 
Let us point out that when evaluating the sphere partition in geometric phases there is some ambiguity when it comes to identifying the pairing and the complex conjugation operator. In the definitions we have given, the grading operator $\mathrm{Gr}$ that acts on the state space apprears twice: one in the definition of the Mukai pairing and once in $(-1)^{\mathrm{Gr}}\frac{\widehat{\Gamma}}{\widehat{\Gamma}^*}$ in (\ref{idual}). This means that all the signs coming from $(-1)^{\mathrm{Gr}}$ actually cancel and it would be consistent, at least from the point of view of the sphere partition function, to use a pairing $\langle\alpha,\beta\rangle=\int_X\alpha\wedge\beta$ instead of the Mukai pairing and to define complex conjugation via $\frac{\widehat{\Gamma}}{\widehat{\Gamma}^*}$ instead of (\ref{idual}). 

With $\mathcal{H}_{nar}=H^2(X,\mathbb{C})$ (where we have excluded the primitive cohomolgy) and $H^{0}(X,\mathbb{C})$ singling out a distinguished component, the flat coordinates are defined by the corresponding components $I_i$ ($i=1,\ldots,h$) and $I_0$ of the $I$-function: 
\begin{equation}
  t_i(u)=\frac{I_{i}}{I_0},
\end{equation}
and the $J$-function is defined by
\begin{equation}
  J_X(t)=\frac{I_X(u(t))}{I_0(u(t))}. 
  \end{equation}
%%%%%%%%%%%%%%%%%%%%%%%%%%%%%%%%%%%%%%%%%%%%%%%%%%%%%%%%
\subsection{Hybrid phases}
\label{sec-hybgeneral}
A further non-trivial test for (\ref{zs2formula}) is to study regions in the moduli space that are more exotic than geometric and Landau-Ginzburg phases. A class of such examples are hybrid models that are fibrations of Landau-Ginzburg orbifolds over some base manifold $B$. In the physics literature they have been studied for instance in \cite{Bertolini:2013xga,Bertolini:2017lcz,Bertolini:2018now}. In the mathematics literature there is a generalisation of FJRW theory that captures a class of one-parameter hybrid models \cite{MR3590512,MR3842378,clader2018wallcrossing,zhao2019landauginzburgcalabiyau,Chiodo_2018}. In the examples below we will recover the mathematics results for the $I$-functions and the Gamma class from the sphere partition function and conjecture new ones in the multi-parameter cases.

The class of models we are considering consists of fibrations of Landau-Ginzburg orbifolds over certain base manifolds. To give a more precise definition we follow \cite{Bertolini:2013xga}. We consider a K\"ahler manifold $Y_0$ together with a holomorphic function $W$ whose critical locus defines a compact subset $B$ such that $dW^{-1}(0)=B\subset Y_0$. In the case of a Landau-Ginzburg model $B$ is a point, whereas a compact $Y_0$ (and hence trivial $W$) leads to a nonlinear sigma model.

To obtain an action for the hybrid model, one introduces $Y$, which is the total space of a rank $N$ vector bundle $X\rightarrow B$ where we assume that $B$ is compact, smooth and K\"ahler of dimension $r$. It is possible to write down an $\mathcal{N}=(2,2)$ supersymmetric action for the hybrid model on $Y$ \cite{Bertolini:2013xga} whose kinetic term describes a non-linear sigma model on $Y$ and which includes a potential term involving the superpotential $W$ satisfying the superpotential condition $dW^{-1}(0)=B$. Given a suitable choice of K\"ahler metric on $Y$, the superpotential condition ensures that at low energies the field fluctuations will be localised on $B$.

The IR theory is an $\mathcal{N}=(2,2)$ superconformal theory characterised by the massless ground states of the hybrid theory. It is the IR behaviour that determines the distinction between a ``good'' hybrid model and a ``pseudo-hybrid'' \cite{Aspinwall:2009qy,Bertolini:2013xga}. To this end, one has to consider the $U(1)_L\times U(1)_R$-symmetry. If there is no potential, these symmetries exist due to an integrable, metric-compatible complex structure on $Y$. To guarantee that these symmetries are also present, at least classically, when there is a non-zero potential, there must be a holomorphic Killing vector field $F$ satisfying $\mathcal{L}_FW=W$. At the quantum level, $U(1)_L$ exhibits a chiral anomaly unless $c_1(T_Y)=0$. This is satisfied if the canonical bundle $K_Y$ is trivial which will be assumed. A consequence of this is that $B$ has to be Fano, which is indeed the case for all the examples that we consider, where $B=\mathbb{P}^r$ for $r=1,2,4$.

In order for the UV R-symmetry to lead to a well-defined $R$-symmetry in the IR, it is required that all forms $\omega\in\Omega(B)$ satisfy $\mathcal{L}_F\pi^{\ast}(\omega)=0$ and that $U(1)_L\times U(1)_R$ fixes $B$ point-wise. Such models are referred to as good hybrids and it is possible to write down an explicit expression for $F$ \cite{Bertolini:2013xga}. These conditions ensure that the local picture of a Landau-Ginzburg model fibered over every point in $B$ is valid.

In order for the $U(1)_L\times U(1)_R$-charges of all (NS,NS)-sector states to be integral, one has to orbifold by the discrete symmetry generated by $e^{2\pi i J_0}$, where $J_0$ is the conserved $U(1)_L$-charge. As in the Landau-Ginzburg case, we denote the orbifold group by $G$. Due to the properties of $F$, the orbifold only acts on the fibre coordinates. Hybrids of this type arise in the context of type II string compactifications on Calabi-Yaus that we are considering here, and also in heterotic settings. All the good hybrids we will discuss are of this type. Note that in the context of hybrids arising from GLSMs there could be more general orbifold actions arising as discrete unbroken subgroups of the GLSM gauge group. This has been discussed, for instance, in a Landau-Ginzburg context in \cite{Knapp:2020oba}. While we expect the structures discussed in this work to appear in this more general context as well, we will not consider this more general setting here.

The massless spectrum for good hybrids arising from the cohomology of the right-moving supersymmetry generator was computed in \cite{Bertolini:2013xga} and interpreted in the context of heterotic string compactifications. These results provide techniques to obtain the $(c,c)$- and $(a,c)$-rings of the internal Calabi-Yau CFT.  In \cite{Bertolini:2018now} the elements of the $(c,c)$-ring in the untwisted sector of in the B-twisted good hybrids have been computed explicitly. These works use spectral sequences that arise from the structure of the supercharges of the hybrid models to obtain representatives of the states in terms of the matter content of the hybrid model. In our approach the state space only enters via its dimension and the existence of a pairing. Therefore we find it more convenient to use a definition of the state space as it can be found in the mathematics literature \cite{MR3590512,Chiodo_2018}, even though it appears to be less general than the physics prescription. In \cite{MR3590512} the state space has been defined for two hybrid models that arise in the same moduli space as certain one-parameter complete intersections in toric ambient spaces. In Section~\ref{sec-onepar} these two examples are labelled K1 and M1. Our results imply, however, that this prescription applies in a more general setting and we expect it to hold for all good hybrids.

We need to identify the subset of the $(a,c)$-ring that corresponds to the narrow sectors $\delta\in G$. These turn out to be precisely those sectors whose cohomology is determined by the cohomology classes of the base $B$ so that we can characterise the narrow state space as
\begin{equation}
  \label{hybstate}
  \mathcal{H}=\bigoplus_{\delta}H^{*}(B,\mathbb{C})^{(\delta)}.
\end{equation}
In other words, there is a copy of $H^{*}(B,\mathbb{C})$ for every narrow sector. Following \cite{MR3590512}, the narrow sectors can be identified as follows. Let us consider a good hybrid model that is a $G=\mathbb{Z}_d$-orbifold over $\mathbb{P}^r$ with fibre coordinates $x_1,\ldots,x_N$. By definition, the base coordinates do not transform under the orbifold action. Let $q_1,\ldots,q_N$ be the $U(1)_L$-charges of the fibre coordinates so that the $\mathbb{Z}_{d}$-orbifold is generated by $\langle J\rangle$ with $J=(e^{2\pi i q_1},\ldots,e^{2\pi i q_N})$, in complete analogy the the Landau-Ginzburg case. The $\ell$-th twisted sector is referred to as narrow if there is no $j\in \{1,\ldots,N\}$ such that $e^{2\pi i \ell q_j}=1$. %We note that in the special case of a Landau-Ginzburg orbifold this definition of a narrow sectors singles out those twisted sectors that only contain the vacuum state.
In all the examples we discuss in the subsequent sections the definition (\ref{hybstate}) is consistent with the results from the sphere partition function. In particular, the counting of narrow states for hybrids phases matches with the counting in the geometric and Landau-Ginzurg phases. Note that a more abstract definition of narrow sectors in hybrids arising in moduli spaces of complete intersection Calabi-Yaus has been given in \cite{Chiodo_2018}. 

% no closed for for I-function and Gamma class, refer to examples
With these structures in mind, we can evaluate the GLSM sphere partition function in models with good hybrid phases where we recover the form advertised in (\ref{zs2formula}). This allows us to confirm the mathematics results for the $I$-functions and the Gamma class from the sphere partition function and to conjecture new ones in the multi-parameter cases. While it seems possible to give a general expression of the $I$-function and the Gamma class for a rather general class of multiparamter good hybrid models, one expects technical complications similar to those encountered in the Landau-Ginzburg case \cite{Knapp:2020oba}. From the GLSM perspective, this reflects the often complicated symmetry breaking pattern that occurs in phases of GLSMs. The standard examples of hybrid models that we also study here are very simple and reading off the (conjectural) expressions for the $I$-functions and the Gamma class on a case-by-case basis is fairly obvious. In contrast to the Landau-Ginzburg and geometry cases, we do not have a vast amout of literature to build upon, nor is there a classification of good hybrids at our disposal to apply any general statements to. We therefore leave finding general expressions for the Gamma class and the $I$-function for good hybrids for future work.

A final remark concerns the definition of the paring that is implicit in  (\ref{zs2formula}). In the hybrid case, this expression includes an integral over the base manifold $B$ that is not Calabi-Yau. For an algebraic variety $B$ there are the following relations between characteristic classes:
\begin{equation}
  \label{classes}
  \mathrm{Td}(B)=e^{\frac{c_1(B)}{2}}\widehat{A}(B)=e^{\frac{c_1(B)}{2}}\widehat{\Gamma}_B\widehat{\Gamma}^*_B,
\end{equation}
where $\mathrm{Td}$ is the Todd class, $c_1$ is the first Chern class, $\widehat{A}$ is the $A$-roof genus, and $\widehat{\Gamma}$ is the Gamma class. Using such identities we can show that the sphere partition function indeed takes the form (\ref{zs2formula}). The results from the sphere partition function are not enough to deduce the correct definition of the pairing. If we, following \cite{MR2141853,Halverson:2013qca}, interpret the integral over $B$ as an artifact of the Mukai pairing, then we have to modify the definition of $\alpha^{\vee}$ in (\ref{mukai}) to be $\alpha^{\vee}=(-1)^{\mathrm{Gr}}e^{\frac{c_1(B)}{2}}\alpha$. Consistency with the result of the sphere partition function would then further imply that $(-1)^{\mathrm{Gr}}\frac{\widehat{\Gamma}}{\widehat{\Gamma}^*}$ in the conjugation operation (\ref{idual}) would have to be modified to $(-1)^{\mathrm{Gr}}e^{-\frac{c_1(B)}{2}}\frac{\widehat{\Gamma}}{\widehat{\Gamma}^*}$. It would be interesting to study this further.
%%%%%%%%%%%%%%%%%%%%%%%%%%%%%%%%%%%%%%%%%%%%%%%%%%%%%%%%%%%%%%%%%%%%%%%
\subsection{Pseudo-hybrid phases}
A class of hybrids that are not good hybrids habe been termed pseudo-hybrids in \cite{Aspinwall:2009qy}. They are associated to singular CFTs. One of the properties that follows from the violation of the conditions for being a good hybrid is there is no unique R-charge assignment in the IR\footnote{In the language of variation of GIT quotients this is referred to as a ``lack of good lift''.}. This is related to the fact that there is no known enumerative problem in the sense of FJRW theory. %A state space isomorphism between general hybrid models and CY complete intersections has been proven in \cite{Chiodo_2018}.
Still, it is possible to evaluate the sphere partition function of a given GLSM in a pseudo-hybrid phase and there is at least some understanding of the low-energy physics \cite{Aspinwall:2009qy}. A further feature of pseudo-hybrids is that the solutions of the D-term and F-term equations in the GLSM have several components. This structure is also reflected in the sphere partition function. Below, we present some results that indicate that the components of the sphere partition function that correspond to a specific component of the GLSM vacuum also display a factorisation along the lines of (\ref{zs2formula}). The one-parameter examples we consider in this context and the associated GLSMs have already been discussed in \cite{Erkinger:2019umg} to which we refer for details. 
%%%%%%%%%%%%%%%%%%%%%%%%%%%%%%%%%%%%%%%%%%%%%%%%%%%%%%%%%
\section{One-parameter examples}
\label{sec-onepar}
A canonical class to test the general expression for the sphere partition function is a set of well-studied one-parameter Calabi-Yaus that also has received some recent attention in the context of swampland conjectures \cite{Joshi:2019nzi,Erkinger:2019umg,Andriot:2020lea}. The associated GLSMs have gauge group $\pG=U(1)$ and the following field content\footnote{By abuse of notation we will denote
the chiral superfield and its scalar component by the same lower case letter.}
\begin{equation}
\begin{array}{c|c|c|c|c|c|c}
	& p_1 & p_{2_1,\dots,2_k} & x_{1,\dots,5-n-j+k} & x_{\alpha_1, \dots, \alpha_n }
	& x_{\beta_1, \dots, \beta_j} &\mathrm{FI} \\
	\hline
	U(1) & -d_1 & -d_2  & 1 & \alpha & \beta &\zeta \\
        \hline
	{U(1)}_{V} & 2-2d_1q & 2- 2d_2 q & 2 q & 2 \alpha q & 2 \beta q&
\end{array}
\label{eqn:1paramGLSMs}
\end{equation}
with the following restrictions
\begin{align}
0 &\leq k \leq 3, &  0 &\leq n \leq 2, & 0&\leq j \leq 2,
\end{align}
\begin{equation}
5+k-n-j + \alpha n + j\beta= d_1+k d_2, \label{eqn:1paramCYcon}
\end{equation}
where the last equation is the  Calabi-Yau
condition. The \(U(1)_V\) charges satisfy \(0\leq q\leq 2\) if 
\begin{align}
0 \leq q \leq \frac{1}{\max[d_1,d_2]}.
\end{align}
The explicit values of these parameters for all 14 abelian one-parameter models 
can be found in\footnote{Compared to \cite{Erkinger:2019umg} we changed the labels of some models.} Table \ref{tab:1ParamModels}.
The models have a superpotential of the form
\begin{equation}
W = p_1 G_{d_1}(x_n) +\sum_{i=1}^k p_{2_i} G_{i,d_2}(x_n),
\end{equation}
where $G_{d_1}$ is a weighted homogeneous polynomial of degree $d_1$ and
similarly for $G_{i,d_2}$.
The large volume phases ($\zeta\gg0$) are complete intersections in weighted projective space: 
\begin{equation}
	\mathbb{P}^{5+k-1}_{1^{5+k-n-j} \alpha^n \beta^j }[d_1,\underbrace{d_2,\dots,d_2}_{k\text{-times}}]. \label{eqn:phm1}
\end{equation}
In the above formula we denote by a superscript the dimension and 
by a subscript the weights of the coordinates. In the brackets we 
give the weighted homogeneous degree of the defining equations.
There are four types of small volume phases ($\zeta\ll 0$) that can be classified according to their monodromy around the limiting point.  They are labeled by M, F, K, and C \cite{MR3822913}. The M-points have monodromy similar to large volume points. There is only a single model with this property and it turns out that the two phases are not birational, much like in non-abelian GLSMs. This has been studied in \cite{Caldararu:2007tc}, see also \cite{Sharpe:2012ji} for the computation of the sphere partition function. Type C points are  pseudo-hybrid phases. The points of type F have Landau-Ginzburg or pseudo-hybrid phases, type K corresponds to (good) hybrid theories, i.e.~fibrations of Landau-Ginzburg orbifolds over some base manifold.

\begin{table}
	\centering
	\begin{tabular}{ccccc|cc}
		\multicolumn{5}{c}{model-data}& \multicolumn{2}{c}{IR-description} \\
		\toprule
		label&\(\alpha^n\) &\(\beta^j\)&\(d_1\)& \( d_2^k\) & \(\zeta \gg 0 \) & \(\zeta \ll 0\)  \\
		\midrule
		\multicolumn{7}{c}{F-type}\\
		\midrule 
		F1&-&- &5 & -&\(\mathbb{P}_{1^5}[5]\)&LG orbifold\\
		F2&-&2 &6 & -&\(\mathbb{P}_{1^4,2}[6]\)&LG orbifold\\
		F3&-&4 &8 & -&\(\mathbb{P}_{1^4,4}[8]\)&LG orbifold\\
		F4&2&5 &10 & -&\(\mathbb{P}_{1^3,2,5}[10]\)&LG orbifold\\
		F5&-&2 &4 & 3&\(\mathbb{P}_{1^5,2}[4,3]\)&Pseudo-Hybrid\\
		F6&\(2^2\)& 3 &6 & 4&\(\mathbb{P}_{1^3,2^2,3}[6,4]\)&Pseudo-Hybrid\\
		F7&4&6 &12 & 2&\(\mathbb{P}_{1^4,4,6}[12,2]\)&Pseudo-Hybrid\\
		\multicolumn{7}{c}{C-type}\\
		\midrule 
		C1&-&- &4 & 2&\(\mathbb{P}_{1^6}[4,2]\)&Pseudo-Hybrid\\
		C2&-&3 &6 & 2&\(\mathbb{P}_{1^5,3}[6,2]\) &Pseudo-Hybrid\\
		C3&-&- &3 & \(2^2\)&\(\mathbb{P}_{1^7}[3,2,2]\)& Pseudo-Hybrid\\
		\multicolumn{7}{c}{K-type}\\
		\midrule 
		K1&-&- &3 & 3&\(\mathbb{P}_{1^6}[3,3]\)&Hybrid\\
		K2&-&\(2^2\) &4 & 4&\(\mathbb{P}_{1^4,2^2}[4,4]\)&Hybrid\\
		K3&\(2^2\)&\(3^2\) &6 & 6&\(\mathbb{P}_{1^2,2^2,3^2}[6,6]\)&Hybrid\\
		\multicolumn{7}{c}{M-type}\\
		\midrule 
		M1&-&- &\(2\) & \(2^3\)&\(\mathbb{P}_{1^8}[2,2,2,2]\)&Non-linear \(\sigma\)\\
	\end{tabular}
	\caption{Model data of one-parameter abelian GLSMs.}
	\label{tab:1ParamModels}
\end{table}

%%%%%%%%%%%%%%%%%%%%%%%%%%%%%%%%%%%%%%%%%%%%%%%%%%%%%%%%%%%%%%%%%%%%%%%%%%%%%%
\subsection{Evaluation of the sphere partition function}
The sphere partition function in our 
GLSMs reads
\begin{align}
\begin{split}
  Z_{S^2}=\frac{e^{-4 \pi \zeta q}}{2\pi} \sum_{m \in \mathbb{Z}}
  \int_{-\infty+iq}^{\infty+iq}
  \mathrm{d} \sigma  Z_{p_1}Z^k_{p_2} Z^{5+k-n-j}_{1}
  Z^n_{\alpha}  Z^j_{\beta}
  e^{(-2 \pi \zeta - i \theta)\left(i \sigma +\frac{m}{2}\right)} e^{(-2 \pi \zeta + i\theta)\left(i \sigma -\frac{m}{2}\right)},
\end{split}
\label{eqn:spherepart1param}
\end{align}
with
\begin{equation}
\begin{gathered}
Z_{p_1} =\frac{\Gamma \left(\frac{1}{2} (m+2 i \sigma ) d_1+1\right)}{
	\Gamma \left(\frac{1}{2} (m-2 i \sigma ) d_1\right)}, \quad
 Z_{p_2}= \frac{\Gamma \left(\frac{1}{2} (m+2 i \sigma ) d_2+1\right)}{
	\Gamma \left(\frac{1}{2} (m-2 i \sigma ) d_2\right)},  \quad
 Z_{1} = \frac{\Gamma \left(-\frac{m}{2}-i \sigma \right)}{
	\Gamma \left(-\frac{m}{2}+i \sigma +1\right)} ,\\
Z_{\alpha}=\frac{\Gamma \left(-\frac{1}{2} \alpha  (m+2 i \sigma )\right)}{
	\Gamma \left(i \sigma  \alpha -\frac{m \alpha }{2}+1\right)}, \quad
  Z_{\beta}=\frac{\Gamma \left(-\frac{1}{2} \beta  (m+2 i \sigma )\right)}{
	\Gamma \left(i \sigma  \beta -\frac{m \beta }{2}+1\right)}.
\label{eqn:pshzx1poles}
\end{gathered}
\end{equation}
Observe that in (\ref{eqn:spherepart1param}) we have transformed \(\sigma \rightarrow -iq+ \sigma\).
We evaluate the sphere partition function by application of the residue theorem. The result depends on the phase of the GLSM. Much of this has already been done in \cite{Erkinger:2019umg} to which we refer for details on how to determine the contributing poles. The most important
steps in the evaluation are also summarized in Appendix 
\ref{sec:spherePart1ParamEval}.
We observe that in all examples of this class the contributing poles in 
a phase are associated to fields that get 
a non-zero VEV in the given phase. 
\subsubsection{\(\zeta \gg 0\) Phase} 
In this phase the poles of \(Z_1, Z_\alpha\) and \(Z_\beta\) 
contribute. It is sufficient to sum over the poles 
of \(Z_\beta\). The contributions from the missed poles of \(Z_\alpha\) vanish in all models, as we show in
Appendix \ref{sec:spherePart1ParamEval}. The final result 
is given by:
\begin{align}
Z_{S^2}^{\zeta \gg 0}=-\frac{1}{2\pi}
\oint_0 \mathrm{d} \varepsilon \mathcal{Z}_{1,sing}(\varepsilon)
|\mathcal{Z}_{1,reg}(\varepsilon,\pt)|^2,
\label{eqn:zs21ParamLarge1}
\end{align} 
with
\begin{align}
\begin{split}
\mathcal{Z}_{1,reg}(\varepsilon)&=
\sum_{a=0}^{\infty}  (-1)^{a (5+k-n-j+\alpha n+ j\beta )}
e^{-\pt(i \varepsilon +a+q)} \\
& \quad \cdot \frac{
	\Gamma \left(a d_1
	+i \varepsilon  d_1+1\right)
}{
	\Gamma \left(a+i \varepsilon +1\right)^{5+k-n-j}
	\Gamma \left(a \alpha 
	+i \varepsilon  \alpha +1\right)^n
} 
\frac{\Gamma \left(a d_2+i \varepsilon  d_2+1\right)^k}
{\Gamma (a \beta +i \varepsilon  \beta  +1)^j},
\end{split}
\label{eqn:zs21ParamLarge1Reg}
\end{align} 
and
\begin{align}
\mathcal{Z}_{1,sing}(\varepsilon)&=
\frac{
	\pi^{4}
	\sin \left(\pi \left(i \varepsilon d_1 \right)\right)
	\sin \left(\pi \left(i \varepsilon d_2 \right)\right)^k
}{
	\sin \left(\pi \left(i \varepsilon \right)\right)^{5+k-n-j}
	\sin \left(\pi \left(i \varepsilon \alpha \right)\right)^n
	\sin \left(\pi \left(i \varepsilon \beta\right)\right)^j
}.
\label{eqn:zs21ParamLarge1Sing}
\end{align}

\subsubsection{\(\zeta\ll0\) Phase}
For this phase the sphere partition function gets two contributions. In the first contribution one sums over the poles of \(Z_{p_1}\). In the second contribution one accounts for previously missed poles of \(Z_{p_2}\), if there are any. One gets:  
\begin{equation}
  Z_{S^2}^{\zeta\ll0}=Z^{\zeta \ll 0}_{S^2,1}+Z^{\zeta \ll 0}_{S^2,2},
  \label{eqn:zs21ParamSmallTotal}
\end{equation} 
where details on \(Z^{\zeta \ll 0}_{S^2,1}\) are  
given in (\ref{eqn:zs21ParamSmall1}).
The \(Z^{\zeta \ll 0}_{S^2,2}\) contribution is only non-zero in
 models with a pseudo-hybrid phase. Because the focus of 
this work lies on models with  Landau-Ginzburg and  
hybrid phases we discuss the features
 of \(Z^{\zeta \ll 0}_{S^2,2}\) and pseudo-hybrids  in the Appendix \ref{sec:spherePart1ParamEval}.
In models with a Landau-Ginzburg or hybrid phase we can
further simplify  \(Z^{\zeta \ll 0}_{S^2,1}\), because in these cases we have \(d_1 = d_2\). Typically in 
these phases \(Z^{\zeta \ll 0}_{S^2,1}\) is a sum of different contributions, which we label by \(\delta\),
where \(\delta \in \mathbb{Z}_{>0}\).
The integrand depends on \(\delta \) in such a way, that \(Z^{\zeta \ll 0}_{S^2,1}\) vanishes unless
\begin{align}
\left\langle  \frac{\delta}{d_1} \right\rangle & \not = 0,
&\left\langle \alpha \frac{\delta}{d_1} \right\rangle & \not = 0 ,
& \left\langle \alpha \frac{\beta}{d_1} \right\rangle & \not = 0.
\label{eqn:1paramRestOnDelta}
\end{align}
The possible \(\delta\) values are restricted from above by \(\delta< d_1\) and we will denote the set of \(\delta \) values which fulfil (\ref{eqn:1paramRestOnDelta})
by \(narrow\), because (\ref{eqn:1paramRestOnDelta}) corresponds to the narrow sectors discussed in Section \ref{sec:lgFJRWBackground}. 
For the models of interest we 
summarize the contributing sectors and the order of the  poles in Table \ref{tab:1ParamSectorsHybLg}. In the narrow sector we can show
\begin{equation}
\left\langle \alpha - \alpha \frac{k}{d} \right\rangle = 1- \left\langle \alpha
\frac{k}{d} \right\rangle.
\label{eqn:narrowFrac}
\end{equation}
Therefore we can use the identity:
\begin{align}
\sin \left( \pi \left(i \beta \varepsilon + \alpha \frac{k}{d} \right)\right)&=
\sin \left( \pi \left(i \beta \varepsilon
+ \left\langle \alpha\frac{k}{d} \right\rangle
+ \left\lfloor \alpha \frac{k}{d} \right\rfloor
\right)\right), \nonumber \\
&=(-1)^{\left\lfloor \alpha \frac{k}{d} \right\rfloor}
\frac{\pi}{
	\Gamma\left(i \beta \varepsilon
	+ \left\langle \alpha\frac{k}{d} \right\rangle\right)
	\Gamma \left( -i \beta \varepsilon
	+ \left\langle \alpha\frac{d-k}{d} \right\rangle\right)
},  
\label{eqn:reflectionFormula}
\end{align}
which is useful in rewriting \(\mathcal{Z}_{1,sing}\) (\ref{eqn:zs21ParamSmall1}).
After the variable transformation \( \varepsilon \rightarrow \frac{i \varepsilon}{d_1},\) 
(\ref{eqn:zs21ParamSmall1}) can be written in the 
following form:
\begin{equation}
Z^{\zeta \ll 0}_{S^2,1} =\frac{1}{2\pi i d_1} \sum_{\delta \in
	\mathit{narrow}}
\oint_0 \mathrm{d} \varepsilon \frac{(-1)^{\operatorname{Gr}}}
{\varepsilon^{k+1}}
\frac{\hat\Gamma_\delta (\varepsilon)}{\hat\Gamma^\ast_\delta(\varepsilon)}
 | I^{\zeta \ll 0}_\delta
( \pt,\varepsilon)|^2,
\label{eqn:1ParamSmallSpherePartHybLg}
\end{equation}
with 
\begin{align}
\begin{split}
I^{\zeta \ll 0}_\delta(\pt,\varepsilon) &= \sum_{a=0}^\infty
e^{\pt(\frac{\varepsilon}{d_1} +a + \frac{\delta}{d_1}-q)}
(-1)^{a (5+k-n-j+ \alpha n + j\beta)}\\
& \quad \cdot \frac{\Gamma\left(1+ \varepsilon \right)^{k+1}}{
	\Gamma\left(\frac{\varepsilon}{d_1} +  \left\langle \frac{\delta}{d_1} \right\rangle\right)^{5+k-n-j}
	\Gamma\left(\alpha \frac{\varepsilon}{d_1}+ \left\langle\alpha\frac{\delta}{d_1}\right\rangle\right)^{n}
	\Gamma\left(\beta \frac{\varepsilon}{d_1} + \left\langle\beta \frac{\delta}{d_1} \right\rangle \right)^{j}}\\
& \quad \cdot \frac{
	\Gamma \left(a+ \frac{\varepsilon}{d_1} +\frac{\delta}{d_1}\right)^{5+k-n-j}
	\Gamma \left(a \alpha +   \alpha \frac{\varepsilon}{d_1}
	+\frac{\alpha }{d_1}\delta\right)^n
	\Gamma \left(a \beta + \beta \frac{\varepsilon}{d_1}+\frac{\beta }{d_1}\delta\right)^j
}{
	\Gamma \left(\delta +a d_1+\varepsilon \right)^{k+1}},
\end{split}
\label{eqn:1ParamHybLgIfunc}
\end{align}
and
\begin{align}
(-1)^{\operatorname{Gr}}&= (-1)^{\delta (k+1) }(-1)^{\left(5+k-n-j\right)\left\lfloor  \frac{\delta}{d_1} \right\rfloor}(-1)^{n\left\lfloor \alpha \frac{\delta}{d_1} \right\rfloor}(-1)^{j\left\lfloor \beta \frac{\delta}{d_1} \right\rfloor}.
\end{align}
Here we introduced
\begin{align}
\hat\Gamma_\delta (\varepsilon)=\Gamma\left(1- \varepsilon \right)^{k+1}&
\Gamma\left(\frac{\varepsilon}{d_1} +  \left\langle \frac{\delta}{d_1} \right\rangle\right)^{5+k-n-j} \nonumber \\
& \cdot \Gamma\left(\alpha \frac{\varepsilon}{d_1}+ \left\langle\alpha\frac{\delta}{d_1}\right\rangle\right)^{n}
\Gamma\left(\beta \frac{\varepsilon}{d_1} + \left\langle\beta \frac{\delta}{d_1} \right\rangle \right)^{j}, 
\label{eqn:1ParamHybLgGamma}\\
\hat\Gamma^\ast_\delta(\varepsilon)=\Gamma\left(1+ \varepsilon \right)^{k+1}&
\Gamma\left(-\frac{\varepsilon}{d_1} + \left\langle \frac{d_1-\delta}{d_1}\right\rangle\right)^{5+k-n-j} \nonumber \\
& \cdot \Gamma\left(- \alpha \frac{\varepsilon}{d_1}  +
\left\langle \alpha\frac{d_1-\delta}{d_1}\right\rangle\right)^{n}
\Gamma\left(-\beta \frac{\varepsilon}{d_1} + \left\langle \beta \frac{d_1-\delta}{d_1}\right\rangle\right)^{j}.
\label{eqn:1ParamHybLgGammaBar}
\end{align}
It is possible to obtain \(\widehat\Gamma^\ast_\delta(\varepsilon)\) from \(\widehat\Gamma_\delta(\varepsilon)\) 
by applying the following transformations
\begin{align}
\varepsilon &\rightarrow - \varepsilon,
& \langle  \cdot \rangle &\rightarrow 1-\langle \cdot\rangle,
\label{eqn:gammaRelations}
\end{align} 
and as final step (\ref{eqn:narrowFrac}) is used.
For later convenience we also introduce
\begin{align}
\gamma_{\delta}(H) = (-1)^{\operatorname{Gr}}\frac{\hat\Gamma_{\delta}(H)}{\hat\Gamma^\ast_{\delta}(H)}
\label{eqn:gammmaQuotient}.
\end{align}
Below we will show that (\ref{eqn:1ParamHybLgIfunc}),
(\ref{eqn:1ParamHybLgGamma}), and (\ref{eqn:1ParamHybLgGammaBar})
exactly match the expression known from FJRW theory in Landau-Ginzburg and hybrid models.
\begin{table}
\centering
\begin{tabular}{c|cccc|cccc|cccc|cccc|cc|cc|cc|c}
	\toprule
	& \multicolumn{4}{c}{F1} &\multicolumn{4}{c}{F2}
	&\multicolumn{4}{c}{F3}&\multicolumn{4}{c}{F4} 
	&\multicolumn{2}{c}{K1} & \multicolumn{2}{c}{K2}
	&\multicolumn{2}{c}{K3}& M1\\
	\midrule
	\(\delta\) & 1&2 &3 &4 &1 &2 & 4 &5& 1 &3 &5 &7
	&1&3&7&9 & 1 &2 & 1 &3 & 1 &5& 1\\
	pole order & \multicolumn{4}{|c|}{1} &\multicolumn{4}{|c|}{1}
	&\multicolumn{4}{|c|}{1}&\multicolumn{4}{|c|}{1} 
	&\multicolumn{2}{|c|}{2} & \multicolumn{2}{|c|}{2}
	&\multicolumn{2}{|c|}{2}& 4\\
	
	\bottomrule
\end{tabular}
\caption{Pole order and contributing sectors for Landau-Ginzburg and hybrid models.}
\label{tab:1ParamSectorsHybLg}
\end{table}

%%%%%%%%%%%%%%%%%%%%%%%%%%%%%%%%%%%%%%%%%%%%%%%%%%%%%%%%%%%%%%%%%%%%%%%%%%%%%%
\subsection{Landau-Ginzburg phases}
We begin with those models of type F, which are Landau-Ginzburg orbifold models. Consulting Table \ref{tab:1ParamModels} these are the models F1, F2, F3 and F4.  The matrix $q$ that determines the $I$-function and the Gamma class is obtained by dividing  the GLSM charge vectors by the charge of the (single) $p$-field: 
\begin{equation}
  \label{1parqmat}
  q=\begin{pmatrix}
  1 & -\frac{1}{d_1}
  & -\frac{1}{d_1}
  & -\frac{1}{d_1}
  &-\frac{\alpha}{d_1}
  & -\frac{\beta}{d_1}
  \end{pmatrix}.
\end{equation}
    In these cases it is very easy to evaluate the sphere partition function because only first order poles contribute. This is a consequence of the fact that \(k=0\) in these models (see Table \ref{tab:1ParamModels}). Then
    (\ref{eqn:1ParamSmallSpherePartHybLg}) reads:
    \begin{equation}
      Z_{S^2}^{\zeta \ll 0}=\frac{1}{d_1} \sum_{\delta  \in narrow}
      (-1)^{\operatorname{Gr}} \frac{\widehat{\Gamma}_\delta(0)}{\widehat{\Gamma}^\ast_\delta(0)}
      \left|I^{\zeta \ll 0}_\delta(\pt,0)\right|^2,
      \label{eqn:1paramLgSpf}
    \end{equation}
    where the explicit \(\delta \) values can be read off from Table \ref{tab:1ParamSectorsHybLg} and it 
    can be shown that these values correspond to the narrow sectors as introduced in Section \ref{sec:lgFJRWBackground}.
	 Expressions (\ref{eqn:1ParamHybLgGamma}) and 
	 (\ref{eqn:1ParamHybLgGammaBar}) read:
    \begin{align}
    \widehat{\Gamma}_\delta(0) &= 
    	\Gamma \left( \left\langle  \frac{\delta}{d_1}
    	\right\rangle \right)^{3}
    	\Gamma \left(\left\langle  \alpha \frac{\delta}{d_1}
    	\right\rangle \right)
    	\Gamma \left( \left\langle  \beta \frac{\delta}{d_1}
    	\right\rangle \right)   
    ,\label{eqn:1paramGammaLgSpf1}\\
    \widehat{\Gamma}^\ast_\delta(0)
    &= \Gamma \left(
    	\left\langle  \frac{d_1 -\delta}{d_1}
    	\right\rangle \right)^{3}
    	\Gamma \left(
    	\left\langle  \alpha \frac{d_1 -\delta}{d_1}
    	\right\rangle \right)
    	\Gamma \left(
    	\left\langle  \beta \frac{d_1 -\delta}{d_1}
    	\right\rangle \right),
    	\label{eqn:1paramGammaLgSpf2}
    \end{align} 
    and inserting into (\ref{eqn:1ParamHybLgIfunc}) gives
    \begin{align}
    \begin{split}
    I^{\zeta \ll 0}_\delta( \pt,0 )
    &= \sum_{a=0}^\infty
   \frac
    {
    	 e^{\pt(a + \frac{\delta}{d_1}-q)}
    	(-1)^{a (3+ \alpha+\beta)} 
    }
    {
    	\Gamma \left( \left\langle  \frac{\delta}{d_1}
    	\right\rangle \right)^{3}
    	\Gamma \left(\left\langle  \alpha \frac{\delta}{d_1}
    	\right\rangle \right)
    	\Gamma \left( \left\langle  \beta \frac{\delta}{d_1}
    	\right\rangle \right) 
    }\\
    & \qquad \cdot   \frac{
    	\Gamma \left(a +\frac{\delta}{d_1}\right)^{3}
    	\Gamma\left(a \alpha +\frac{\alpha }{d_1}\delta\right)
    	\Gamma \left(a \beta  +\frac{\beta }{d_1}\delta\right)
    }{
    	\Gamma \left(\delta +a d_1\right)
    }.
    \end{split}
    \label{eqn:1paramLGIeval1}
    \end{align}
    The next step is to show that (\ref{eqn:1paramLgSpf}) matches (\ref{lgproposal}), which means in particular that the $I$-function, the Gamma class and the pairing matches with the definitions given in Section \ref{sec:lgFJRWBackground}. Since this is rather tedious we have relegated this discussion to Appendix \ref{sec:lgMatchOneParam}. By expanding 
     (\ref{eqn:1paramLgSpf}) in terms of \(\delta\) we can read off the matrix \(M\) introduced in (\ref{zs2matrix}):
     \begin{align}
     M=
     \begin{pmatrix}
     \frac{\gamma_{\delta_1}(0)}{d_1} & 0 & 0 &0 \\
     0& \frac{\gamma_{\delta_2}(0)}{d_1} & 0 & 0 \\
     0& 0 & -\frac{1}{d_1 \gamma_{\delta_2}(0)} & 0\\
     0& 0&  0&-\frac{1}{d_1 \gamma_{\delta_1}(0)}
     \end{pmatrix},
     \end{align}
     where we used (\ref{eqn:gammmaQuotient}) to write the result in 
     a compact way.

%%%%%%%%%%%%%%%%%%%%%%%%%%%%%%%%%%%%%%%%%%%%%%%%%%%%%%%%%%%%%%%%%%%%%%%%%%%%%%
    \subsection{Geometry}
    Next we consider the geometric phases $\zeta\gg0$. To evaluate the sphere partition function we follow the steps outlined in \cite{Hori:2013ika} in the context of the hemisphere partition function.
    The first step is to rewrite the contribution in the large
    radius phase, given in (\ref{eqn:zs21ParamLarge1}). We apply the   transformation
    \[
    \varepsilon \rightarrow -\frac{H}{2 \pi}
    \] 
    in (\ref{eqn:zs21ParamLarge1}) and 
    introduce 
    \begin{equation}
    \widehat{\Gamma}(H)= \frac{
    	\Gamma \left(1-\frac{H}{2\pi i} \right)^{5-n-j+k}
    	\Gamma \left(1- \alpha \frac{H}{2\pi i} \right)^n
    	\Gamma \left(1-  \beta \frac{H}{2\pi i} \right)^j
    }{
    	\Gamma \left(1-d_1 \frac{H}{2\pi i}\right)
    	\Gamma \left(1- d_2\frac{H}{2\pi i}\right)^k
    }. \label{eqn:1ParamLargeGamma1}
    \end{equation}
    Let us denote by  \(\widehat{\Gamma}^\ast\) the conjugate of
    \(\widehat{\Gamma}\) obtained by setting \(i\rightarrow -i\).
    Also we can  normalize
    the first summand in (\ref{eqn:zs21ParamLarge1Reg})
    to $1$ if we define\footnote{We observe 
    	that the alternating sign in the 
    	summation can be removed by a
    	$\theta$-angle shift between IR and 
    	UV theory
    	 (see e.g \cite{Herbst:2008jq}). 
    	 We will drop this, because it gets
    	   cancelled in the sphere partition
    	    function.}
    \begin{align}
    \begin{split}
    I^{\zeta \gg 0 }(\pt,H) &=
    \widehat{\Gamma}(H)^\ast \mathcal{Z}_{1,reg}\left(\frac{-H}{2\pi}\right) \\
    &=\frac{
    	\Gamma \left(1+\frac{H}{2\pi i} \right)^{5-n-j+k}
    	\Gamma \left(1+ \alpha \frac{H}{2\pi i} \right)^n
    	\Gamma \left(1+  \beta \frac{H}{2\pi i} \right)^j
    }{
    	\Gamma \left(1+d_1 \frac{H}{2\pi i}\right)
    	\Gamma \left(1+ d_2\frac{H}{2\pi i}\right)^k
    } \\
	& \quad \cdot  \sum_{a=0}^{\infty}  (-1)^{a (5+k-n-j+\alpha n+ j\beta )}
    u(\pt)^{(\frac{H}{2\pi i} +a+q)} \\
    &\qquad \cdot \frac{
    	\Gamma \left(1+a d_1
    	+  d_1 \frac{H}{2\pi i}\right)
    	\Gamma \left(1+a d_2+d_2 \frac{H}{2\pi i}\right)^k
    }{
    	\Gamma \left(1+ a+\frac{H}{2\pi i}\right)^{5+k-n-j}
    	\Gamma \left(1+a \alpha
    	+\alpha \frac{H}{2\pi i}\right)^n
    	\Gamma \left(1+a \beta + \beta \frac{H}{2\pi i}\right)^j
    }, \label{eqn:zs21ParamLargeI1}
    \end{split}
    \end{align}
    we introduced \(u(\pt)= e^{-\pt}\).
    We can now write the sphere partition function in the large
    radius phase as
    \begin{align}
    Z_{S^2}^{\zeta \gg 0}=(2\pi i)^3\frac{d_2^k d_1}{\alpha^n \beta^j}
    \oint_0 \frac{\mathrm{d} H}{2 \pi i} \frac{1}{H^4}
    \frac{\widehat{\Gamma}(H)}{\widehat{\Gamma}^\ast (H )}
    I^{\zeta \gg 0}(u(\pt),H)I^{\zeta \gg 0}(\overline{u}(\overline{\pt}),H).
    \label{eqn:1paramGeoResInt}
    \end{align}
     The crucial observation is now that the infrared 
    description of all one-parameter models in the large radius phase is given by a non-linear sigma model on a complete intersection Calabi-Yau \(X\) in weighted projective space of 
    type (\ref{eqn:phm1}).
	Recall that the total Chern class of the normal bundle \(\xi\) of \(X\) is given by
	\begin{equation}
	    c(\xi) = (1+ d_1 H)(1+d_2 H)^k,
	\end{equation}
	 where \(H\) is the hyperplane class of the ambient weighted projective space \(X_\Sigma\). 
    The normal bundle \(\xi\) has rank \(k+1\) and we get for the
    top Chern class:
	\begin{equation}
	    c_{k+1} (\xi) = d_1 d_2^k H^{k+1}.
	\end{equation}
    An integration along \(X\) can be pulled back 
    from the embedding space with the help of the  top Chern class of \(\xi\):
    \begin{align}
    \int_{X} g(H) &= \int_{X_\Sigma}
    c_{k+1} (\xi) \wedge g(H) \nonumber \\
    &= \frac{d_1 d_2^k}{3!} \frac{\partial^3}{\partial H^3} g(H)|{}_{H=0} = d_1 d_2^k \oint \frac{\mathrm{d} z}{2\pi i } \frac{1}{z^4} g(z).
    \label{eqn:intPullback}
    \end{align}
	We see that (\ref{eqn:1ParamLargeGamma1}) matches  (\ref{geomgamma}) and 
	by (\ref{eqn:intPullback}) we can write 
    \begin{equation}
    Z_{S^2}^{\zeta \gg 0}=\frac{(2\pi i)^3 }{\alpha^n \beta^j}
    \int_{X}
    \frac{\widehat{\Gamma}_{X}(H)}{
    	\widehat{\Gamma}^\ast_{X}(H)}
    I^{\zeta \gg 0}(u(\pt),H)I^{\zeta \gg 0}(\overline{u}(\overline{\pt}),H).
    \end{equation}
    To read off the matrix \(M\) introduced in (\ref{zs2matrix}) we expand the 
    different components in the integrand in powers of \(H\) and extract the 
    \(H^3\) coefficient. We obtain\footnote{We divide $M$ by $8 \pi^3$ in order to 
    get a canonically normalised $\zeta(3)$ term  in the geometric phase. See also  \cite{Morrison:2016bps,Jockers:2012dk} where similar normalisations have been applied. }
    \begin{align}
    \frac{M}{8 \pi^3}=
    \begin{pmatrix}
     \frac{\chi(X)  \zeta(3)}{4 \pi^3} & 0 & 0 & - i    \kappa \\
    0 & 0 &- i   \kappa & 0 \\
    0 &  - i  \kappa  & 0 & 0 \\
    - i  \kappa & 0 & 0& 0
    \end{pmatrix},
    \label{eqn:zs2InGeo1Param}
    \end{align}
    where \(\kappa = \frac{d_1 d_2^k}{ \alpha^n \beta^j }\) is
    the triple intersection number and \(\chi(X)\) the Euler number of the
    Calabi-Yau \(X\).
    In the pairing matrix (\ref{eqn:zs2InGeo1Param}) one can see the expected \(\zeta(3)\) coefficient.

%%%%%%%%%%%%%%%%%%%%%%%%%%%%%%%%%%%%%%%%%%%%%%%%%%%%%%%%%%%%%%%%%%
\subsection{K-type hybrid models}
\label{sec:kTypeHybrid}
Now we consider the models K1, K2 and K3 in Table \ref{tab:1ParamModels}, in the phase of a Landau-Ginzburg
orbifold with orbifold groups 
\(G=\mathbb{Z}_3,\mathbb{Z}_4,\mathbb{Z}_6\) fibered over \(\mathbb{P}^1\).
For these models \(k=1\)  and so we can bring (\ref{eqn:1ParamSmallSpherePartHybLg}), into the following 
form after the transformation 
\(\varepsilon \rightarrow \frac{H}{2 \pi i}\)
\begin{equation}
Z^{\zeta \ll 0}_{S^2,1} =\frac{2\pi i }{d_1} \sum_{\delta \in  Narrow}
\oint \frac{\mathrm{d} H}{2\pi i} \frac{1}{H^2} (-1)^{\operatorname{Gr}}\frac{\Gamma_\delta(H)}{\Gamma_\delta^\ast(H)}
I^{\zeta \ll 0}_\delta(\pt,H)I^{\zeta \ll 0}_\delta(\bar{\pt},H),
\label{eqn:kTypeZs2}
\end{equation} 
with
\begin{align}
\Gamma_\delta(H) &= \Gamma \left(1 -   \frac{H}{2 \pi i} \right)^2
\Gamma \left( \frac{H}{2 \pi i d_1}
+\left\langle  \frac{\delta}{d_1}
\right\rangle \right)^{6-n-j} \nonumber \\
& \quad \cdot \Gamma \left(\alpha \frac{H}{2 \pi i d_1}
+\left\langle  \alpha \frac{\delta}{d_1}
\right\rangle \right)^{n}
\Gamma \left(  \beta \frac{H}{2 \pi i d_1}
+\left\langle  \beta \frac{\delta}{d_1}
\right\rangle \right)^{j}  , \label{eqn:1ParamHybGamma}\\
\Gamma^\ast_\delta(H) &=
\Gamma \left(1 +  
\frac{H}{2 \pi i}\right)^2
\Gamma \left( -\frac{H}{2 \pi i d_1}
+\left\langle  \frac{d_1 -\delta}{d_1}
\right\rangle \right)^{6-n-j} \nonumber \\
& \quad \cdot  \Gamma \left(-\alpha \frac{H}{2 \pi i d_1}
+\left\langle  \alpha \frac{d_1 -\delta}{d_1}
\right\rangle \right)^{n}
\Gamma \left(-  \beta \frac{H}{2 \pi i d_1}
+\left\langle  \beta \frac{d_1 -\delta}{d_1}
\right\rangle \right)^{j} ,
\label{eqn:1ParamHybGammaBar}
\end{align}
and
\begin{align}
\begin{split}
\mathcal{I}^{\zeta \ll 0 }_\delta(\pt,H) &=
\frac
{\Gamma \left(1 +  
\frac{H}{2 \pi i}\right)^2
}
{
	\Gamma \left( \frac{H}{2 \pi i d_1}
	+\left\langle  \frac{\delta}{d_1}
	\right\rangle \right)^{6-n-j}
	\Gamma \left(\alpha \frac{H}{2 \pi i d_1}
	+\left\langle  \alpha \frac{\delta}{d_1}
	\right\rangle \right)^{n}
	\Gamma \left(  \beta\frac{H}{2 \pi i d_1}
	+\left\langle  \beta \frac{\delta}{d_1}
	\right\rangle \right)^{j}   
}\\
&\quad \cdot \sum_{a=0}^\infty
e^{\pt (\frac{H}{ 2 \pi i d_1} +a + \frac{\delta}{d_1}-q)}
(-1)^{a (6-n-j+ \alpha n + j\beta)} \\
&\qquad \cdot \frac{
	\Gamma\left(a+\frac{H}{2 \pi i d_1} +\frac{\delta}{d_1}\right)^{6-n-j} 
	\Gamma \left(a \alpha + \alpha \frac{H}{2 \pi i d_1} +\frac{\alpha }{d_1}\delta\right)^n
	\Gamma \left(a \beta +\beta \frac{H}{2 \pi i d_1}+\frac{\beta }{d_1}\delta\right)^j
}{
	\Gamma \left(\delta +a d_1 +\frac{H}{2 \pi i}\right)^2
}.
\end{split}
\label{eqn:1ParamHybKI}
\end{align} 
The vacuum manifold is \(B=\mathbb{P}^1\) and similar to (\ref{eqn:intPullback})  we can write the sphere partition function as 
\begin{equation}
Z^{\zeta \ll 0}_{S^2,1} =\frac{2 \pi i}{d_1} \sum_{\delta \in  Narrow}
\int_{\mathbb{P}^1} (-1)^{\operatorname{Gr}}\frac{\Gamma_\delta(H)}{\Gamma_\delta^\ast(H)}
I^{\zeta \ll 0}_\delta(\pt,H)I^{\zeta \ll 0}_\delta(\bar{\pt},H).
\label{eqn:1paramHybridSpfFinal}
\end{equation} 
As in the previous examples this can be rewritten in a matrix notation  (\ref{zs2matrix}). Therefore 
we expand each \(\delta\) sector in (\ref{eqn:1paramHybridSpfFinal}) in \(H\) and 
extract the \(H^1\) component. By inserting
(\ref{eqn:1ParamHybGamma}) and (\ref{eqn:1ParamHybGammaBar}) into (\ref{eqn:gammmaQuotient}) the matrix  \(M\) takes the form
\begin{align}
M&=
\begin{pmatrix}
-\frac{\nu}{d_1^2} \gamma_{\delta_1}(0) & 2\pi i \frac{1}{d_1}\gamma_{\delta_1}(0) &0 & 0\\
2\pi i\frac{1}{d_1} \gamma_{\delta_1}(0) & 0 & 0 & 0 \\
0 & 0 & -\frac{\nu}{d_1^2}\frac{1}{ \gamma_{\delta_1}(0)} & 2\pi i\frac{1}{d_1} \frac{1}{ \gamma_{\delta_1}(0)} \\
0 & 0 & 2\pi i\frac{1}{ d_1} \frac{1}{ \gamma_{\delta_1}(0)}  & 0
\end{pmatrix}
\label{eqn:1ParamPariK}.
\end{align}
Evaluating \(\nu\) for the  K type models
gives
\begin{align}
\begin{array}{c|c|c|c}
& \mathrm{K1} & \mathrm{K2} & \mathrm{K3}\\
\hline
\nu & \log 3^{18} &  \log 2^{40} &
\log\left( 2^{32} 3^{18}\right)
\end{array}.
\end{align}
Hybrid models have also been studied in mathematics and therefore
we want to match our results with those in the literature.
We focus on the K1 model which was studied in \cite{MR3590512,zhao2019landauginzburgcalabiyau} in the context of 
FJRW theory.
The definition of the  \(I\) function can be found in\footnote{We are using the same notation as \cite{MR3590512} here. The parameter $t$ is not the flat coordinate but is, as we will show, related to the FI-theta parameter $\pt$. } 
\cite{MR3590512}: 
\begin{align}
I_{hyb} &= z \sum_{d > 0 \atop d  \not\equiv -1 \mod 3}
e^{\left(d +1 +\frac{H^{(d+1)}}{z}\right)t} z^{
	- 6 \langle\frac{d}{3} \rangle} 
\frac{\Gamma \left(\frac{H^{(d+1)}}{3z} + \frac{d}{3} + \frac{1}{3}\right)^6}{
	\Gamma \left(\frac{H^{(d+1)}}{3z}
	+\langle \frac{d}{3} \rangle+ \frac{1}{3}\right)^6
}
\frac{\Gamma \left(\frac{H^{(d+1)}}{z} + 1\right)^2}{
	\Gamma \left(\frac{H^{(d+1)}}{z}+d+1\right)^2
}.
\end{align} 
We can simplify the above sum by replacing \(
d = 3n + \delta \), with \( \delta=0,1\). 
In this case we always have
\(\lfloor \frac{\delta}{3} \rfloor = 0\), so we can drop the  \(\langle \cdot \rangle\) operations in the above formulas. Further we note that the label in the superscript of
\(H^{(3n+\delta)}\) is defined modulo 3:
\begin{equation}
H^{(3n+\delta)} = H^{(\delta)}.
\end{equation}
After performing the shift \(\delta +1 \rightarrow \delta\) we find:
\begin{align}
I_{hyb}&= z \sum_{\delta=1}^2\sum_{n = 0}^{\infty}
e^{\left(3n + \delta  +\frac{H^{(\delta)}}{z}\right)t} z^{
	- 2(\delta-1) } 
&\frac{\Gamma \left(\frac{H^{(\delta)}}{3z} + \frac{\delta}{3} + n\right)^6}{
	\Gamma \left(\frac{H^{(\delta)}}{3z}
	+ \frac{\delta}{3}  \right)^6
}
\frac{\Gamma \left(\frac{H^{(\delta)}}{z} + 1\right)^2}{
	\Gamma \left(\frac{H^{(\delta)}}{z}+3n+\delta\right)^2
}.\label{eqn:CubicMath}
\end{align}
Specialising (\(\ref{eqn:1ParamHybKI}\)) to the K1 model we obtain
\begin{align}
\begin{split}
I^{\zeta \ll 0 }_\delta(\pt,H) &=
\frac
{\Gamma \left(1 +  
	\frac{H}{2 \pi i}\right)^2
}
{
	\Gamma \left(\frac{H}{3\cdot 2 \pi i }
	+ \frac{\delta}{3}
	\right)^{6}
}
\sum_{a=0}^\infty
e^{\pt\left( \frac{H}{3\cdot 2 \pi i} +a + \frac{\delta}{3}-q\right)}
(-1)^{6a}
\frac{
	\Gamma^{6} \left(a+\frac{H}{3 \cdot 2 \pi i} +\frac{\delta}{3}\right)
}{
	\Gamma^2 \left(\delta +3 a +\frac{H}{2 \pi i}\right)
}.\label{eqn:1ParamHybCubic}
\end{split}
\end{align} We can match (\(\ref{eqn:1ParamHybCubic}\)) and
(\(\ref{eqn:CubicMath}\)) if we identify\footnote{With our approach we cannot unambiguously fix the value of the parameter $z$, because the sphere partition function is not affected by overall signs. Both, $z=1$ and $z=-1$ are consistent. To resolve this, one would have to analyse the $J$-function and the enumerative invariants.}: 
\begin{align}
 q&= 0, & H^{(\delta)}&= \frac{H}{2 \pi i},  &z &= 1, &
e^{3t} & = e^{\pt}.
\label{eqn:kmatchI}
\end{align}
The superscript of \(H^{(\delta)}\) in (\ref{eqn:CubicMath}) 
labels the sector of the narrow state space. 
We do not see this label explicitly in the 
sphere partition function, because the pairing is partially evaluated. 
%%%%%%%%%%%%%%%%%%%%%%%%%%%%%%%%%%%%%%%%%%%%%%%%%%%%%%%%%%%%%%%%%%%%%%%%%
\subsection{M-type model}
\label{sec:mTypeHybrid}
There is only one model that has M-type monodromy in the $\zeta \ll 0$-phase. This model has been studied in detail in \cite{Caldararu:2007tc}. The sphere partition function and Gromov-Witten invariants have been computed in \cite{Sharpe:2012ji}. The interesting feature of this model is that the moduli space has two points that behave like large volume phases and that the two Calabi-Yaus associated to these points are not birational. In this sense this model shares many features with non-abelian GLSMs. While the $\zeta \ll 0$-phase turns out to be geometric, the analysis of the phase of the GLSM is  much closer to a hybrid model. The vacuum manifold is a $\mathbb{P}^3$ defined by the $p$-fields. Turning on fluctuations of the $x$-fields gives a theory with potential of the form
\begin{equation}
  W=\sum_{i,j}x_i A^{ij}(p)x_j.
\end{equation}
The $x$s are massive except when $\det A=0$. It has been shown in \cite{Caldararu:2007tc} that the $\zeta \ll 0$-phase is the non-commutative resolution of a singular branched double cover over $\mathbb{P}^3$ with branching locus $\det A=0$. 

Many steps in the calculation of the sphere partition function are  
similar to the models of K-type. The only difference is that the vacuum manifold is now a \(\mathbb{P}^3\).  From Table 
\ref{tab:1ParamModels} we can read off that \(k=3\) and \(d_1=d_2=2\). Again we apply \(\varepsilon 
\rightarrow \frac{H}{2\pi i}\) whereupon (\ref{eqn:1ParamSmallSpherePartHybLg}) takes the form
\begin{equation}
Z^{\zeta \ll 0}_{S^2,1} =\frac{(2 \pi i )^3}{2} 
\int_{\mathbb{P}^3}  (-1)^{\operatorname{Gr}}
\frac{\Gamma_1 (H)}{\Gamma^\ast_1(H)}
| I^{\zeta \ll 0}_1
( \pt,H)|^2,
\label{eqn:mTypePartFunc}
\end{equation}
with (\ref{eqn:1ParamHybLgIfunc}):
\begin{align}
\begin{split}
I^{\zeta \ll 0}_1(\pt,H) &= \frac{	\Gamma\left(1+ \frac{H}{2\pi i} \right)^{4}}{\Gamma\left(\frac{H}{2 \cdot 2 \pi i} +   \frac{1}{2} \right)^{8}} \sum_{a=0}^\infty
e^{\pt(\frac{H}{2 \cdot 2 \pi i} +a + \frac{1}{2}-q)}
(-1)^{8a }
\frac{ \Gamma \left(a+ \frac{H}{2 \cdot 2 \pi i} +\frac{1}{2}\right)^{8}
}{
	\Gamma \left(1 +2a +\frac{H}{2\pi i} \right)^{4}},
\end{split}
\label{eqn:M1IFunc}
\end{align}
and
(\ref{eqn:1ParamHybLgGamma}), \ref{eqn:1ParamHybLgGammaBar}) are given by:
\begin{align}
\Gamma_1 (H)=\Gamma\left(1- \frac{H}{ 2 \pi i}\right)^{4}&
\Gamma\left(  \frac{1}{2}+\frac{H}{2 \cdot 2 \pi i}  \right)^{8}, &
\Gamma^\ast_1(H)=\Gamma\left(1+ \frac{H}{2 \pi i} \right)^{4}&
\Gamma\left( \frac{1}{2}-\frac{H}{2 \cdot 2 \pi i}\right)^{8} .
\end{align} 
Here we used the fact that \(\delta\) only takes the value
\(1\) for M1. The  matrix \(M\) (\ref{zs2matrix}) is given by 
\begin{align}
M=
\begin{pmatrix}
-\frac{\tau^3}{12} - \zeta (3) &i \pi \frac{\tau^2}{2}&2 \pi^2 \tau &-4 i \pi^3 \\
i \pi \frac{\tau^2}{2} & 2 \pi^2 \tau &-4 i \pi^3  & 0 \\
2 \pi^2 \tau & -4 i \pi^3  &0& 0 \\
-4 i \pi^3  & 0 &0 & 0
\end{pmatrix},
\label{eqn:1ParamPariM}
\end{align}
with 
\begin{align}
\tau =  \log 2^{16}.
\end{align}
We can now compare (\ref{eqn:zs2InGeo1Param})
and (\ref{eqn:1ParamPariM}). Although both 
points are points of maximal unipotent monodromy 
the structure of (\ref{eqn:1ParamPariM}) differs from
the structure of \(M\) in geometry.

This model was also studied in \cite{MR3590512}, where
the \(I\)-function was shown to be 
\begin{align}
I_{hyb}(t)
&=\sum_{d>0 \atop d \not \equiv -1 \mod 2}
\frac{z e^{(d+1 + \frac{H^{(d+1)}}{z})t}}{2^{8 \lfloor
		\frac{d}{2}\rfloor}}
\frac{\prod\limits_{1 \leq b \leq d \atop b \equiv d+1 \mod 2}\left(H^{(d+1)}+b z\right)^8}{\prod\limits_{1 \leq b \leq d}\left(H^{(d+1)}+b z\right)^4}.
\end{align}
We can explicitly take into account the restriction on \(d\)
by writing \(d=2n\) and by  simplifying the products over \(b\) one gets
\begin{align}
I_{hyb}(t) = \sum_{n=0}^\infty
\frac{z e^{(2n+1 + \frac{H^{(2n+1)}}{z})t}}{2^{8 \lfloor
		\frac{2n}{2}\rfloor}}
\frac{\prod_{s=1}^{n}\left(H^{(2n+1)}+2nz +z -2sz \right)^8}{\prod_{b=1}^{2n}\left(H^{(2n+1)}+b z\right)^4}.
\end{align}
We use the identity: 
\begin{align}
z^l \frac{\Gamma \left(1 + \frac{x}{z}+l\right)}
{ \Gamma \left(1+ \frac{x}{z} \right)} &=
\prod_{k=1}^l \left( x+ kz \right)
\end{align}
and find 
\begin{align}
I_{hyb}(t) 
&=\frac{\Gamma \left(1+ \frac{H^{(1)}}{z} \right)^4}{\Gamma \left(\frac{1}{2}+ \frac{H^{(1)}}{2z}
	\right)^8} \sum_{n=0}^\infty
z e^{(2n+1 + \frac{H^{(1)}}{z})t}
\frac{\Gamma \left(\frac{1}{2}+
	\frac{H^{(1)}}{2z}+n\right)^8
	}
{ 
	\Gamma\left(1+ \frac{H^{(1)}}{z}+2n\right)^4}.
\end{align}
The exponent on \(H^{(1)}\) labels the state space sector, 
see also the sentence bellow (\ref{eqn:kmatchI}).
We can match the above result with
(\ref{eqn:M1IFunc}) if we identify: 
\begin{align}
q&= 0,  & H^{(1)}&= \frac{H}{2\pi i }, & z&=1, &
e^{2t} &= e^{\pt}.
\end{align}

As a final remark, note the factor $2$ in the overall normalisation of the sphere partition function (\ref{eqn:mTypePartFunc}) that must come from the pairing. This is consistent with the $\mathbb{Z}_2$ that encodes the information about the double cover in this phase \cite{Caldararu:2007tc}.   	
%%%%%%%%%%%%%%%%%%%%%%%%%%%%%%%%%%%%%%%%%%%%%%%%%%%%%%%%%%%%%%%%%%%%%%%%%%
\subsection{Pseudo-Hybrid-Models}
The pseudo-hybrid phases of this class of models have been discussed in \cite{Erkinger:2019umg}. One distinguishing feature of these models is that the phases have several components in the sense that the vacuum equations of the GLSM allow for different types of solutions. The existence of these components is also responsible for the fact that there is no unique R-charge assignment in the IR theory. The properties of the different components is reflected in the pole structure of the sphere partition function.  

Pseudo-hybrid phases appear in the models with 
a C-type singularity and also for the F-type singularity 
models F1, F6 and F7 (see Table \ref{tab:1ParamModels}).
The sphere partition functions of C-type models have a mixture of first order pole contributions
and a second order pole contribution. F-type models have 
only first order pole contributions. Therefore we 
will study this two types separately.
Details of the evaluation are given in Appendix 
\ref{sec:pseudoHybridDetails} and we will only present the  
final results here. In all models the main task is 
to rewrite (\ref{eqn:pseudoGeneralSing1}) and (\ref{eqn:pseudoGeneralSing2})
by using (\ref{eqn:reflectionFormula}).

Our results indicate that there may be a sensible definition for pairings, $I$-functions and Gamma classes for each individual component. It would be interesting to see if this also makes sense mathematically.

\subsubsection{F-type models}
As discussed in \cite{Erkinger:2019umg}, the pseudo-hybrid phase has features of two different Landau-Ginzburg models with orbifold groups $\mathbb{Z}_{d_1}$ and $\mathbb{Z}_{d_2}$. Consistently, the two contributions to the sphere partition functions only have first order poles, and also the twisted sectors associated to the corresponding orbifold groups make an appearance.

Because we only have first order poles we can directly
evaluate the sphere partition function and  get 
\begin{align}
\begin{split}
Z_{S^2}^{\zeta \ll 0 }&=  \frac{1}{d_1} \sum_{\delta=1}^{d_1 -1 }
(-1)^{\operatorname{Gr}}\frac{\widehat\Gamma_{\delta}(0)}{\widehat\Gamma_{\delta}^\ast(0)}
I_{\delta}(\pt,0)I_{\delta}(\bar{\pt},0) \\
& \quad+ \frac{1}{d_2} \sum_{\delta=1}^{\tau_{d_2}-1}\sum_{\gamma=0}^{\kappa_2-1}
(-1)^{\widetilde{\operatorname{Gr}}} \frac{\widetilde{\widehat\Gamma}_{\delta}(0)}{\widetilde{\widehat\Gamma}^\ast_{\delta}(0)}
\widetilde{I}_{\delta,\gamma}(\pt,0)\widetilde{I}_{\delta,\gamma}(\bar{\pt},0),
\end{split}
\label{eqn:f5f6f7Spf}
\end{align}
with  parameters defined in (\ref{eqn:kappaTau}).
Here we  introduced
\begin{align}
\widehat{\Gamma}_{\delta}(0)&=
\Gamma\left(  \left\langle
\tau_{d_2}\frac{\tau_{d_1}-\delta}{\tau_{d_1}}\right\rangle\right)^k
%\frac{\tau_{d_2}}{\tau_{d_1}} \delta\right\rangle\right)^k
\Gamma\left( \left\langle
\frac{\delta}{d_1}\right\rangle\right)^{5+k-n-j}
\Gamma\left(  \left\langle
\alpha \frac{\delta}{d_1}\right\rangle\right)^{n}
\Gamma\left( \left\langle
\beta \frac{\delta}{d_1}\right\rangle\right)^j
\label{eqn:pseudoFtypeGamma1},\\
(-1)^{\operatorname{Gr}}&= (-1)^\delta
(-1)^{ k\left\lfloor \tau_{d_2}\frac{\delta}{\tau_{d_1}} \right\rfloor}
(-1)^{(5+k-n-j)\left\lfloor \frac{\delta}{d_1} \right\rfloor}
(-1)^{n\left\lfloor \alpha \frac{\delta}{d_1} \right\rfloor}
(-1)^{j\left\lfloor \beta\frac{\delta}{d_1} \right\rfloor}
\label{eqn:pseudoFtypeGrading1}.
\end{align}
Taking into account that \(k=1\) for all F-type models,
 \begin{align}
\widetilde{\widehat{\Gamma}}_{\delta}(0)
&=   \Gamma \left(
%\left\langle \tau_{d_1} \frac{\delta }{\tau_{d_2}}\right\rangle \right)
\left\langle \tau_{d_1} \frac{\tau_{d_2}-\delta }{\tau_{d_2}}\right\rangle \right)
\Gamma\left(\left\langle \frac{\delta + \tau_{d_2} \gamma}{d_2} \right\rangle\right)^{6-n-j} 
 \Gamma\left(  \left\langle \alpha \frac{\delta + \tau_{d_2} \gamma}{d_2} \right\rangle\right)^{n}
\Gamma\left( \left\langle \beta
\frac{\delta + \tau_{d_2} \gamma}{d_2} \right\rangle \right)^j,
\label{eqn:pseudoFtypeGamma2} \\
(-1)^{\widetilde{\operatorname{Gr}}} 
&=(-1)^\delta (-1)^{\gamma (\tau_{d_2}+\tau_{d_1})}(-1)^{ \left\lfloor \frac{d_2}{d_1}\delta \right\rfloor}
(-1)^{(6-n-j)\left\lfloor\frac{\delta + \tau_{d_2} \gamma}{d_2}\right\rfloor}\
(-1)^{n\left\lfloor \alpha \frac{\delta + \tau_{d_2} \gamma}{d_2} \right\rfloor}
(-1)^{j\left\lfloor \beta \frac{\delta + \tau_{d_2} \gamma}{d_2}\right\rfloor },
\end{align}
where \(\gamma\) is introduced in the process of rewriting the sum over the poles (see (\ref{eqn:zs21ParamLarge2Sing})).
The conjugate expressions follow from (\ref{eqn:gammaRelations}).
 Next we define: 
 \begin{align}
\begin{split}
I_{\delta}(\pt,0) &= \frac{
	\Gamma\left(  \left\langle
	\frac{\tau_{d_2}}{\tau_{d_1}} \delta\right\rangle\right)^k
	\Gamma\left(
	\left\langle \tau_{d_2}\frac{\tau_{d_1}-\delta}{\tau_{d_1}} \right\rangle \right)^k
}{\widehat\Gamma_{\delta}(0)}\sum_{a=0}^\infty
e^{\pt(a + \frac{\delta}{d_1}-q)}
(-1)^{a (5+k-n-j+ \alpha n + j\beta)} \\
& \quad \cdot \frac{
	\Gamma \left(a +\frac{\delta}{d_1}\right)^{5+k-n-j}
	\Gamma\left(a \alpha +\frac{\alpha }{d_1}\delta\right)^n
	\Gamma \left(a \beta  +\frac{\beta }{d_1}\delta\right)^j
}{
	\Gamma \left(\delta +a d_1 \right)
	\Gamma \left(a d_2
	+\frac{\tau_{d_2}}{\tau_{d_1}}\delta\right)^k},
\label{eqn:pseudoFtypeI1}
\end{split}
\end{align} 
and
\begin{align}
\begin{split}
\widetilde{I}_{\delta}(\pt,0)
&=\frac{ \Gamma \left(
	\left\langle \tau_{d_1} \frac{\delta }{\tau_{d_2}}\right\rangle \right)
	\Gamma \left( \left\langle \tau_{d_1} \frac{\tau_{d_2}-\delta }{\tau_{d_2}}\right\rangle  
	\right)
}{\widetilde{\widehat\Gamma}_{\delta}(0)}\sum_{a=0}^{\infty}(-1)^{a (6-n-j+ \alpha n + j \beta)}
e^{\pt(a + \frac{\tau_{d_2}\gamma+ \delta}{d_{2}}-q)} \\
& \quad \cdot \frac{
	\Gamma\left(a+\frac{\tau_{d_2}\gamma+ \delta}{d_{2}}\right)^{6-n-j} 
	\Gamma \left(a \alpha
	+ \alpha \frac{\tau_{d_2}\gamma+ \delta}{d_{2}}\right)^n
	\Gamma \left(a \beta +
	\beta\frac{\tau_{d_2}\gamma+ \delta}{d_{2}}\right)^j
}{
	\Gamma \left(\frac{\tau_{d_1}}{\tau_{d_2}}\delta
	+ d_1 a  + \tau_{d_1} \gamma \right)
	\Gamma \left(\delta+d_2 a +\tau_{d_2}\gamma\right)
}.
\end{split}
\end{align} 

The structure of (\ref{eqn:f5f6f7Spf}) highly resembles the result in the
Landau-Ginzburg phases (\ref{eqn:1paramLgSpf}), except there are now two contributions. Additionally,
expressions (\ref{eqn:pseudoFtypeGamma1}) and (\ref{eqn:pseudoFtypeGamma2}) that we would like to identify with the Gamma class, come with an extra term compared to the pure
Landau-Ginzburg phases (see (\ref{eqn:1paramGammaLgSpf1}) and 
(\ref{eqn:1paramGammaLgSpf2})). The is also
visible in the \(I\)-function whose structure is more along the lines of hybrid models (\ref{eqn:1ParamHybKI}). Note that the
second contribution is absent for the F7 model, consistent with the observation that one of the Landau-Ginzburg models appearing as a component is massive.

\subsubsection{C-type models}
The C-type phases are closer to good hybrid models in the sense that there is a base manifold $B$ of non-zero dimension. In all three cases there is a component with one-dimensional $B$ and a Landau-Ginzburg component. More details can be found in \cite{Erkinger:2019umg}. This structure is also reflected in the sphere partition function, where we encounter first and second order poles.  

Here we will first discuss the C1 and C2 models before we come to the C3
model. The models differ in the structure of the sphere partition function. This again seems to relate to the different ways the two components emerge in C1 and C2, compared to C3.

\subsubsection*{C1 and C2}

For both models \(Z_{S^2,2}^{\zeta \ll 0 }=0\), while 
\(Z_{S^2,1}^{\zeta \ll 0}\) splits into two components with first 
and second order poles, respectively. The part with the
second order poles is given for  \(\delta= \tau_{d_1}\).
This allows to split \(Z_{S^2,1}^{\zeta \ll 0}\) into \begin{align}
\begin{split}
Z_{S^2,1}^{\zeta \ll 0} &=  \frac{1}{d_1}\left.\sum_{\delta}
\right|_{\delta  \not = \tau_{d_1}}
(-1)^{\operatorname{Gr}} \frac{\widehat \Gamma_{\delta}(0)}{\widehat\Gamma_{\delta}^\ast(0)}
I_{\delta}(\pt,0)I_{\delta}(\bar{\pt},0) \\
& \quad +\frac{2 \pi i}{d_2} \oint \frac{\mathrm{d}  \varepsilon}{2\pi i}
\frac{(-1)^{\widetilde{\operatorname{Gr}}}}{\varepsilon^2 } \frac{\widetilde{\widehat\Gamma}(\varepsilon)}{\widetilde{\widehat\Gamma}^\ast(\varepsilon)}
\widetilde{I}(\pt,\varepsilon) \widetilde{I}(\bar{\pt},\varepsilon).
\end{split}
\label{eqn:c1c2Spf}
\end{align} 
In the above equation \((-1)^{\operatorname{Gr}}\), the
\(\widehat\Gamma_{\delta}(0), \widehat\Gamma^\ast_{\delta}(0)\) functions and
\(I_{\delta}(\pt,0)\) have the same structure as in the F-type examples (see
(\ref{eqn:pseudoFtypeGrading1}),(\ref{eqn:pseudoFtypeGamma1})
and (\(\ref{eqn:pseudoFtypeI1}\)) respectively). In the second
contribution we used the following quantities: 
\begin{align}
\widetilde{\widehat\Gamma}(\varepsilon) &= \Gamma \left(1-\frac{\varepsilon}{2 \pi i}\right)
\Gamma \left(1- \frac{\tau_{d_2}}{\tau_{d_1}}\frac{\varepsilon}{2 \pi i}\right)
\Gamma \left(\frac{\varepsilon}{2 \pi i d_1}+ \left\langle \frac{1}{k_2} \right\rangle\right)^{6-n-j}
\nonumber \\
&\quad \cdot \Gamma \left(\alpha \frac{\varepsilon}{2 \pi i d_1}+ \left\langle \alpha\frac{1}{k_2} \right\rangle\right)^{n}
\Gamma \left(\beta \frac{\varepsilon}{2 \pi i d_1}+ \left\langle \beta\frac{1}{k_2} \right\rangle\right)^{j},\\
(-1)^{\widetilde{\operatorname{Gr}}}&=(-1)^{\tau_{d_1}}(-1)^{\tau_{d_2}}
(-1)^{6-n-j \left\lfloor \frac{1}{k_2}\right\rfloor}
(-1)^{n\left\lfloor \frac{\alpha}{k_2}\right\rfloor}
(-1)^{j\left\lfloor \frac{\beta}{k_2}\right\rfloor},
\end{align}
where one can obtain the conjugate expressions by using 
(\ref{eqn:gammaRelations})  and
 \begin{align}
\begin{split}
\widetilde{I}(\pt,\varepsilon)&= \frac{\Gamma\left(1-\frac{\varepsilon}{2 \pi i}\right)
	\Gamma\left(1+\frac{\varepsilon}{2 \pi i}\right)
	\Gamma \left(1- \frac{\tau_{d_2}}{\tau_{d_1}}\frac{\varepsilon}{2 \pi i}\right)
	\Gamma \left(1+ \frac{\tau_{d_2}}{\tau_{d_1}}\frac{\varepsilon}{2 \pi i}\right)
}{\widetilde{\widehat\Gamma}(\varepsilon)} \\
& \quad \cdot  \sum_{a=0}^\infty
e^{\pt( \frac{\varepsilon}{2 \pi i d_1} +a + \frac{1}{\kappa_2}-q)}
(-1)^{a (6-n-j+ \alpha n + j\beta)} \\
&\qquad \cdot \frac{
	\Gamma\left(a+\frac{\varepsilon}{2 \pi i d_1}  +\frac{1}{\kappa_2}\right)^{6-n-j} 
	\Gamma \left(a \alpha + \alpha \frac{\varepsilon}{2 \pi i d_1}  +\frac{\alpha }{\kappa_2}\right)^n
	\Gamma \left(a \beta + \beta \frac{\varepsilon}{2 \pi i d_1}  +\frac{\beta }{\kappa_2}\right)^j
}{
	\Gamma \left(\tau_{d_2} +a d_1+ \frac{\varepsilon}{2 \pi i}\right)
	\Gamma \left(a d_2+\tau_{d_2} \frac{\varepsilon}{2 \pi i \tau_{d_1}}
	+\tau_{d_1}\right)}.
\end{split}
\end{align} 
Comparing with (\ref{eqn:c1c2Spf}) we
see that the first line resembles the result in the Landau-Ginzburg case (\ref{eqn:1paramLgSpf}) and 
the second line is similar to the result for the  hybrid models (\ref{eqn:1paramHybridSpfFinal}) .

\subsubsection*{C3}
In contrast to the C1 and C2 model we now have \(Z_{S^2,2}^{\zeta \ll 0} \not = 0\),
whereas \(Z_{S^2,1}^{\zeta \ll 0}\) has only first order poles. Making use of Table \ref{tab:1ParamSmallConclusion} we  can bring  the sphere partition function into the
following form 
\begin{align}
\begin{split}
Z_{S^2}^{\zeta \ll 0 }&=\frac{1}{d_1} \sum_{\delta} (-1)^{\operatorname{Gr}}
\frac{\widehat\Gamma_{\delta}(0)}{\widehat\Gamma^\ast_{\delta}(0)} I_{\delta}(\pt,0)
I_{\delta}(\bar{\pt},0) \\
&\quad  +\frac{2 \pi i }{ d_2} \oint \frac{\mathrm{d}  \varepsilon}{2\pi i}   
\frac{(-1)^{\widetilde{\operatorname{Gr}}}}{\varepsilon^2}
\frac{\widetilde{\widehat\Gamma}(\varepsilon)}{\widetilde{\widehat\Gamma}^\ast(\varepsilon)} \widetilde{I}(\pt,\varepsilon)
I(\bar{\pt},\varepsilon),
\label{eqn:c3Spf}
\end{split}
\end{align} 
where \((-1)^{\operatorname{Gr}}\), the
\(\widehat\Gamma_{\delta}(0)\), \(\widehat\Gamma^\ast_{\delta}(0)\) functions and
\(I_{\delta}(\pt,0)\), similar to the C1 and C2 model, are given by the F-type
expressions
(\ref{eqn:pseudoFtypeGrading1}), (\ref{eqn:pseudoFtypeGamma1}), 
and (\ref{eqn:pseudoFtypeI1}), respectively. In the second term we have
introduced the following quantities
\begin{align}
\widetilde{\widehat\Gamma}(\varepsilon) &= \Gamma \left(1- \frac{\varepsilon}{2 \pi i} \right)^2  \Gamma \left( - \tau_{d_1}\frac{\varepsilon}{2 \pi i\tau_{d_2}}
+\left\langle \tau_{d_1} \frac{\tau_{d_2}-1 }{\tau_{d_2}}\right\rangle  
\right)
\Gamma\left(\frac{\varepsilon}{2 \pi i d_2} +\left\langle \frac{1 }{d_2}
\right\rangle\right)^{7-n-j} \nonumber \\
& \quad \cdot \Gamma\left(  \alpha \frac{\varepsilon}{2 \pi i d_2}
+ \left\langle \frac{\alpha }{d_2}
\right\rangle\right)^{n}
\Gamma\left(\beta \frac{\varepsilon}{2 \pi i d_2}
+ \left\langle \frac{ \beta }{d_2}
\right\rangle \right)^j
\\
(-1)^{\widetilde{\operatorname{Gr}}}&=(-1)^{ \left\lfloor \frac{d_2}{d_1} \right\rfloor}
(-1)^{(7-n-j)\left\lfloor\frac{1 }{d_2} \right\rfloor}
(-1)^{n\left\lfloor  \frac{\alpha }{d_2} \right\rfloor}
(-1)^{j\left\lfloor  \frac{ \beta }{d_2}
	\right\rfloor },
\end{align} 
and 
\begin{align}
\begin{split}
\widetilde{I}(\varepsilon,\pt)&=\frac{\Gamma \left(1- \frac{\varepsilon}{2 \pi i} \right)^2
	\Gamma \left(1+ \frac{\varepsilon}{2 \pi i} \right)^2
	\Gamma \left( - \tau_{d_1}\frac{\varepsilon}{2 \pi i\tau_{d_2}}
	+\left\langle \tau_{d_1} \frac{\tau_{d_2}-1 }{\tau_{d_2}}\right\rangle  
	\right)
	\Gamma \left(\tau_{d_1}\frac{\varepsilon}{2 \pi i\tau_{d_2}}
	+\left\langle \tau_{d_1} \frac{1}{\tau_{d_2}}\right\rangle \right)}{\widetilde{\widehat\Gamma}(\varepsilon)}
\\
& \quad \cdot \sum_{a=0}^{\infty}(-1)^{a (7-n-j+ \alpha n + j \beta)}
e^{\pt(\frac{\varepsilon}{2 \pi i d_2}+a  + \frac{1}{d_2}-q)} \\
&\qquad \cdot \frac{
	\Gamma \left(a+\frac{\varepsilon}{2 \pi i d_2} +\frac{1}{d_2}
	\right)^{7-n-j}
	\Gamma \left(a \alpha + \alpha \frac{\varepsilon}{2 \pi i d_2}
	+\frac{\alpha }{d_2}
	\right)^n
	\Gamma \left(a \beta + \beta \frac{\varepsilon}{2 \pi i d_2}
	+\frac{ \beta }{d_2} \right)^j
}{
	\Gamma \left(\frac{\tau_{d_1}}{\tau_{d_2}}
	+ d_1 a + \tau_{d_1}\frac{\varepsilon}{2 \pi i\tau_{d_2}} \right)
	\Gamma \left(1+d_2 a+ \frac{\varepsilon}{2 \pi i} \right)^2
}.
\end{split}
\end{align}
Again we see that the sphere partition function (\ref{eqn:c3Spf}) has a part which looks Landau-Ginzburg-like and a second contribution which resembles the 
hybrid case.

%%%%%%%%%%%%%%%%%%%%%%%%%%%%%%%%%%%%%%%%%%%%%%%%%%%%%%%%%%%%%%%%%%
\section{Two-parameter example}
\label{sec-twopar}
The results discussed in this article also apply to examples with more than one K\"ahler parameter. We consider one of the standard examples of a two-parameter model \cite{Candelas:1993dm,Hosono:1993qy}. The GLSM has $\pG=U(1)^2$ with field content
\begin{equation}
  \begin{array}{c|rr|rrrrr|c}
    &p&x_6&x_3&x_4&x_5&x_1&x_2&\mathrm{FI}\\
    \hline
    U(1)_1&-4&1&1&1&1&0&0&\zeta_1\\
    U(1)_2&0&-2&0&0&0&1&1&\zeta_2\\
    \hline
    U(1)_V&2-8q_1&2q_1-4q_2&2q_1&2q_1&2q_1&2q_2&2q_2,
    \end{array}
\end{equation}
where $0\leq q_1\leq\frac{1}{4}$ and $0\leq q_2\leq \frac{1}{8}$. The superpotential is $W=pG_{(4,0)}(x_1,\ldots,x_6)$. The sphere partition function is
\begin{equation}
  \label{twoparzs2def}
  Z_{S^2}=\frac{1}{(2\pi)^2}\sum_{m\in\mathbb{Z}^2}\int_{-\infty}^{\infty}\mathrm{d} ^2\sigma Z_pZ_6Z^3_aZ^2_b e^{-4\pi i(\zeta_1\sigma_1+\zeta_2\sigma_2)-i(\theta_1m_1+\theta_2m_2)},
\end{equation}
where
\begin{align}
  Z_p&=\frac{\Gamma\left(1-4q_1+4i\sigma_1+2m_1\right)}{\Gamma\left(4q_1-4i\sigma_1+2m_1 \right)}\quad 
  Z_6=\frac{\Gamma\left(q_1-2q_2-i\sigma_1+2i\sigma_2-\frac{m_1}{2}+m_2 \right)}{\Gamma\left(1-q_1+2q_2+i\sigma_1-2i\sigma_2-\frac{m_1}{2}+m_2 \right)}\nonumber\\
  Z_a&=\frac{\Gamma\left(q_1-i\sigma_1-\frac{m_1}{2} \right)}{\Gamma\left(1-q_1+i\sigma_1-\frac{m_1}{2} \right)}\qquad
  Z_b=\frac{\Gamma\left(q_2-i\sigma_2-\frac{m_2}{2} \right)}{\Gamma\left(1-q_2+i\sigma_2-\frac{m_2}{2} \right)}.
\end{align}
The model has four phases: a geometric phase ($\zeta_1 \gg0,\zeta_2 \gg0$) which is a hypersurface $G_{(4,0)}(x_1,\ldots,x_6)=0$ in the toric ambient space defined by the $U(1)^2$-charges of $x_1,\ldots,x_6$, a Landau-Ginzburg orbifold phase ($2\zeta_1+\zeta_2 \ll 0,\zeta_2 \ll 0$) with $G=\mathbb{Z}_8$ and $W_{LG}=G_{(4,0)}(x_1,\ldots,x_5,1)$, a hybrid phase ($\zeta_1 \ll 0,\zeta_2 \gg0$) which is a fibration of a Landau-Ginzburg orbifold with $G=\mathbb{Z}_4$ over $B=\mathbb{P}^1$, and an orbifold phase ($2\zeta_1+\zeta_2 \gg 0,\zeta_2 \ll 0$) which is a singular hypersurface $G_{(4,0)}(x_1,\ldots,x_5,1)=0$ in the ambient space defined by the charges of $x_1,\ldots,x_5$ under $2Q_{i,1}+Q_{i,2}$. In the following we will discuss the Landau-Ginzburg, the geometric, and the hybrid phase. In the context of supersymmetric localisation this model has also been discussed in \cite{Benini:2013xpa,Closset:2015rna,Morrison:2016bps}.
%%%%%%%%%%%%%%%%%%%%%%%%%%%%%%%%%%%%%%%%%%%%%%%%%%%%%%%%%%%%%%%%%%%%%%%%
\subsection{Geometric phase}
For a discussion of the sphere partition function of this phase, see also \cite{Morrison:2016bps}. After defining $z_i=i\sigma_i-q_i$, the poles of the sphere partition functions are determined by the following divisors
\begin{equation}
  \begin{array}{lll}
  D_a=z_1-n_1+\frac{m_1}{2}& \qquad& n_1\geq\mathrm{max}[0,m_1]\in\mathbb{Z}_{\geq 0}\\
  D_b=z_2-n_2+\frac{m_2}{2}& \qquad& n_2\geq\mathrm{max}[0,m_2]\in\mathbb{Z}_{\geq 0}\\
  D_P=4z_1+n_P+2m_1+1& \qquad& n_P\geq\mathrm{max}[0,-4m_1]\in\mathbb{Z}_{\geq 0}\\
  D_6=-z_1+2z_2+n_6-\frac{m_1}{2}+m_2& \qquad& n_6\geq\mathrm{max}[0,m_1-2m_2]\in\mathbb{Z}_{\geq 0}.
  \end{array}
\end{equation}
In the geometric phase $D_a\cap D_b$ and $D_b\cap D_6$ contribute, call them $Z_{S^2}^{geom}$ and $\widetilde{Z}_{S^2}^{geom}$, respectively. The former has additional poles from $Z_6$. They contribute for $n_1\geq 2n_2$ (and $n_1'\geq 2n_2'$ where $n_1',n_2'$ are obtained by $m_i=n_i-n_i'$, $i=1,2$). One can show that by a change of summation variable $\widetilde{Z}_{S^2}^{geom}$ can be transformed into $Z_{S^2}^{geom}$ under the condition $n_1\geq 2n_2$. This shows that all contributing poles are accounted for by just computing $Z_{S^2}^{geom}$. We get: 
\begin{align}
  Z_{S^2}^{geom}&=\frac{1}{(2\pi)^2}\sum_{n_1,n_2,n_1',n_2'\geq 0}\oint \mathrm{d}^2\varepsilon Z_pZ_6Z_a^3Z_b^2 \nonumber\\
  &\quad \cdot e^{(-2\pi\zeta_1-i\theta_1)n_1+(-2\pi\zeta_2-i\theta_2)n_2}e^{(-2\pi\zeta_1+i\theta_1)n_1'+(-2\pi\zeta_2+i\theta_2)n_2'}e^{-4\pi(\zeta_1\varepsilon_1+\zeta_2\varepsilon_2)},
\end{align}
where
\begin{align}
  Z_p&=\frac{\Gamma\left(1+4n_1+4\varepsilon_1 \right)}{\Gamma\left(-4n_1'-4\varepsilon_1 \right)} \qquad
  Z_6=\frac{\Gamma\left(-n_1+2n_2-\varepsilon_1+2\varepsilon_2 \right)}{\Gamma\left(1+n_1'-2n_2'+\varepsilon_1-2\varepsilon_2 \right)}\nonumber\\
  Z_a&=\frac{\Gamma\left(-n_1-\varepsilon_1 \right)}{\Gamma\left(1+n_1'+\varepsilon_1 \right)}\qquad\quad
  Z_b=\frac{\Gamma\left(-n_2-\varepsilon_2 \right)}{\Gamma\left(1+n_2'+\varepsilon_2 \right)}.
\end{align}
Here we have chosen $q_1=q_2=0$ in order to comply with the R-charge assignment of the non-linear sigma model. To further evaluate this integral we define $\varepsilon_i=\frac{H_i}{2\pi i}$ ($i=1,2$) with $H_{i}\in H^2(X,\mathbb{C})$. The next step in the calculation is to use the reflection formula on those Gamma-factors whose argument is negative. Collecting all sines and factors of $\pi$ that the reflection formula produces we get
\begin{align}
  &-(2\pi i)^3\frac{\sin\pi\frac{H_1}{2\pi i}}{\sin^3\pi\frac{H_1}{2\pi i}\sin^2\pi\frac{H_2}{2\pi i}\sin\pi\left(\frac{H_1}{2\pi i}-2\frac{H_2}{2\pi i}\right)}\left\{\begin{array}{ll}
  \pi^2&\quad n_1\geq 2n_2\\
  (-1)^{n_1+n_1'}\sin^2\pi\left(\frac{H_1}{2\pi i}-2\frac{H_2}{2\pi i}\right)&\quad n_1<2n_2
  \end{array} \right.\nonumber\\
 =& -(2\pi i)^3\mathrm{Td}(X)\frac{4H_1}{H_1^3H_2^3(H_1-2H_2)}\left\{\begin{array}{ll}
  (2\pi i)^2&\quad n_1\geq 2n_2\\
  (-1)^{n_1+n_1'}(2i)^2\sin^2\pi\left(\frac{H_1}{2\pi i}-2\frac{H_2}{2\pi i}\right)&\quad n_1<2n_2,
  \end{array} \right.
\end{align}
where we have used
\begin{equation}
  \mathrm{Td}(X)=\frac{(1-e^{-4H_1})}{(1-e^{-H_1})^3(1-e^{-H_2})^2(1-e^{-(H_1-2H_2)})}\frac{H_1^3H_2^2(H_1-2H_2)}{4H_1}.
\end{equation}
This implies the definition of the following $I$-function:
\begin{align}
  \label{twoparifun}
  I_X(\pt,H)&=\frac{\Gamma\left(1+\frac{H_1}{2\pi i}\right)^3\Gamma\left(1+\frac{H_2}{2\pi i}\right)^2}{\Gamma\left(1+4\frac{H_1}{2\pi i}\right)}\sum_{n_1,n_2\geq 0}e^{-\pt_1 n_1}e^{-\pt_2 n_2}e^{-\pt_1\frac{H_1}{2\pi i}}e^{-\pt_2\frac{H_2}{2\pi i}}\nonumber\\
  &\quad\cdot \frac{\Gamma\left(1+4n_1+4\frac{H_1}{2\pi i}\right)}{\Gamma\left(1+n_1+\frac{H_1}{2\pi i}\right)^3\Gamma\left(1+n_2+\frac{H_2}{2\pi i}\right)^2}\left\{\begin{array}{ll}
  \frac{\Gamma\left(1+\frac{H_1}{2\pi i}-2\frac{H_2}{2\pi i} \right)}{\Gamma\left(1+n_1-2n_2+\frac{H_1}{2\pi i}-2\frac{H_2}{2\pi i} \right)}&\quad n_1\geq 2n_2\\
  (-1)^{n_1}\frac{\Gamma\left(-n_1+2n_2-\frac{H_1}{2\pi i}+2\frac{H_2}{2\pi i} \right)}{\Gamma\left(-\frac{H_1}{2\pi i}+2\frac{H_2}{2\pi i} \right)}&\quad n_1<2n_2.
\end{array}
  \right.
  \end{align}
The Gamma class  is
\begin{equation}
  \widehat{\Gamma}=\frac{\Gamma\left(1-\frac{H_1}{2\pi i}\right)^3\Gamma\left(1-\frac{H_2}{2\pi i}\right)^2\Gamma\left(1-\frac{H_1}{2\pi i}+2\frac{H_2}{2\pi i}\right)}{\Gamma\left(1-\frac{4H_1}{2\pi i}\right)}.
\end{equation}
The whole expression for the sphere partition function can then be written as
\begin{align}
  Z_{S^2}^{geom}&=-\frac{1}{(2\pi)^2}\oint \frac{\mathrm{d}^2H}{
    (2\pi i)^2}(2\pi i)^3\mathrm{Td}(X) \frac{4H_1}{H_1^3H_2^2(H_1-2H_2)}  \frac{\Gamma\left(1+\frac{4H_1}{2\pi i}\right)^2}{\Gamma\left(1+\frac{H_1}{2\pi i}\right)^3\Gamma\left(1+\frac{H_1}{2\pi i}\right)^2}\nonumber\\
  &\quad\cdot\left\{
  \begin{array}{ll}
    \frac{(2\pi i)^2}{\Gamma\left(1+\frac{H_1}{2\pi i}-2\frac{H_2}{2\pi i}\right)^2}I_X(\pt,H)I_X(\overline{\pt},H)&\quad n_1\geq 2n_2\\
   (2i)^2\sin^2\pi\left(\frac{H_1}{2\pi i}-2\frac{H_2}{2\pi i}\right)\Gamma\left(-\frac{H_1}{2\pi i}+2\frac{H_2}{2\pi i}\right)^2I_X(\pt,H)I_X(\overline{\pt},H)&\quad n_1< 2n_2
    \end{array}
  \right.\nonumber\\
  &=-\frac{(2\pi i)^5}{(2\pi)^2}\oint\frac{d^2H}{(2\pi i)^2} \frac{4H_1}{H_1^3H_2^2(H_1-2H_2)}\frac{\widehat{\Gamma}}{\widehat{\Gamma}^*}I_X(\pt,H)I_X(\overline{\pt},H)
  \end{align}
In the second step we have used $\mathrm{Td}=\widehat{\Gamma}\widehat{\Gamma}^*$.

Next, we have to rewrite the integral as an integral over the Calabi-Yau $X$. Consider a power series $h(H_1,H_2)=\sum_{i,j\geq 0}a_{i,j}H_1^iH_2^j$. Then
\begin{equation}
  \int_X h(H_1,H_2)=8a_{3,0}+4 a_{2,1}=\int_{X_{\Sigma}}(4H_1)h(H_1,H_2)=\oint_0\frac{d^2H}{(2\pi i)^2}\left[\frac{8}{H_1^4H_2}+\frac{4}{H_1^3H_2^2}\right]h(H_1,H_2)
\end{equation}
where we have used that the non-zero triple intersection numbers of $X$ are
\begin{equation}
  H_1^3=8, \qquad H_1^2H_2=4. 
\end{equation}
To show this we have to transform the integral by using the following property of multidimensional residues (see for instance \cite{GriffithsPhillip2011PoAG}). Consider a residue integral in $n$ variables $z_1,\ldots,z_n$ and holomorphic functions $\{f_1(z_i),\ldots,f_n(z_i)\}$ and $\{g_1(z_i),\ldots,g_n(z_i)\}$ satisfying
\begin{equation}
  g_k(z_i)=T_{kj}f_j(z_i),
  \end{equation}
where $T$ is a holomorphic matrix. Then
\begin{equation}
  \mathrm{Res}\left(\frac{h(z_i)\mathrm{d}z_1\wedge\ldots\wedge\mathrm{d}z_n}{f_1(z_2)\cdot\ldots\cdot f_n(z_i)}\right)=\mathrm{Res}\left(\mathrm{\det}T\frac{h(z_i)\mathrm{d}z_1\wedge\ldots\wedge\mathrm{d}z_n}{g_1(z_i)\cdot\ldots\cdot g_n(z_i)}\right).
\end{equation}
In our case we find the following transformation:
\begin{align}
\begin{pmatrix}
H_2^2 \\ H_1^4
\end{pmatrix}
&=
\begin{pmatrix}
 1 & 0\\
4 H_1^2 & H_1+2 H_2
\end{pmatrix}
\begin{pmatrix}
H_2^2 \\
H_1^2(H_1-2H_2)
\end{pmatrix}
\end{align} 
and so 
\begin{align}
\det T = H_1+2H_2.
\end{align}
This transforms the sphere partition function into the expected form:
\begin{equation}
  \label{geommatrix}
  Z_{S^2}^{geom}=-\frac{(2\pi i)^5}{(2\pi)^2}\oint\left[\frac{8}{H_1^4H_2}+\frac{4}{H_1^3H_2^2}\right]\frac{\widehat{\Gamma}}{\widehat{\Gamma}^*}I(\pt)I(\overline{\pt})=(2\pi i)^3\int_X\frac{\widehat{\Gamma}}{\widehat{\Gamma}^*}I(\pt)I(\overline{\pt})
  \end{equation}
The result can be rewritten as
\begin{align}
  \label{geommatrix2}
\frac{Z_{S^2}}{8 \pi^3}
&=
\begin{pmatrix}
\overline{I}^{(0,0)}, & \dots
\end{pmatrix}
\begin{pmatrix}
 -\frac{168 \zeta (3)}{4\pi ^3} & 0 & 0 & 0 & 0 & -4 i \\
 0 & 0 & 0 & 0 & -4 i & 0 \\
 0 & 0 & 0 & -4 i & -8 i & 0 \\
 0 & 0 & -4 i & 0 & 0 & 0 \\
 0 & -4 i & -8 i & 0 & 0 & 0 \\
 -4 i & 0 & 0 & 0 & 0 & 0 \\
\end{pmatrix}
\begin{pmatrix}
 I^{(0,0)} \\
 I^{(0,1)} \\
 I^{(1,0)} \\
 I^{(1,1)} \\
 I^{(2,0)} \\
I^{(2,1)}+2I^{(3,0)} \\
\end{pmatrix},
\end{align}
where by $I^{(i,j)}$ we denote the coefficient of $H_1^iH_2^j$ in the expansion of the $I$-function with respect to $H_1,H_2$.

The $I$-function and the Gamma class match with (\ref{geomi}) and (\ref{geomgamma}), respectively. As a further consistency check it is not hard to verify that the Picard-Fuchs operators annihilate the components of the $I$-function appearing in (\ref{geommatrix2}). The differential operators are \cite{Hosono:1993qy}
\begin{align}
  \label{deg8lvpf}
  \mathcal{L}_1&=\theta_1^2(\theta_1-2\theta_2)-4z_1(4\theta_1+3)(4\theta_1+2)(4\theta_1+1)\nonumber\\
  \mathcal{L}_2&=\theta_2^2-z_2(2\theta_2-\theta_1+1)(2\theta_1-\theta_1),
\end{align}
where $z_i=e^{-\pt_i}$ and $\theta_i=z_i\frac{\partial_i}{\partial z_i}$.
%%%%%%%%%%%%%%%%%%%%%%%%%%%%%%%%%%%%%%%%%%%%%%%%%%%%%%%%%%%%%%%%%%%%%%
\subsection{Landau-Ginzburg phase}
This phase  has also been considered in 
\cite{Knapp:2020oba} in the context of the hemisphere partition function. The orbifold group is $G=\mathbb{Z}_8$. Labelling its elements by $\gamma\in\{0,\ldots,8\}$, the sectors $\gamma=0,4$ are broad. We will show below how the remaining six narrow sectors labelled by $\delta$ emerge from the sphere partition function. We start off with (\ref{twoparzs2def}) and the following coordinate change:
\begin{align}
\sigma_1 &= i \frac{z_1}{4} & \sigma_2 &= i\frac{z_1+4z_2}{8}.
\end{align}
The location of the poles is given by the divisors
\begin{equation}
\begin{aligned}
	D_a &=\frac{1}{4} \left(-2 m_1+z_1+1\right)+n_1,&\qquad&& n_1 &\geq \max \left[0,m_1 \right], \\
	D_b &=\frac{1}{8} \left(-4 m_2+z_1+4 z_2+1\right)+n_2, &\qquad&& n_2 &\geq \max \left[0,m_2 \right], \\
		D_P&= 2 m_1+n_P-z_1,&\qquad && n_P &\geq \max \left[0,-4m_1 \right],\\
	D_6&=-\frac{m_1}{2}+m_2+n_6-z_2, &\qquad&&
	 n_6 &\geq \max \left[0,m_1-2m_2 \right].
\end{aligned} 
\end{equation}
The only contributing poles in this phase are given by 
\(D_6 \cap D_P\) and therefore we perform the transformations
\begin{align}
z_1 &\to 2 m_1+n_P+\varepsilon_1, &
z_2&\to \frac{1}{2} \left(-m_1+2 m_2+2 n_6\right)+\varepsilon_2.
\end{align}
The sums in the partition function can be simplified in 
two steps. First we introduce:
\begin{equation}
\begin{aligned}
a&=  n_P +4n_6 + 8 m_2, & c &= n_P +4 n_6, &
b&= 4m_1 +n_P, & d&= n_P.
\end{aligned}
\label{eqn:p112228newSumVar1}
\end{equation}
The new summation variables are interrelated and are 
constrained by 
\begin{equation}
\begin{aligned}
a-c &\in 8 \mathbb{Z}, &
b-d &\in 4 \mathbb{Z}, &
c-d & \in 4 \mathbb{Z}_{\geq 0}, &
a-b &\in 4 \mathbb{Z}_{\geq 0},
\label{eqn:p112228LgSumConst}
\end{aligned}
\end{equation}
as one can show by inserting the definitions (\ref{eqn:p112228newSumVar1}) and taking into account 
that \(n_P,n_6,m_1,m_2\in\mathbb{Z}\). In the second step we introduce
\begin{equation}
\begin{aligned}
a&= 8l+ \delta_1 & c &= 8k+\delta_1 & \delta_1 &= 0,1,\dots,7,\\
b&= 4p + \delta_2 & d&= 4q + \delta_2 & \delta_2 &= 0,1, \dots,3.
\end{aligned}
\label{eqn:p112228newSumVar2}
\end{equation}
The constraints (\ref{eqn:p112228LgSumConst}) are 
fulfilled if we restrict to the following \(\delta_1,
\delta_2\) combinations: 
\begin{equation}
\begin{array}{c|c|c|c|c|c|c|c|c}
\delta_1 & 0 & 1 &2 &3 & 4 &5 & 6 &7 \\
\hline
\delta_2 & 0 &1 &2 &3 & 0 &1 & 2 & 3 \\
\hline
\kappa=\delta_1-\delta_2 & 0 & 0 &0 &0 &4 &4 & 4& 4.
\end{array}
\end{equation}
This result shows that we can express \(\delta_1=\delta_2+\kappa\) and as consequence we
can write the sphere partition function, with \(\delta_2 \equiv 
\delta\), in the following form:
\begin{equation}
\begin{aligned}
Z_{S^2}^{LG}&=-\frac{1}{8(2\pi i)^2} \sum_{\kappa \in \{0,4\}}
\left(
\sum_{\delta=0}^3\oint_{(0,0)} \mathrm{d}^2\varepsilon
\frac{1}{\pi^3}
\frac{\sin \left(\pi\left(\frac{\delta+1}{4}+
	\frac{\varepsilon_1}{4}\right) \right)^3
	\sin\left(\pi\left(\frac{\delta+1+\kappa}{8}+
	\frac{\varepsilon_1+4 \varepsilon_2}{8}\right) \right)^2}{
	\sin\left(\pi \varepsilon_1\right)\sin
	\left(\pi \left(\frac{\kappa }{4}+\varepsilon_2\right)\right)}\right. \\
& \quad \cdot 
\left|e^{\pt_1\frac{\varepsilon_1}{4}}
e^{\pt_2\frac{\varepsilon_1+4\varepsilon_2}{8}}\sum_{l=0}^\infty \sum_{p=0}^{2l+\frac{\kappa}{4}}
(-1)^{p} e^{\frac{\pt_1}{4}\left(4p+\delta\right)} e_2^{\frac{\pt_2}{8}\left(8l+\delta+ \kappa\right)} \right. \\
  &   \left.\hphantom{e^{-\pt_1\frac{\varepsilon_1}{4}}
  		e^{-\pt_2\frac{\varepsilon_1+4\varepsilon_2}{8}}\sum_{l=0}^\infty}\cdot
  	\left.\frac{\Gamma \left(p+\frac{\delta+1}{4}+\frac{\varepsilon_1}{4}\right)^3
\Gamma \left(l+\frac{\delta+1+\kappa}{8}+\frac{\varepsilon_1+4 \varepsilon_2}{8}\right)^2}{
\Gamma\left(1+4p+\delta+\varepsilon_1\right)
\Gamma \left(1+2l-p +\frac{\kappa}{4}+\varepsilon_2\right)} 
\right|^2 \right). 
\end{aligned}
\end{equation} 
In the above equation we see that only first order 
poles occur and therefore a direct evaluation is 
possible. Furthermore $\delta=3$ gives no contribution. 
This is expected, because these terms correspond to a 
broad sector.
After evaluation of the residues and 
application of the transformations 
\(\kappa  \rightarrow 4 \kappa\), and \( \delta \rightarrow \delta-1\),
the sphere partition functions reads
\begin{align}
\begin{split}
Z_{S^2}^{LG}=\frac{1}{8} \sum_{k \in \{0,1\}} 
&\left(\sum_{\delta=1}^3 (-1)^\delta(-1)^\kappa \frac{1}{\pi^5}
\sin \left(\pi\frac{\delta}{4}\right)^3
\sin\left(\pi\frac{\delta+4\kappa}{8}\right)^2 \right.\\
&\quad \cdot \left. \left|
\sum_{l=0}^\infty \sum_{p=0}^{2l+\kappa}
(-1)^{p} e^{\frac{\pt_1}{4}(4p+\delta-1)}
 e^{\frac{\pt_2}{8}(8l+\delta-1+4\kappa)} 
\frac{\Gamma \left(p+\frac{\delta}{4}\right)^3
	\Gamma \left(l+\frac{\delta+4\kappa}{8}\right)^2}{
	\Gamma\left(4p+\delta\right)
	\Gamma \left(1+2l-p +\kappa\right)}\right|^2\right).
\end{split}
\end{align}
We  use (\ref{eqn:reflectionFormula}) and introduce
\begin{align}
\begin{split}
(-1)^{\operatorname{Gr}_\kappa}&= (-1)^{\delta}(-1)^\kappa
(-1)^{3 \left\lfloor \frac{\delta}{4} \right\rfloor}
(-1)^{2 \left\lfloor \frac{\delta+4\kappa}{8} \right\rfloor},\\
\widehat\Gamma_{\delta,\kappa}(0)&=
\Gamma \left(\left\langle \frac{\delta}{4} \right\rangle\right)^3
\Gamma \left(\left\langle \frac{\delta+4\kappa}{8} \right\rangle\right)^2,
\end{split}\label{eqn:p112228Gamma}
\end{align}
where  \(\widehat\Gamma^\ast_{\delta,\kappa}(0)\) follows from similar manipulations as in the 
 one parameter Landau-Ginzburg phases. By defining
\begin{align}
\begin{split}
I_{\delta,\kappa}(\pt_1,\pt_2,0) =
&\frac{1}{
	\Gamma \left(\left\langle \frac{\delta}{4} \right\rangle\right)^3
	\Gamma \left(\left\langle \frac{\delta+4\kappa}{8} \right\rangle\right)^2
} \\
&\sum_{l=0}^\infty \sum_{p=0}^{\infty}
(-1)^{p} e^{\frac{\pt_1}{4}(4p+\delta-1)} e^{\frac{\pt_2}{8}(8l+\delta-1+4\kappa)}   
\frac{\Gamma \left(p+\frac{\delta}{4}\right)^3
	\Gamma \left(l+\frac{\delta+4\kappa}{8}\right)^2}{
	\Gamma\left(4p+\delta\right)
	\Gamma \left(1+2l-p +\kappa\right)},
\end{split}
\label{eqn:p112228Ifunc}
\end{align} 
\(Z_{S^2}^{LG}\) can be written compactly:
 \begin{align}
\begin{split}
Z_{S^2}^{LG}=  \frac{1}{8} \sum_{\delta=1}^3
&\left((-1)^{\operatorname{Gr}_0}
\frac{\widehat\Gamma_{\delta,0}(0)}{\widehat\Gamma^\ast_{\delta,0}(0)}
I_{\delta,0}(\pt_1,\pt_2,0)I_{\delta,0}(\overline{\pt}_1,
\overline{\pt}_2,0) \right. \\
& \qquad +\left.(-1)^{\operatorname{Gr}_1}
\frac{\widehat\Gamma_{\delta,1}(0)}{\widehat\Gamma^\ast_{\delta,1}(0)}
I_{\delta,1}(\pt_1,\pt_2,0)I_{\delta,1}(\overline{\pt}_1.
\overline{\pt}_2,0)\right).
\end{split}
\label{eqn:p112228zs2final}
\end{align}
We can rewrite (\ref{eqn:p112228zs2final}) into matrix form (see (\ref{zs2matrix})) by 
inserting (\ref{eqn:p112228Gamma}) into (\ref{eqn:gammmaQuotient}) for \(\kappa =0\).
Let us point out that we do not need (\ref{eqn:p112228Gamma}) for \(\kappa=1\)  to extract \(M\) from (\ref{eqn:p112228zs2final}).
We find that 
\begin{align}
M=
\begin{pmatrix}
\frac{\gamma_{1}(0)}{8} & 0 & 0 & 0 &0 &0 \\
0 & \frac{\gamma_{2}(0)}{8} & 0 & 0 &0 &0 \\
0 & 0 &\frac{\gamma_{3}(0)}{8} & 0 &0 &0 \\
0 & 0 & 0 & -\frac{1}{8\gamma_{3}(0) } &0 &0 \\
0 & 0 & 0 & 0 &-\frac{1}{8\gamma_{2}(0) } &0 \\
0 & 0 & 0 & 0 &0 &-\frac{1}{8\gamma_{1}(0)}  \\
\end{pmatrix}
\end{align}
The last expression can be matched to (\ref{lgproposal}) as we show in Appendix \ref{sec:p112228LgMatch}.
%%%%%%%%%%%%%%%%%%%%%%%%%%%%%%%%%%%%%%%%%%%%%%%%%%%%%%%%%%%%%%%%%%%%%%%%
\subsection{Hybrid phase}
Let us briefly recall the structure of the hybrid phase. The D-terms are
\begin{align}
  -4|p|^2+|x_6|^2+\sum_{i=3}^5|x_i|^2&=\zeta_1\nonumber\\
  -2|x_6|^2+|x_1|^2+|x_2|&=\zeta_2.
\end{align}
The vacuum equations for $\zeta_1 \ll 0,\zeta_2 \gg0$ are
\begin{equation}
  p=\sqrt{-\frac{\zeta_1}{4}},\qquad |x_1|^2+|x_2|^2=\zeta_2.
  \end{equation}
The first $U(1)$ is broken to a $\mathbb{Z}_4$, the second $U(1)$ is completely broken, and the vacuum manifold is a $\mathbb{P}^1$. The low energy theory is a $\mathbb{Z}_4$ Landau-Ginzburg orbifold fibered over this $\mathbb{P}^1$. To compute the sphere partition function using a standardised approach we change coordinates to 
\begin{equation}
  z_1=-1+4q_1-4i\sigma_1, \qquad z_2=-q_2+i\sigma_2.
\end{equation}
Finding out which poles contribute following \cite{MR1631772,Gerhardus:2015sla} is rather tedious. The discussion depends on the sign of $2\zeta_1+\zeta_2$ (even though there is no phase boundary when $\zeta_1 \ll 0$ and $\zeta_2 \gg0$). The upshot of this lengthy calculation is that only the poles associated to $D_b\cap D_P$ contribute, consistent with the observation that only poles associated to fields that obtain a VEV in the given phase contribute. Making a shift $n_P'=n_P+4m_1,n_2'=n_2-m_2$ and choosing $q_1=\frac{1}{4},q_2=0$ the sphere partition function becomes
\begin{align}
  Z_{S^2}&=\frac{1}{4(2\pi )^2}\sum_{n_i,n_i'=0}^{\infty}\oint \mathrm{d}^2\varepsilon \frac{\Gamma\left(-n_P-\varepsilon_1 \right)}{\Gamma\left(1+n_P'+\varepsilon_1\right)}\frac{\Gamma\left(\frac{1}{4}+\frac{n_P}{4}+\frac{\varepsilon_1}{4}+2n_2+2\varepsilon_2 \right)}{\Gamma\left(1-\frac{1}{4}-\frac{n_P'}{4}-\frac{\varepsilon_1}{4}-2n_2'-2\varepsilon_2 \right)}\nonumber\\
  &\quad\cdot\left[\frac{\Gamma\left(\frac{1}{4}+\frac{n_P}{4}+\frac{\varepsilon_1}{4} \right)}{\Gamma\left(1-\frac{1}{4}-\frac{n_P'}{4}-\frac{\varepsilon_1}{4} \right)}\right]^3\left[\frac{\Gamma\left(-n_2-\varepsilon_2 \right)}{\Gamma\left(1+n_2'+\varepsilon_2 \right)}\right]^2\nonumber\\
  &\qquad\cdot e^{\frac{2\pi\zeta_1+i\theta_1}{4}n_P} e^{\frac{2\pi\zeta_1-i\theta_1}{4}n_P'}e^{-(2\pi\zeta_2+i\theta_2)n_2}e^{-(2\pi\zeta_2-i\theta_2)n_2'}e^{\pi\zeta_1\varepsilon_1}e^{-4\pi\zeta_2\varepsilon_2}.
\end{align}
The $\varepsilon_1$-integral can be easily evaluated because the poles are only first order. Defining
\begin{equation}
  n_P+1=4a+\delta, \quad n_P'+1=4b+\delta, \qquad a,b\in\mathbb{Z}_{\geq 0},\quad \delta=1,2,3,4,
\end{equation}
and using the reflection formula we get
\begin{align}
  Z_{S^2}&=-\frac{2 \pi i}{4(2 \pi)^2}\sum_{a,b,n_2,n_2'}\sum_{\delta=1}^4\oint \mathrm{d} \varepsilon_2 (-1)^{\delta}\frac{1}{\pi^2}\frac{\sin\pi\left(\frac{\delta}{4}+2\varepsilon_2\right)\sin^3\pi\frac{\delta}{4}}{\sin^2\pi\varepsilon_2}\nonumber\\
  &\quad\cdot\frac{\Gamma\left(a+\frac{\delta}{4}+2n_2+2\varepsilon_2 \right)\Gamma\left(b+\frac{\delta}{4}+2n_2'+2\varepsilon_2 \right)\Gamma\left(a+\frac{\delta}{4} \right)^3\Gamma\left(b+\frac{\delta}{4} \right)^3}{\Gamma\left(4a+\delta\right)\Gamma\left(4b+\delta\right)\Gamma\left(1+n_2+\varepsilon_2 \right)^2\Gamma\left(1+n_2'+\varepsilon_2 \right)^2}\nonumber\\
  &\quad\cdot e^{\frac{2\pi\zeta_1+i\theta_1}{4}(4a+\delta-1)} e^{\frac{2\pi\zeta_1-i\theta_1}{4}(4b+\delta-1)}e^{-(2\pi\zeta_2+i\theta_2)n_2}e^{-(2\pi\zeta_2-i\theta_2)n_2'}e^{-4\pi\zeta_2\varepsilon_2}.
\end{align}
Now we evaluate the $\varepsilon_2$-integral. Writing $\varepsilon_2=\frac{H}{2\pi i}$ we note that
\begin{equation}
  \frac{\sin\pi\left(\frac{\delta}{4}+2\varepsilon_2\right)}{\sin^2\pi\varepsilon_2}=(2i)e^{i\pi\frac{\delta}{4}}\frac{1-e^{-2\pi i\frac{\delta}{4}-2H}}{(1-e^{-H})^2}=(2i)e^{i\pi\frac{\delta}{4}}\left(1-e^{-2\pi i\frac{\delta}{4}-2H}\right)\frac{\mathrm{Td}(\mathbb{P})^1}{H^2}.
\end{equation}
Then we can write
\begin{align}
  Z_{S^2}&=-\frac{2 \pi i}{4(2 \pi)^2}\sum_{a,b,n_2,n_2'}\sum_{\delta=1}^4\int_{\mathbb{P}^1}(-1)^{\delta}\frac{(2\pi i)}{\pi^3}e^{i\pi\frac{\delta}{4}}\left(1-e^{-2\pi i\frac{\delta}{4}-2H}\right)\mathrm{Td}(\mathbb{P})^1 \sin^3\pi\frac{\delta}{4}\nonumber\\
    &\quad\cdot\frac{\Gamma\left(a+\frac{\delta}{4}+2n_2+2\frac{H}{2\pi i} \right)\Gamma\left(b+\frac{\delta}{4}+2n_2'+2\frac{H}{2\pi i} \right)\Gamma\left(a+\frac{\delta}{4} \right)^3\Gamma\left(b+\frac{\delta}{4} \right)^3}{\Gamma\left(4a+\delta\right)\Gamma\left(4b+\delta\right)\Gamma\left(1+n_2+\frac{H}{2\pi i} \right)^2\Gamma\left(1+n_2'+\frac{H}{2\pi i} \right)^2}\nonumber\\
  &\qquad\cdot e^{\frac{2\pi\zeta_1+i\theta_1}{4}(4a+\delta-1)} e^{\frac{2\pi\zeta_1-i\theta_1}{4}(4b+\delta-1)}e^{-(2\pi\zeta_2+i\theta_2)n_2}e^{-(2\pi\zeta_2-i\theta_2)n_2'}e^{-4\pi\zeta_2\frac{H}{2\pi i}}.
\end{align}
For $\delta=4$ we observe that the expression is zero because $\sin\pi=0$. We expect this to correspond to a broad sector. Since this is a two-parameter model we expect the $I$-function to have six components, two of which will lead to $\log$-periods. So we expect that all three remaining values for $\delta$ contribute.
We write the first line above as
\begin{equation}
  (-1)^{\delta}(2\pi i)e^{i\pi\frac{\delta}{4}}(1-e^{-2\pi i\frac{\delta}{4}-2H})\frac{\Gamma\left(1+\frac{H}{2\pi i}\right)^2\Gamma\left(1-\frac{H}{2\pi i}\right)^2}{\Gamma\left(\frac{\delta}{4}\right)^3\Gamma\left(1-\frac{\delta}{4}\right)^3}.
\end{equation}
Furthermore we use
\begin{align}
  e^{i\pi\frac{\delta}{4}}(1-e^{-2\pi i\frac{\delta}{4}-2H})
  &=e^{-H}\frac{(2\pi i)}{\Gamma\left(\frac{\delta}{4}+\frac{H}{\pi i}\right)\Gamma\left(1-\frac{\delta}{4}-\frac{H}{\pi i}\right)}.
\end{align}
Then the whole first line in the sphere partition function reads
\begin{equation}
  (2\pi i)^2(-1)^{\delta}e^{-H}\frac{\Gamma\left(1+\frac{H}{2\pi i}\right)^2\Gamma\left(1-\frac{H}{2\pi i}\right)^2}{\Gamma\left(\frac{\delta}{4}+\frac{H}{\pi i}\right)\Gamma\left(\frac{\delta}{4}\right)^3\Gamma\left(1-\frac{\delta}{4}-\frac{H}{\pi i}\right)\Gamma\left(1-\frac{\delta}{4}\right)^3}.
  \end{equation}
Now it is tempting to define
\begin{equation}
  I_{\delta}(\pt_1,\pt_2,H)=\frac{\Gamma\left(1+\frac{H}{2\pi i}\right)^2}{\Gamma\left(\frac{\delta}{4}+\frac{H}{\pi i}\right)\Gamma\left(\frac{\delta}{4}\right)^3}e^{-\pt_2\frac{H}{2\pi i}}\sum_{a,n\geq 0}\frac{\Gamma\left(a+\frac{\delta}{4}+2n+2\frac{H}{2\pi i} \right)\Gamma\left(a+\frac{\delta}{4} \right)^3}{\Gamma\left(4a+\delta\right)\Gamma\left(1+n+\frac{H}{2\pi i} \right)^2}e^{\frac{\pt_1}{4}(4a+\delta-1)}e^{-\pt_2n}.
\end{equation}
Then one can write the sphere partition function as
\begin{equation}
  Z_{S^2}=\frac{2\pi i}{4}\sum_{\delta=1}^3\int_{\mathbb{P}^1}(-1)^{\delta}
  \frac{\Gamma\left(\frac{\delta}{4}+\frac{H}{\pi i}\right)\Gamma\left(\frac{\delta}{4}\right)^3\Gamma\left(1-\frac{H}{2\pi i}\right)^2}{\Gamma\left(1-\frac{\delta}{4}-\frac{H}{\pi i}\right) \Gamma\left(1-\frac{\delta}{4}\right)^3\Gamma\left(1+\frac{H}{2\pi i}\right)^2}I_{\delta}(\pt_1,\pt_2,H)I_{\delta}(\overline{\pt}_1,\overline{\pt}_2,H),
  \label{eqn:2paramHybridZs2}
\end{equation}
which implies
\begin{align}
  \widehat{\Gamma}_{\delta}(H)&=\Gamma\left(\frac{\delta}{4}+\frac{H}{\pi i}\right)\Gamma\left(\frac{\delta}{4}\right)^3\Gamma\left(1-\frac{H}{2\pi i}\right)^2\nonumber\\
  \widehat{\Gamma}^*_{\delta}(H)&=\Gamma\left(1-\frac{\delta}{4}-\frac{H}{\pi i}\right)\Gamma\left(1-\frac{\delta}{4}\right)^3\Gamma\left(1+\frac{H}{2\pi i}\right)^2.
  \label{eqn:2paramHybGammas}
\end{align}
The factor $e^{-H}$ is the factor $e^{-\frac{c_1(B)}{2}}$, that we need to relate the Todd class to the Gamma class via (\ref{classes}). Then there is also an extra $(-1)^{\delta}$ that we identify with $(-1)^{\mathrm{Gr}}$. So we find a match with (\ref{zs2formula}). To rewrite this in the form (\ref{zs2matrix}) we can use the definition (\ref{eqn:gammmaQuotient}) of $\gamma_n(H)$, with (\ref{eqn:2paramHybGammas}) inserted, to 
extract the  matrix \(M\) from (\ref{eqn:2paramHybridZs2}):
\begin{align}
M=
\begin{pmatrix}
 \gamma_{1}(0)  \log 2^3 & -\frac{i\pi}{2} \gamma_{1}(0) & 0 & 0 &0 &0 \\
-\frac{i\pi}{2} \gamma_{1}(0) &0& 0 & 0 &0 &0 \\
0 & 0 & \gamma_{2}(0)  \log 2^2   &   -\frac{i\pi}{2} \gamma_{2}(0) &0 &0 \\
0 & 0 &   -\frac{i\pi}{2}\gamma_{2}(0) & 0 &0 &0 \\
0 & 0 & 0 & 0 &  \frac{1}{\gamma_{1}(0)}  \log 2^3 & -\frac{i\pi}{2} \frac{1}{\gamma_{1}(0)} \\
0 & 0 & 0 & 0 & -\frac{i\pi}{2} \frac{1}{\gamma_{1}(0)} &0  \\
\end{pmatrix}.
\label{eqn:p8Deg2hypPair}
\end{align}

In order to test our result we check that the proposed $I$-function is annihilated by the Picard-Fuchs system (\ref{deg8lvpf}) transformed to local coordinates of the hybrid phase. For this purpose we define
\begin{equation}
  y_1=z_1^{-\frac{1}{4}}, \qquad y_2=z_2.
\end{equation}
In the $y$-variables, the Picard-Fuchs operators read
\begin{align}
  \label{deg8hybpf}
  \mathcal{L}_1&=4(\theta_1-1)(\theta_1-2)(\theta_1-3)-\frac{y_1^4}{64}\theta_1^2(\theta_1+8\theta_2)\nonumber\\
  \mathcal{L}_2&=\theta_2^2-\frac{y_2}{16}(\theta_1+8\theta_2)(\theta_1+8\theta_2+4).
\end{align}
We identify
\begin{equation}
  e^{-\pt_1}= y_1^{-4}, \qquad e^{-\pt_2}= y_2.
  \end{equation}
The $I$-function encodes six periods. For this purpose we expand it in terms of a power series in $H$. The coefficient of $H^0$ encodes three power series $\varpi_{0,\delta}$ for $\delta=1,2,3$. The coefficient of $H^1$ encodes three series $\varpi_{1,\delta}$ involving logarithms in $y_2$. All these expressions are annihilated by the Picard-Fuchs system.
%%%%%%%%%%%%%%%%%%%%%%%%%%%%%%%%%%%%%%%%%%%%%%%%%%%%%%%%%%%%%%%%%%%%%%%
\subsubsection*{Comment on further hybrid examples}
So far, we have only considered hybrid models that are Landau-Ginzburg fibrations over $\mathbb{P}^1$, but not all hybrids have a $\mathbb{P}^1$-base. A well-known two-parameter example within the same class is the $U(1)^2$ GLSM defined by
\begin{equation}
  \begin{array}{c|rr|rrrrr|c}
    &p&x_6&x_4&x_5&x_1&x_2&x_3&\mathrm{FI}\\
    \hline
    U(1)_1&-6&1&2&3&0&0&0&\zeta_1\\
    U(1)_2&0&-3&0&0&1&1&1&\zeta_2\\
    \hline
    U(1)_V&2-12q_1&2q_1-6q_2&4q_1&6q_1&2q_2&2q_2&2q_2
    \end{array}
\end{equation}
where $0\leq q_1\leq\frac{1}{6}$ and $0\leq q_2\leq \frac{1}{18}$ and $W=pG_{(6,0)}(x_1,\ldots,x_6)$. The phase structure is the same as in the previous example. The hybrid phase in $\zeta_1 \ll 0,\zeta_2 \gg0$ is a $G=\mathbb{Z}_6$ Landau-Ginzburg orbifold fibered over $\mathbb{P}^2$. The calculation of the sphere partition function is almost identical to the two-parameter example presented here and the results are similar to the previous hybrid cases and therefore we refrain from giving more details.   
%%%%%%%%%%%%%%%%%%%%%%%%%%%%%%%%%%%%%%%%%%%%%%%%%%%%%%%%%%%%%%%%%%%%%%%%
\section{Outlook}
In this work we have studied the GLSM sphere partition function in a large class of phases of abelian GLSMs. We have found that the exact result can be written in terms of a general expression that has the same structure in different kinds of phases. There are several obvious directions for further research.

We expect that our results also hold in the more general case of non-abelian GLSMs. The sphere partition function has been computed for many examples of non-abelian GLSMs, including the R{\o}dland model \cite{rodland98,Hori:2006dk,Jockers:2012dk}. The Gamma class for simple non-abelian models has also been addressed in \cite{Hori:2013ika}. 
We hope to return to this in future work.

While we could show that the sphere partition function in hybrid phases reduces to the proposed form and that the result is consistent with results of the mathematics literature, a better understanding of the physics of the hybrid models would be desirable. See for instance \cite{Bertolini:2013xga,Bertolini:2017lcz,Bertolini:2018now} for recent results. Furthermore it would be interesting to see if the (conjectural) $I$-functions and Gamma classes we computed for two-parameter hybrid models and one-parameter pseudo-hybrid models are consistent with FJRW theory. A better understanding of the state spaces and pairings would also be desirable. 

Another direction contains enumerative invariants for hybrid models. The invariants, the $I$-function, the $J$-function and the mirror map have been defined in \cite{MR3590512}. It would be interesting to compute them explicitly. 

While we have focused on the sphere, one can consider other results from supersymmetric localisation in GLSMs and see if they also evaluate to something that has the same structure in every phase. For the hemisphere partition function this has already been shown for geometric and Landau-Ginzburg phases \cite{Knapp:2020oba}. It would be interesting to show explicitly that this also holds for more general hybrid models. This in particular requires a better understanding for D-branes in hybrid phases. For instance, it would be interesting to study D-branes and the results of \cite{zhao2019landauginzburgcalabiyau} via GLSM and localisation techniques.

Finally, there are fascinating connections between 2D supersymmetric gauge theories and gauge theories in higher dimensions. It is certainly worthwhile to explore this further in the context of this article. 
%%%%%%%%%%%%%%%%%%%%%%%%%%%%%%%%%%%%%%%%%%%%%%%%%%%%%%%%%%%%%%%%%%%%%%%%%%%
\appendix

\section{Sphere Partition function in one-parameter models}\label{sec:spherePart1ParamEval}
Here we give more details on the evaluation of the 
sphere partition function (\ref{eqn:spherepart1param}) in one-parameter models. Subsequently we 
outline the main steps in the calculation of 
(\ref{eqn:zs21ParamLarge1}) and (\ref{eqn:zs21ParamSmallTotal}).
The parameters for a specific model can be found
in Table \ref{tab:1ParamModels}. 

\subsection{Location of the poles and contour of integration}
In order to determine the  position of the poles we follow the procedure outlined in \cite{Gerhardus:2015sla}.
The position of the poles of the \(\Gamma\) functions are interpreted as divisors \(D_i\) in 
\(\mathbb{C}\). For our models of interest  we can 
read off from (\ref{eqn:pshzx1poles}) that the divisors are:
\begin{align}
\begin{aligned}
D_{p_1}&=\frac{1}{2} d_1 (m+2 i \sigma )+n_1+1
&\qquad n_1 &\geq \max [0,-d_1 m], \\
D_{p_2}&=\frac{1}{2} d_2 (m+2 i \sigma )+n_2+1
&\qquad n_2 & \geq \max [0,-d_2 m],\\
D_1&= -\frac{m}{2}+n_3-i \sigma
&\qquad n_3 &\geq \max [0,m],\\
D_{\alpha}&=n_4-\frac{1}{2} \alpha  (m+2 i \sigma ) 
&\qquad  n_{\alpha} & \geq \max [0,\alpha m], \\
D_{\beta}&= n_5-\frac{1}{2} \beta  (m+2 i \sigma )
&\qquad n_{\beta} & \geq \max [0,\beta m].
\end{aligned}
\label{eqn:1paramPoleDivisors}
\end{align} 
Having determined the position of the poles it remains 
to study to convergence properties of the integral. 
For large  \(\zeta\) values the integrand is
dominated by 
\begin{equation}
e^{-4\pi i \zeta \sigma} = e^{-4 \pi i \zeta \mathrm{Re} (\sigma)}
e^{4 \pi \zeta \mathrm{Im} (\sigma)}.
\end{equation}
To obtain a convergent result we have to close the contour as indicated below. Then the following divisors contribute:
\begin{equation}
\zeta =
\begin{cases}
\gg 0: & \mathrm{Im}(\sigma) < 0 \quad D_1,D_\alpha, D_\beta ,\\
\ll 0: & \mathrm{Im}(\sigma)  > 0 \quad D_{p_1},D_{p_2} ,
\end{cases}.
\end{equation}

\subsection{Counting of poles}
It is possible that certain divisors encode the same poles. 
Therefore in the summation over the contributing poles an 
 over-counting has to be avoided.
We introduce:
\begin{equation}
\begin{gathered}
\gcd(\beta, \alpha) = \kappa_{1}, \qquad 
\frac{\alpha}{\kappa_1} = \tau_\alpha,\qquad 
\frac{\beta}{\kappa_1} = \tau_\beta, \\
\gcd(d_1, d_2) = \kappa_{2},  \qquad
\frac{d_1}{\kappa_2} = \tau_{d_1},\qquad 
\frac{d_2}{\kappa_2} = \tau_{d_2}.
\end{gathered}
\label{eqn:kappaTau}
\end{equation}
In the large radius phase we find that we can sum over the poles of 
\(Z_\beta\) and thereby get all poles of \(Z_1\) and some of the poles 
of \(Z_\alpha\). The poles of \(Z_\alpha\) we miss are of the form
\begin{align}
n_{\alpha} &= \tau_\alpha n + \delta & \delta  &=1, \dots, \tau_\alpha-1
\quad n \in \mathbb{Z}_{\geq 0}.
\label{eqn:remAlpha}
\end{align} 
A similar discussion shows that in the small radius phase we can first sum over
the poles of \(Z_{p_1}\) and in a second summation sum over the remaining poles of 
\(Z_{p_2}\) which are of the form:
\begin{align}
n_2 = \tau_{d_2} n + \delta \quad \delta = 0,1,2,\dots, d_2 -2.
\label{eqn:remP1}
\end{align}

\subsection{Manipulations of the integrand}
Here we simplify (\ref{eqn:spherepart1param}) by 
manipulations  which are applicable in all phases. Phase dependent specifics are 
discussed in the main text. We apply the following steps:
\begin{enumerate}
	\item Write  (\ref{eqn:spherepart1param}) as sum over poles. The contributing poles depend
	on the phase and their location  is  determined by  (\ref{eqn:1paramPoleDivisors}).
	\item We shift the locations of the poles by a variable transformation
	\begin{align}
		\sigma \rightarrow \varepsilon + const
	\end{align}
	 so that the poles are now 
	at \(\varepsilon=0\).
	\item We simplify the sums over the magnetic charge lattice (parametrized by \(m\))  and 
	the sum over the different poles \(n_i\) (\(i \in \{1,2,3, \alpha, \beta\}\)),  see (\ref{eqn:1paramPoleDivisors}). 
	\item We apply the reflection formula
	\begin{align}
	\Gamma(x) \Gamma(1-x) &= \frac{\pi}{\sin(\pi x)} \label{eqn:gammaIdentites1}
	\end{align}
	to further simplify the integrand.
\end{enumerate}
After the above steps we find that in all cases, (\ref{eqn:spherepart1param})  
can be written as
\begin{equation}
Z_{S^2} = \sum_i Z_{S^2,i}
\end{equation} 
 with contributions of the form 
\begin{align}
Z_{S^2,i} &= - \frac{1}{2\pi} \sum_{finite} (-1)^{sgn} \oint \mathrm{d}  \varepsilon
\mathcal{Z}_{i,sing}(\varepsilon)|\mathcal{Z}_{i,reg}(\pt,\varepsilon)|^2 
\label{eqn:1paramGeneralForm1}.
\end{align}
The exact form of the different components are phase dependent and we will comment on 
their structure below.

\subsection{\(\zeta \ll 0\) phase}

In this phase  (\ref{eqn:spherepart1param})  splits into two contributions 
\begin{align}
Z_{S^2}^{\zeta \ll 0} &= Z_{S^2,1}^{\zeta \ll 0} + Z_{S^2,2}^{\zeta \ll 0}. 
\label{eqn:zs21ParamSmallBlockForm}
\end{align}
The first contribution comes from poles of \(Z_{p_1}\) and 
the second term from the remaining poles of \(Z_{p_2}\), 
of the form (\ref{eqn:remP1}). Both terms are of the from
(\ref{eqn:1paramGeneralForm1}). \(Z_{S^2,1}^{\zeta \ll 0}\) consists of the 
following contributions:
\begin{equation}
\begin{split}
\sum_{finite} (-1)^{sgn} &= \sum_{\delta=1}^{d_1-1}, \\
\mathcal{Z}_{1,reg}(\pt,\varepsilon) &= \sum_{a=0}^\infty
e^{\pt (-i \varepsilon +a + \frac{\delta}{d_1}-q)}
(-1)^{a (5+k-n-j+ \alpha n + j\beta)} \\
&\quad \cdot \frac{
	\Gamma \left(a-i \varepsilon +\frac{\delta}{d_1}\right)^{5+k-n-j}
	\Gamma \left(a \alpha -i \varepsilon  \alpha +\frac{\alpha }{d_1}\delta\right)^n
	\Gamma \left(a \beta -i \varepsilon  \beta +\frac{\beta }{d_1}\delta\right)^j
}{
	\Gamma \left(\delta +a d_1-i \varepsilon  d_1\right)
	\Gamma \left(a d_2-i \varepsilon  d_2+\frac{d_2}{d_1}\varepsilon\right)^k}, \\
\mathcal{Z}_{1,sing}(\varepsilon) &= \frac{1}{\pi^{4}}
\frac{
	\sin\left( \pi \left(-i \varepsilon +\frac{\delta}{d_1}\right)\right)^{5+k-n-j}
	\sin\left(\pi \left(-i \varepsilon  \alpha +\frac{\alpha }{d_1}\delta\right)\right)^n
	\sin\left( \pi \left(-i \varepsilon  \beta +\frac{\beta }{d_1}\delta\right)\right)^j
}{
	\sin \pi \left( i \varepsilon  d_1\right)
	\sin\left( \pi \left(-i \varepsilon  d_2+\frac{d_2}{d_1}\delta\right)\right)^k}.
\end{split}
\label{eqn:zs21ParamSmall1}
\end{equation}
The building blocks of \(Z^{\zeta \ll 0}_{S^2,2}\) are given by 
\begin{equation}
\begin{split}
\sum_{finite} (-1)^{sgn} &=\sum_{\delta=1}^{\tau_{d_2}-1} \sum_{\gamma =0}^{\kappa_2-1}
(-1)^{k\delta} (-1)^{\tau_{d_1} \gamma}
(-1)^{ k\tau_{d_2} \gamma}, \\
\mathcal{Z}_{2,reg}(\pt,\varepsilon)
&=\sum_{a=0}^{\infty}(-1)^{a (5+k-n-j+ \alpha n + j \beta)}
e^{\pt(-i \varepsilon+a + \frac{\gamma}{\kappa_{2}} + \frac{\delta}{d_2}-q)} \\
&\quad \cdot \frac{
	\Gamma \left(a-i \varepsilon +\frac{\delta}{d_2}
	+\frac{\gamma }{\kappa_2}\right)^{5+k-n-j}
	\Gamma \left(a \alpha -i \varepsilon  \alpha
	+\frac{\alpha }{d_2}\delta
	+\frac{\gamma  \alpha }{\kappa_2}\right)^n
}{
	\Gamma \left(\frac{\tau_{d_1}}{\tau_{d_2}}(\delta)
	+ d_1 b -  i \varepsilon d_1  + \tau_{d_1} \gamma \right)
} \\
& \qquad \cdot \frac{\Gamma \left(a \beta -i \varepsilon  \beta
	+\frac{ \beta }{d_2} \delta
	+\frac{\gamma  \beta }{\kappa_2}\right)^j}
{\Gamma \left(d_2 a-i \varepsilon d_2 +\tau_{d_2}\gamma\right)^k}, \\
\mathcal{Z}_{2,sing}(\varepsilon) &=\frac{1}{\pi^4}
\frac{
	\sin\left(\pi \left(-i \varepsilon + \frac{\delta }{d_2} +
	\frac{\gamma}{\kappa_2}\right)\right)^{5+k-n-j}
	\sin \left( \pi \left(-i \varepsilon  \alpha
	+\frac{\alpha }{d_2}\delta
	+\frac{\gamma  \alpha }{\kappa_2}\right)\right)^n
}{
	\sin \left(\pi \left(\frac{\tau_{d_1}}{\tau_{d_2}} \delta  
	- i \varepsilon d_1\right)\right)
} \\
&\qquad \cdot \frac{\sin \left( \pi \left( -i \varepsilon  \beta
	+\frac{ \beta }{d_2} \delta
	+\frac{\gamma  \beta }{\kappa_2} \right)\right)^j}{	\sin \left(\pi i \varepsilon d_2\right)^k}.
\end{split}
\label{eqn:zs21ParamSmall2}
\end{equation}

In the small radius phase it strongly depends on the nature of the 
phase which combinations of the parameters  in (\ref{eqn:zs21ParamSmall1}) and (\ref{eqn:zs21ParamSmall2}) lead to a non-vanishing contribution. In Table \ref{tab:1ParamSmallConclusion} we give an overview of 
the contributing combinations for all 14 one-parameter 
models.

\begin{table}
	\centering
	\begin{tabular}{ccc|cc|ccc}
		\multicolumn{3}{c}{Contribution}& \multicolumn{2}{c}{$Z^{\zeta \ll 0}_{S^2,1}$} & \multicolumn{3}{c}{$Z^{\zeta \ll 0}_{S^2,2}$} \\
		\toprule
		&\(\kappa_{1}\)&\(\kappa_{2}\)& \( \delta\) & Order & \(\delta\) & \(\gamma\) & Order \\
		\midrule
		\multicolumn{6}{c}{F-type}\\
		\midrule 
		F1& -& - & 1,2,3,4 & 1 & - &- &-\\
		F2&-&- & 1,2,4,5 & 1 & - &- &-\\
		F3&-&- & 1,3,5,7 & 1 & - &- &-\\
		F4&1 &- & 1,3,7,9 & 1 & - &- &-\\
		F5&-&1 & 1,3 & 1 & 1,2 &0 &1\\
		F6&1&2 & 1,5 & 1 & 1 &0,1 &1\\
		F7&2&2 & 1,5,7,11 & 1 & - &- &-\\
		\multicolumn{6}{c}{C-type}\\
		\midrule 
		C1&-&2 & 1,3 & 1 & - &- &-\\
		C1&-&2 & 2 & 2 & - &- &-\\
		C2&-&2 & 1,5 & 1 & - &-&-\\
		C2&-&2 & 3 & 2 & - &-&-\\
		C3&-&1 & 1,2 & 1 & 1 &0 &2\\
		\multicolumn{6}{c}{K-type}\\
		\midrule 
		K1&-&3 & 1,2 & 2 & - &- &-\\
		K2&-&4 & 1,3 & 2 & - &- &-\\
		K3&1&6 & 1,5 & 2 & - &- &-\\
		\multicolumn{6}{c}{M-type}\\
		\midrule 
		M1&-&2 & 1 & 4 & - &- &-\\
	\end{tabular}
	\caption{Contributing poles and pole order in the 
		\(\zeta \ll 0\) phase.}
	\label{tab:1ParamSmallConclusion}
\end{table}

\subsection{\(\zeta \gg 0\) phase}

Similar to the small radius phase we find that
(\ref{eqn:spherepart1param}) splits into two parts 
\begin{align}
Z_{S^2}^{\zeta \gg 0} &= Z^{\zeta \gg 0}_{S^2,1} + Z^{\zeta \gg 0}_{S^2,2}
\label{eqn:zs21ParamLargeBlockForm},
\end{align}
with the contributions of the form (\ref{eqn:1paramGeneralForm1}).
\( Z^{\zeta \gg 0}_{S^2,1}\) comes from the poles of \(Z_\beta\)
and \(Z^{\zeta \gg 0}_{S^2,2}\) originates from the
leftover poles (\ref{eqn:remAlpha})  of \(Z_\alpha\).
The form of \( Z^{\zeta \gg 0}_{S^2,1}\) is given in 
(\ref{eqn:zs21ParamLarge1}), but let us comment why no 
alternating sign appears. The sign appearing in \( Z^{\zeta \gg 0}_{S^2,1}\) always is $1$ because
\begin{align}
(-1)^{5+k-n-j+n+j+k+1}= (-1)^{6+2k}=1.
\end{align}
For  \(Z^{\zeta \gg 0}_{S^2,2}\) we only give 
\begin{equation}
\begin{split}
\sum_{sing} (-1)^{sgn} &= \sum_{\delta =1}^{\tau_\alpha -1}\sum_{\gamma =0}^{\kappa_1-1}
(-1)^{n \delta} (-1)^{n \gamma \tau_\alpha}(-1)^{j \gamma \tau_\beta}, \\
\mathcal{Z}_{2,sing}(\varepsilon) &=
\pi^4
\frac{
	\sin \left( \pi \left(\frac{\delta  d_1}{\alpha }+i \varepsilon d_1
	+\frac{\gamma d_1}{\kappa_1 } \right)\right)
	\sin \left( \pi \left(\frac{\delta  d_2}{\alpha }+i \varepsilon d_2
	+\frac{\gamma  d_2}{\kappa_1 } \right)\right)^k
}{
	\sin \left(\pi \left(
	\frac{\delta }{\alpha }+i \varepsilon
	+\frac{\gamma  }{\kappa_1 }\right)\right)^{5+k-n-j}
	\sin \left(\pi \left( i \varepsilon \alpha\right)\right)^n
	\sin \left(\pi \left(\frac{\delta \tau_\beta }{\tau_\alpha }
	+i \varepsilon  \beta\right)\right)^j 
}.
\end{split}
\label{eqn:zs21ParamLarge2Sing}
\end{equation}
From the structure of these expression one can conclude that 
for all one-parameter models
\begin{equation}
Z^{\zeta \gg 0}_{S^2,2} = 0,
\end{equation}
because there are always sine-contributions in the numerator of $\mathcal{Z}_{2,sing}(\varepsilon)$ that are zero. 

\section{Pseudo-hybrid models}
\label{sec:pseudoHybridDetails}

Here we will discuss the one parameter pseudo hybrid phases in more
detail.
 We start from the contributions to 
\(Z^{\zeta \ll 0}_{S^2,1}\) (\ref{eqn:zs21ParamSmall1})  and apply the shift 
\(
\varepsilon \rightarrow \frac{i \varepsilon}{d_1}
\).
 In \(Z^{\zeta \ll 0}_{S^2,2}\) (\ref{eqn:zs21ParamSmall2})
 we apply \(
\varepsilon \rightarrow \frac{i \varepsilon}{d_2}
\).
We only get a non-zero
contribution if (see Table \ref{tab:1ParamSmallConclusion})
\begin{equation}
\begin{aligned}
\left\langle \frac{\delta}{d_1} \right\rangle & \not = 0,
&\left\langle \alpha \frac{\delta}{d_1} \right\rangle & \not = 0,
&\left\langle \beta \frac{\delta}{d_1} \right\rangle & \not = 0, \\
\left\langle \frac{\delta}{d_2}+ \frac{\gamma}{\kappa_2} \right\rangle & \not = 0
&\left\langle \alpha\left(\frac{\delta}{d_2}+ \frac{\gamma}{\kappa_2} \right)\right\rangle & \not = 0
&\left\langle \beta \left(\frac{\delta}{d_2}+ \frac{\gamma}{\kappa_2}\right) \right\rangle & \not = 0\\
&& \left\langle \frac{\tau_{d_1}}{\tau_{d_2}} \delta  
\right\rangle  & \not= 0
\end{aligned}
\end{equation}
This allows to rewrite \(\mathcal{Z}_{1,sing}(\varepsilon)\) (\ref{eqn:zs21ParamSmall1}) and \(\mathcal{Z}_{2,sing}(\varepsilon)\) (\ref{eqn:zs21ParamSmall2}) in the following form
 \begin{align}
\begin{split}
\mathcal{Z}_{1,sing}(\varepsilon)&= \frac{
	(-1)^{k \left\lfloor \tau_{d_2}\frac{\delta}{\tau_{d_1}} \right\rfloor}}{\varepsilon}\\
&\quad \cdot \Gamma \left(1- \varepsilon \right)
\Gamma \left(1+ \varepsilon \right)
\Gamma\left(  \tau_{d_2}\frac{\varepsilon}{\tau_{d_1}}+ \left\langle
\frac{\tau_{d_2}}{\tau_{d_1}} \delta\right\rangle\right)^k
\Gamma\left(1-  \tau_{d_2}\frac{\varepsilon}{\tau_{d_1}}-
\left\langle \frac{\tau_{d_2}}{\tau_{d_1}}\delta \right\rangle \right)^k
\\
& \quad \cdot \frac{(-1)^{(5+k-n-j)\left\lfloor \frac{\delta}{d_1} \right\rfloor}}{
	\Gamma\left( \frac{\varepsilon}{d_1} + \left\langle
	\frac{\delta}{d_1}\right\rangle\right)^{5+k-n-j}
	\Gamma\left(-\frac{\varepsilon}{d_1} + \left\langle
	\frac{d_1-\delta}{d_1}\right\rangle\right)^{5+k-n-j}} \\
& \quad \cdot \frac{(-1)^{n\left\lfloor \alpha \frac{\delta}{d_1} \right\rfloor}}{
	\Gamma\left( \alpha \frac{\varepsilon}{d_1} + \left\langle
	\alpha \frac{\delta}{d_1}\right\rangle\right)^{n}
	\Gamma\left(- \alpha \frac{\varepsilon}{d_1} + \left\langle
	\alpha \frac{d_1-\delta}{d_1}\right\rangle\right)^{n}} \\
&\quad \cdot  \frac{(-1)^{j\left\lfloor \beta\frac{\delta}{d_1} \right\rfloor}}{
	\Gamma\left( \beta \frac{\varepsilon}{d_1} + \left\langle
	\beta \frac{\delta}{d_1}\right\rangle\right)^j
	\Gamma\left(- \beta \frac{\varepsilon}{d_1} + \left\langle
	\beta \frac{d_1-\delta}{d_1}\right\rangle\right)^j}, \\
\label{eqn:pseudoGeneralSing1}
\end{split}
\end{align} 
and 
\begin{align}
\begin{split}
\mathcal{Z}_{2, sing} (\varepsilon) &= \frac{
	(-1)^{ \left\lfloor \frac{\tau_{d_2}}{\tau_{d_1}}\delta \right\rfloor}}{\varepsilon^k}\\
& \quad \cdot \Gamma \left(1- \varepsilon \right)^k
\Gamma \left(1+ \varepsilon \right)^k
\Gamma \left( - \tau_{d_1}\frac{\varepsilon}{\tau_{d_2}}
+\left\langle \tau_{d_1} \frac{\tau_{d_2}-\delta }{\tau_{d_2}}\right\rangle  
\right)
\Gamma \left(\tau_{d_1}\frac{\varepsilon}{\tau_{d_2}}
+\left\langle \tau_{d_1} \frac{\delta }{\tau_{d_2}}\right\rangle \right)
\\
& \quad \cdot \frac{(-1)^{(5+k-n-j)\left\lfloor\frac{\delta }{d_2} +
		\frac{\gamma}{\kappa_{2}}\right\rfloor}}{
	\Gamma\left(\frac{\varepsilon}{d_2} +\left\langle \frac{\delta }{d_2} +
	\frac{\gamma}{\kappa_{2}}\right\rangle\right)^{5+k-n-j}
	\Gamma\left(1-\frac{\varepsilon}{d_2} -\left\langle \frac{\delta }{d_2} +
	\frac{\gamma}{\kappa_{2}}\right\rangle\right)^{5+k-n-j}} \\
& \quad \cdot  \frac{(-1)^{n\left\lfloor  \frac{\alpha }{d_2}\delta
		+\frac{\gamma  \alpha }{\kappa_{2}} \right\rfloor}}{
	\Gamma\left(  \alpha \frac{\varepsilon}{d_2}
	+ \left\langle \frac{\alpha }{d_2}\delta
	+\frac{\gamma  \alpha }{\kappa_{2}} \right\rangle\right)^{n}
	\Gamma\left(1- \alpha \frac{\varepsilon}{d_2}
	- \left\langle \frac{\alpha }{d_2}\delta
	+\frac{\gamma  \alpha }{\kappa_{2}} \right\rangle\right)^{n}} \\
& \quad \cdot  \frac{(-1)^{j\left\lfloor  \frac{ \beta }{d_2} \delta
		+\frac{\gamma  \beta }{\kappa_{2}} \right\rfloor }}{
	\Gamma\left(\beta \frac{\varepsilon}{d_2}
	+ \left\langle \frac{ \beta }{d_2} \delta
	+\frac{\gamma  \beta }{\kappa_{2}} \right\rangle \right)^j
	\Gamma\left(1-\beta \frac{\varepsilon}{d_2}
	- \left\langle \frac{ \beta }{d_2} \delta
	+\frac{\gamma  \beta }{\kappa_{2}} \right\rangle \right)^j}.
\label{eqn:pseudoGeneralSing2}
\end{split}
\end{align}

\section{FJRW/Landau-Ginzburg expression for various models}
\label{sec:IfuncOneParamLgSigns}
Here we outline the main steps to match the \(I\)- functions  and \(\widehat{\Gamma}\) 
classes obtained from \(Z_{S^2}\) in the Landau-Ginzburg phases with results 
in the literature. 
 We start from the expressions
(\ref{lgifun}),(\ref{lggamma}), (\ref{lggammaCon}) and
(\ref{lgGrading}) (see section \ref{sec:lgFJRWBackground}).
\subsection{One parameter models}
\label{sec:lgMatchOneParam}
The \(q\) matrix of the models of interest is given in
(\ref{1parqmat}) and evaluation of (\ref{lggamma}) gives
\begin{align}
\widehat{\Gamma}_\delta &= \Gamma \left(1- \left\langle - \frac{k}{d}- \frac{1}{d}\right\rangle\right)^3  \Gamma \left(1- \left\langle - \frac{k \alpha}{d}- \frac{\alpha}{d}\right\rangle\right)\Gamma \left(1- \left\langle - \frac{k \beta}{d}- \frac{\beta}{d}\right\rangle\right).
\end{align}
By inserting \(q\) into (\ref{eqn:fjrwIfuncGeneral}) we find 
\begin{align}
\begin{split}
I_{LG}(u)&=- \sum_{k \geq 0}\frac{u^k}{\Gamma \left( k +1 \right)}
(-1)^{3 \left\langle\frac{k+1}{d}\right\rangle+\left\langle \alpha \frac{k +1 }{d}\right\rangle+\left\langle  \beta \frac{k +1}{d} \right\rangle} \\
&\quad \cdot \frac{\Gamma\left(\left\langle - \frac{k}{d} - \frac{1}{d} \right\rangle\right)^3 \Gamma \left(\left\langle -\frac{k \alpha}{d} -\frac{\alpha}{d}\right\rangle\right)\Gamma\left(\left\langle - \frac{k \beta}{d} - \frac{\beta}{d} \right\rangle\right)}{\Gamma\left(1- \frac{k}{d} - \frac{1}{d}\right)^3 \Gamma \left(1- \frac{k \alpha}{d} -\frac{\alpha}{d}\right)\Gamma \left(1- \frac{k \beta}{d} -\frac{\beta}{d}\right)}.
\end{split}
\end{align}
Applying the shift $k+1 \rightarrow k$ we get 
\begin{align}
\widehat{\Gamma}_\delta = \Gamma \left(1- \left\langle - \frac{k}{d}\right\rangle\right)^3  \Gamma \left(1- \left\langle - \frac{k \alpha}{d}\right\rangle\right)\Gamma \left(1- \left\langle - \frac{k \beta}{d}\right\rangle\right),
\end{align}
and
\begin{align}
I_{LG}(u)=- \sum_{k \geq 1}\frac{u^{k-1}}{\Gamma \left( k  \right)} \frac{(-1)^{3 \left\langle \frac{k}{d} \right\rangle+\left\langle \frac{k \alpha }{d} \right\rangle+\left\langle  \frac{k \beta}{d}\right\rangle}\Gamma\left(\left\langle - \frac{k}{d}\right\rangle\right)^3 \Gamma \left(\left\langle -\frac{k \alpha}{d} \right\rangle\right)\Gamma\left(\left\langle - \frac{k \beta}{d} \right\rangle\right)}{\Gamma\left(1- \frac{k}{d} \right)^3 \Gamma \left(1- \frac{k \alpha}{d}\right)\Gamma \left(1- \frac{k \beta}{d}\right)}.
\end{align}
Next we transform 
\(
k \rightarrow dn + \delta \quad \delta = 1, \dots, d-1
\),
and use 
\begin{align}
\left\langle - \rho n - \frac{\delta \rho}{d} \right\rangle  
&= 1 - \left\langle \frac{\delta \rho}{d} \right\rangle, &
\left\langle \rho n + \frac{\delta \rho}{d}\right\rangle &=\left\langle \frac{\delta \rho}{d} \right\rangle,
\end{align} 
to arrive at the following expressions:
\begin{align}
\widehat{\Gamma}_\delta &= \Gamma \left( \left\langle  \frac{k}{d}\right\rangle\right)^3  \Gamma \left( \left\langle  \frac{k \alpha}{d}\right\rangle\right)\Gamma \left(\left\langle\frac{k \beta}{d}\right\rangle\right),
\label{eqn:1ParamLgGammaEval} \\
I_{LG}(u)&=- \sum_{\delta =1}^{d-1} \sum_{n \geq 0}\frac{u^{dn + \delta -1}}{\Gamma \left( dn + \delta\right)} \frac{(-1)^{3\left\langle\frac{ \delta }{d} \right\rangle+ \left\langle \frac{ \alpha \delta }{d} \right\rangle+\left\langle  \frac{ \beta \delta}{d}\right\rangle}\Gamma\left(1-\left\langle  \frac{\delta}{d}\right\rangle\right)^3 \Gamma \left(1-\left\langle\frac{ \alpha \delta}{d} \right\rangle\right)\Gamma\left(1-\left\langle \frac{\beta \delta }{d} \right\rangle\right)}{\Gamma\left(1- n-\frac{\delta}{d} \right)^3 \Gamma \left(1- \alpha n - \frac{ \alpha \delta }{d}\right)\Gamma \left(1- \beta n - \frac{ \beta \delta}{d}\right)}.
\end{align}
The next identity we apply is 
\begin{align}
3\left\langle\frac{ \delta }{d} \right\rangle+ \left\langle \frac{ \alpha \delta }{d} \right\rangle+\left\langle  \frac{ \beta \delta}{d}\right\rangle 
&= \delta- 3 \left\lfloor \frac{\delta}{d} \right\rfloor -\left\lfloor \frac{ \alpha \delta}{d} \right\rfloor - \left\lfloor \frac{\beta \delta }{d} \right\rfloor.
\end{align}
By using (\ref{eqn:gammaIdentites1}) we get
\begin{align}
\begin{split}
I_{LG}(u)&= -\sum_{\delta =1}^{d-1} \sum_{n \geq 0}  (-1)^\delta(-1)^{dn }\frac{u^{dn + \delta -1}}{\Gamma \left( dn + \delta\right)} \frac{\Gamma \left( n+ \frac{\delta}{d}\right)^3\Gamma \left( \alpha n+ \frac{ \alpha \delta }{d}\right)\Gamma \left( \beta  n+ \frac{\beta \delta }{d}\right) }{ \Gamma\left(\left\langle  \frac{\delta}{d}\right\rangle\right)^3 \Gamma\left(\left\langle  \frac{\alpha \delta}{d}\right\rangle\right)\Gamma\left(\left\langle  \frac{\beta\delta}{d}\right\rangle\right)},\\
&= \sum_{\delta=1}^{d-1} I_\delta(u).
 \label{eqn:IfuncOneParamLGNewSign2App2}
\end{split}
\end{align}
Similar steps as above lead to the following expressions for (\ref{lggamma}) and 
(\ref{lgGrading}):
\begin{align}
\operatorname{Gr}&=
\delta
- \left(
3\left\lfloor \frac{\delta}{d} \right\rfloor
+\left\lfloor \alpha \frac{\delta}{d} \right\rfloor
+\left\lfloor \beta \frac{\delta}{d} \right\rfloor \right),\label{eqn:1ParamLgGrEval} \\
\widehat\Gamma^\ast_\delta 
&=\Gamma\left(\left\langle
\frac{d-k}{d} \right\rangle\right)^3
\Gamma \left(\left\langle  \alpha
\frac{d-k}{d} \right\rangle\right)
\Gamma \left(\left\langle  \beta
\frac{d-k}{d} \right\rangle\right). \label{eqn:1ParamGammaConEval}
\end{align}
We can now insert (\ref{eqn:IfuncOneParamLGNewSign2App2}),
(\ref{eqn:1ParamLgGammaEval}), (\ref{eqn:1ParamGammaConEval}) and (\ref{eqn:1ParamLgGrEval}) into (\ref{lgproposal}):
\begin{align}
Z_{S^2}^{LG} &= \sum_{\delta,\delta^\prime}(-1)^{
	\delta
	+
	3\left\lfloor \frac{\delta}{d} \right\rfloor
	+\left\lfloor \alpha \frac{\delta}{d} \right\rfloor
	+\left\lfloor \beta \frac{\delta}{d} \right\rfloor
}
\frac{\Gamma \left( \left\langle  \frac{\delta}{d}\right\rangle\right)^3
	\Gamma \left(\left\langle  \frac{\delta \alpha}{d}\right\rangle\right)
	\Gamma \left(\left\langle\frac{\delta \beta}{d}\right\rangle\right)}
{\Gamma\left(\left\langle
	\frac{d-\delta}{d} \right\rangle\right)^3
	\Gamma \left(\left\langle  \alpha
	\frac{d-\delta}{d} \right\rangle\right)
	\Gamma \left(\left\langle  \beta
	\frac{d-\delta}{d} \right\rangle\right)} \nonumber \\
& \quad \cdot  I_{\delta}(\overline{u}(\overline{t}))
I_{\delta^\prime }(u(t))
\left\langle e_{\delta^{-1}} , e_{\delta^\prime} \right\rangle \nonumber \\
&= \frac{1}{d} \sum_{\delta}(-1)^{
	\delta
	+
	3\left\lfloor \frac{\delta}{d} \right\rfloor
	+\left\lfloor \alpha \frac{\delta}{d} \right\rfloor
	+\left\lfloor \beta \frac{\delta}{d} \right\rfloor
}
\frac{\Gamma \left( \left\langle  \frac{\delta}{d}\right\rangle\right)^3
	\Gamma \left(\left\langle  \frac{\delta \alpha}{d}\right\rangle\right)
	\Gamma \left(\left\langle\frac{\delta \beta}{d}\right\rangle\right)}
{\Gamma\left(\left\langle
	\frac{d-\delta}{d} \right\rangle\right)^3
	\Gamma \left(\left\langle  \alpha
	\frac{d-\delta}{d} \right\rangle\right)
	\Gamma \left(\left\langle  \beta
	\frac{d-\delta}{d} \right\rangle\right)} \nonumber \\
& \quad \cdot I_{\delta}(\overline{u}(\overline{t}))
I_{\delta }(u(t)).
\label{eqn:1ParamFinalEval}
\end{align}
The last line follows from (\ref{lgpairing}).
We see that (\ref{eqn:1ParamFinalEval}) matches the 
result from the GLSM calculation (\ref{eqn:1paramLgSpf}).

\subsection{Two parameter model}
\label{sec:p112228LgMatch}
In this model the $q$-matrix reads 
\begin{align}
q=
\begin{pmatrix}
1 & 0 & - \frac{1}{4} & - \frac{1}{4} & - \frac{1}{4}
& - \frac{1}{8} & - \frac{1}{8} \\
0 & 1 & 0 &0 &0 & - \frac{1}{2} & - \frac{1}{2}   
\end{pmatrix}
\end{align}
In \cite{Knapp:2020oba} it was shown that
(\ref{eqn:fjrwIfuncGeneral}) can be rewritten in 
the following form:
\begin{align}
I_{LG}(u) &= \sum_{r=1}^3 \left[ \frac{1}{\Gamma\left(\frac{r}{4}\right)^3 \Gamma\left(\frac{r}{8}\right)^2} \widehat\varpi^{ev}_r e_r
+
\frac{1}{\Gamma\left(\frac{r}{4}\right)^3 \Gamma\left(\frac{r}{8}+\frac{1}{2}\right)^2} \widehat\varpi^{od}_r e_{r+4} \right],
\label{eqn:p112228IwitVarPis}
\end{align} with \begin{align}
\begin{split}
\widehat\varpi^{ev}_r &=
(-1)^{r+1} \sum_{n \in 2
	\mathbb{Z}_{\geq 0}}
\frac{
	\Gamma \left(n+\frac{r}{4}\right)^4}
{\Gamma\left(4n+r\right)}
\left(-2^{12} \psi^4\right)^{n+\frac{r-1}{4}} 
 \sum_m \frac{\Gamma \left(m + \frac{n}{2} +
	\frac{r}{8}\right)^2}{
	\Gamma \left(n +\frac{r}{4}\right)
	\Gamma \left(2m +1 \right)
} \left(2 \phi\right)^{2m}\\
&+(-1)^{r} \sum_{n \in 2
	\mathbb{Z}_{\geq 0}+1}
\frac{
	\Gamma \left(n+\frac{r}{4}\right)^4}
{\Gamma\left(4n+r\right)}
\left(-2^{12} \psi^4\right)^{n+\frac{r-1}{4}} 
 \sum_m \frac{\Gamma \left(m + \frac{n}{2} +
	\frac{r}{8}+\frac{1}{2}\right)^2}{
	\Gamma \left(n +\frac{r}{4}\right)
	\Gamma \left(2m +2 \right)
}
\left(2 \phi\right)^{2m+1},
\end{split}
\label{eqn:lgVarpiEv}
\end{align} and \begin{align}
\begin{split}
\widehat\varpi^{odd}_r &=
(-1)^{r+1} \sum_{n \in 2
	\mathbb{Z}_{\geq 0}+1}
\frac{
	\Gamma \left(n+\frac{r}{4}\right)^4}
{\Gamma\left(4n+r\right)}
\left(-2^{12} \psi^4\right)^{n+\frac{r-1}{4}} 
\sum_m \frac{\Gamma \left(m + \frac{n}{2} +
	\frac{r}{8}\right)^2}{
	\Gamma \left(n +\frac{r}{4}\right)
	\Gamma \left(2m +1 \right)
} \left(2 \phi\right)^{2m}\\
&+(-1)^{r} \sum_{n \in 2
	\mathbb{Z}_{\geq 0}}
\frac{
	\Gamma \left(n+\frac{r}{4}\right)^4}
{\Gamma\left(4n+r\right)}
\left(-2^{12} \psi^4\right)^{n+\frac{r-1}{4}} 
 \sum_m \frac{\Gamma \left(m + \frac{n}{2} +
	\frac{r}{8}+\frac{1}{2}\right)^2}{
	\Gamma \left(n +\frac{r}{4}\right)
	\Gamma \left(2m +2 \right)
}
\left(2 \phi\right)^{2m+1}.
\end{split}
\label{eqn:lgVarpiOdd}
\end{align} 
We apply the following transformations
\begin{align}
 (\ref{eqn:lgVarpiEv})&
 \begin{cases}
 k = m + \frac{n}{2} & n \in 2 \mathbb{Z} \\
  k = m + \frac{n+1}{2} & n \in 2 \mathbb{Z} +1
 \end{cases},
  & (\ref{eqn:lgVarpiOdd})&
 \begin{cases}
 k = m + \frac{n}{2} & n \in 2 \mathbb{Z}\\
 k = m + \frac{n-1}{2} & n \in 2 \mathbb{Z} +1
 \end{cases}.
\end{align}
Observe that we performed a shift by an integer so that the limits of summation are not affected.
By identifying 
\begin{align}
e^{\pt_1}&= -2^{11} \psi^4 \phi^{-1} \\
e^{\pt_2} &= 2^2 \phi^2,
\end{align}
it follows that (\ref{eqn:p112228IwitVarPis}) can 
be written as 
\begin{align}
\begin{split}
I_{LG}(u) &= \sum_{\delta=1}^3 \left[
(-1)^{\delta+1} e_\delta
I_{\delta,0}(\pt_1,\pt_2) 
+ 
(-1)^{\delta} e_{\delta+4} I_{\delta,1}(\pt_1,\pt_2)\right],
\end{split}
\label{eqn:p112228LgIfuncEval}
\end{align}
where (\ref{eqn:p112228Ifunc}) was inserted. 
Next we evaluate (\ref{lggamma}):
\begin{align}
\widehat\Gamma_\delta
&=\Gamma \left(1- \left\langle -\frac{k_1+1}{4}\right\rangle\right)^3
\Gamma \left(1- \left\langle -\frac{k_1+1}{8}-\frac{k_2}{2}\right\rangle\right)^2.
\end{align} 
We apply the reparameterization
 \begin{align}
k_1 &= 4n+r-1 & r&= 1, \dots, 4 & k_2 &= 2m+s & s&=0,1,
\end{align}
 given in \cite{Knapp:2020oba} 
 to get 
\begin{align}
\widehat\Gamma_\delta
&=\Gamma \left(1- \left\langle -\frac{r}{4}\right\rangle\right)^3
\Gamma \left(1- \left\langle -\frac{n+s}{2}-\frac{r}{8}\right\rangle\right)^2,
\end{align} 
where we dropped integer shifts from \(\left\langle \cdot \right\rangle\). 
Next we split the above
formula into two contributions with either
 \(n+s \in 2 \mathbb{Z}\) or not:
\begin{align}
\widehat\Gamma_\delta
&=
\begin{cases}
\Gamma \left(1- \left\langle -\frac{r}{4}\right\rangle\right)^3
\Gamma \left(1- \left\langle -\frac{r}{8}\right\rangle\right)^2
& n+s \in 2 \mathbb{Z}
\\
\Gamma \left(1- \left\langle -\frac{r}{4}\right\rangle\right)^3
\Gamma \left(1- \left\langle -\frac{1}{2}-\frac{r}{8}\right\rangle\right)^2
& n+s  \in 2 \mathbb{Z}+1
\end{cases}.
\end{align}
 We focus on the narrow state space where
 \begin{align}
\left\langle \frac{r}{4}\right\rangle &\not= 0,
&  \left\langle \frac{r}{8}\right\rangle &\not = 0,
&  \left\langle \frac{r}{2}+\frac{r}{8}\right\rangle &\not = 0.
\end{align}
 It follows that we can write 
 \begin{align}
\widehat\Gamma_\delta
&=\begin{cases}
\Gamma \left(\left\langle \frac{r}{4}\right\rangle\right)^3
\Gamma \left( \left\langle \frac{r}{8}\right\rangle\right)^2
& n+s \in 2 \mathbb{Z}
\\
\Gamma \left( \left\langle \frac{r}{4}\right\rangle\right)^3
\Gamma \left( \left\langle \frac{4+r}{8}\right\rangle\right)^2
& n+s \in 2 \mathbb{Z}+1
\end{cases}.
\label{eqn:p112228GammaEval}
\end{align}
By the same steps we can rewrite (\ref{lggammaCon}) as
\begin{align}
\widehat\Gamma^\ast_\delta &=\begin{cases}
\Gamma\left(\left\langle\frac{4-r}{4}
\right\rangle\right)^3
\Gamma\left(\left\langle\frac{8-r}{8} \right\rangle\right)^2
& n+s \in 2 \mathbb{Z}
\\
\Gamma\left(\left\langle\frac{4-r}{4}
\right\rangle\right)^3
\Gamma\left(\left\langle\frac{4-r}{8} \right\rangle\right)^2
& n+s \in 2 \mathbb{Z}+1
\end{cases},
\label{eqn:p112228GammaEvalCon}
\end{align} 
and (\ref{lgGrading}) as 
\begin{align}
\operatorname{Gr} &=
\begin{cases}
r- 3 \left\lfloor\frac{r}{4} \right\rfloor- 2 \left\lfloor \frac{r}{8} \right\rfloor 
& n+s \in 2 \mathbb{Z}
\\
r+1- 3 \left\lfloor\frac{r}{4} \right\rfloor- 2 \left\lfloor \frac{4+r}{8} \right\rfloor
& n+s \in 2 \mathbb{Z}+1
\end{cases}.
\label{eqn:p112228GrEval}
\end{align} 
Inserting (\ref{eqn:p112228LgIfuncEval}), (\ref{eqn:p112228GammaEval}),
 (\ref{eqn:p112228GammaEvalCon}) and (\ref{eqn:p112228GrEval})
 into (\ref{lgproposal}) gives
\begin{align}
\begin{split}
Z_{S^2}^{LG}&= \sum_{\delta,\delta^\prime=1}^3\left(
(-1)^{\delta- 3 \left\lfloor\frac{\delta}{4} \right\rfloor
	- 2 \left\lfloor \frac{\delta}{8} \right\rfloor}
\frac{
	\Gamma \left(\left\langle \frac{\delta}{4}\right\rangle\right)^3
	\Gamma \left( \left\langle \frac{\delta}{8}\right\rangle\right)^2
}
{
	\Gamma\left(\left\langle\frac{4-\delta}{4}
	\right\rangle\right)^3
	\Gamma\left(\left\langle\frac{8-\delta}{8} \right\rangle\right)^2
}
(-1)^{\delta+1}I_{\delta,0}(\bar{t_1}, \bar{t_2})
\left\langle e_{\delta^{-1}} \right| \right. \\
&\left.+
(-1)^{\delta+1- 3 \left\lfloor\frac{\delta}{4} \right\rfloor
	- 2 \left\lfloor \frac{4+\delta}{8} \right\rfloor}
\frac{
	\Gamma \left( \left\langle \frac{\delta}{4}\right\rangle\right)^3
	\Gamma \left( \left\langle \frac{4+\delta}{8}\right\rangle\right)^2
}
{
	\Gamma\left(\left\langle\frac{4-\delta}{4}
	\right\rangle\right)^3
	\Gamma\left(\left\langle\frac{4-\delta}{8} \right\rangle\right)^2
}
(-1)^{\delta} I_{\delta,1}(\bar{t_1}, \bar{t_2})\left\langle
e_{(\delta+4)^{-1}} \right| \right) \\
&\quad \cdot \left( (-1)^{\delta^\prime+1}
I_{\delta^\prime,0}(t_1, t_2)
\left| e_{{\delta^\prime}} \right\rangle +
(-1)^{\delta^\prime} I_{\delta^\prime,1}(t_1, t_2) \left|
e_{(\delta^\prime+4)} \right\rangle\right).
\end{split}
\end{align} 
By (\ref{lgpairing}) 
the above results
give for the sphere partition function: 
\begin{align}
Z_{S^2}^{LG}&= \frac{1}{8} \sum_{\delta=1}^3\left(
(-1)^{\delta- 3 \left\lfloor\frac{\delta}{4} \right\rfloor
	- 2 \left\lfloor \frac{\delta}{8} \right\rfloor}
\frac{
	\Gamma \left(\left\langle \frac{\delta}{4}\right\rangle\right)^3
	\Gamma \left( \left\langle \frac{\delta}{8}\right\rangle\right)^2
}
{
	\Gamma\left(\left\langle\frac{4-\delta}{4}
	\right\rangle\right)^3
	\Gamma\left(\left\langle\frac{8-\delta}{8} \right\rangle\right)^2
}
I_{\delta,0}(\bar{t_1}, \bar{t_2})I_{\delta,0}(t_1, t_2) \right. \nonumber  \\
&\left.+
(-1)^{\delta+1- 3 \left\lfloor\frac{\delta}{4} \right\rfloor
	- 2 \left\lfloor \frac{4+\delta}{8} \right\rfloor}
\frac{
	\Gamma \left( \left\langle \frac{\delta}{4}\right\rangle\right)^3
	\Gamma \left( \left\langle \frac{4+\delta}{8}\right\rangle\right)^2
}
{
	\Gamma\left(\left\langle\frac{4-\delta}{4}
	\right\rangle\right)^3
	\Gamma\left(\left\langle\frac{4-\delta}{8} \right\rangle\right)^2
}
I_{\delta,1}(\bar{t_1}, \bar{t_2}) I_{\delta,1}(t_1, t_2) \right) \nonumber \\
&= \frac{1}{8} \sum_{\kappa =0}^1 \sum_{\delta=1}^3
\left(
(-1)^{\delta+\kappa- 3 \left\lfloor\frac{\delta}{4} \right\rfloor
	- 2 \left\lfloor \frac{4\kappa+\delta}{8} \right\rfloor}
\frac{
	\Gamma \left( \left\langle \frac{\delta}{4}\right\rangle\right)^3
	\Gamma \left( \left\langle \frac{4\kappa+\delta}{8}\right\rangle\right)^2
}
{
	\Gamma\left(\left\langle\frac{4-\delta}{4}
	\right\rangle\right)^3
	\Gamma\left(\left\langle\frac{8-4\kappa-\delta}{8} \right\rangle\right)^2
}
I_{\delta,\kappa}(\bar{t_1}, \bar{t_2}) I_{\delta,\kappa}(t_1, t_2) \right).
\label{eqn:p112228LgZS2eval}
\end{align}
So (\ref{eqn:p112228LgZS2eval}) matches 
the GLSM result (\ref{eqn:p112228zs2final}).

\bibliographystyle{utphys}
\bibliography{zs2phases}
\end{document}